\documentclass[prx,twocolumn,superscriptaddress,showpacs,nobibnotes,notitlepage,nofootinbib]{revtex4-2}
\usepackage[bb=boondox]{mathalfa}
\usepackage{amsfonts}
\usepackage{amsmath}
\usepackage{amsthm}
\usepackage{braket}
\usepackage{graphics}
\usepackage{hyperref}
\usepackage{bbm}
\usepackage[dvipsnames]{xcolor}
\usepackage{boxhandler}
\usepackage{amsmath,amscd}
\usepackage[all,cmtip]{xy}
\usepackage{multirow}

\usepackage{etoolbox,tikz, tikz-cd}
\usetikzlibrary{decorations.pathmorphing}
\usetikzlibrary{cd}
\tikzset{fontscale/.style = {font=\relsize{#1}}
    }
\usepackage{overpic}
\usepackage{mathtools}
\usepackage{adjustbox}

\usepackage[capitalize]{cleveref}

\Crefname{enumi}{Case}{Cases}
\Crefname{subsection}{Subsection}{Subsections}

\definecolor{darkblue}{RGB}{0,0,127} 
\definecolor{darkgreen}{RGB}{0,130,80}
\definecolor{darkred}{RGB}{150,10,10}
\hypersetup{
	colorlinks,
	linkcolor=darkblue,
	citecolor=darkgreen,
	filecolor=red,
	urlcolor=blue,
	pdftitle={Topological defect networks for fracton stabilizer codes},
	pdfauthor={Zijian Song, Arpit Dua, Wilbur Shirley, and Dominic J. Williamson}
}

\graphicspath{{./Figures/}}
\usepackage{tikz,tikz-3dplot,pgfplots}\pgfplotsset{compat=newest}
\usetikzlibrary{external,calc,decorations.pathreplacing,decorations.markings,decorations.pathmorphing,arrows.meta,shapes.geometric}
\usepackage{pgfplots}\pgfplotsset{compat=newest}

\newlength\figureheight
\newlength\figurewidth

\usepackage{graphicx}
\usepackage{tabularx}
\usepackage{booktabs}
\usepackage{array}
\usepackage{amsmath}
\usepackage{amsfonts}
\usepackage{amssymb}
\usepackage{amsthm}
\usepackage{enumitem}
\usepackage{permute}
\usepackage{url}
\usetikzlibrary{shapes}

\DeclareFontFamily{U}{mathb}{\hyphenchar\font45}
\DeclareFontShape{U}{mathb}{m}{n}{
      <5> <6> <7> <8> <9> <10> gen * mathb
      <10.95> mathb10 <12> <14.4> <17.28> <20.74> <24.88> mathb12
      }{}
\DeclareSymbolFont{mathb}{U}{mathb}{m}{n}
\DeclareMathSymbol{\bigast}{2}{mathb}{"06}

\def\XXint#1#2#3{{\setbox0=\hbox{$#1{#2#3}{\int}$}
     \vcenter{\hbox{$#2#3$}}\kern-.5\wd0}}

\theoremstyle{plain}

\theoremstyle{definition}

\newtheoremstyle{remark}
{}   % ABOVESPACE, \topsep for no space
{}   % BELOWSPACE, \topsep for no space
{\normalfont}  % BODYFONT
{}       % INDENT (empty value is the same as 0pt)
{\itshape} % HEADFONT
{.}         % HEADPUNCT
{5pt plus 1pt minus 1pt} % HEADSPACE
{}          % CUSTOM-HEAD-SPEC

\theoremstyle{remark}

\newlist{alternative}{enumerate}{4}     % this creates a dedicated counter named 'subtaski'
\setlist[alternative,1]{label=\arabic*., ref=\arabic*}
\setlist[alternative,2]{label=(\alph*), ref=\thealternativei.(\alph*)}
\setlist[alternative,3]{label=\roman*., ref=\thealternativei.(\thealternativeii).\roman*}
\setlist[alternative,4]{label=\Alph*., ref=\thealternativei.(\thealternativeii).\thealternativeiii.\Alph*}

\setlist[enumerate,1]{label=\arabic*., ref=\arabic*}
\setlist[enumerate,2]{label=(\alph*), ref=\theenumi.(\alph*)}
\setlist[enumerate,3]{label=\roman*., ref=\theenumi.(\theenumii).\roman*}
\setlist[enumerate,4]{label=\Alph*., ref=\theenumi.(\theenumii).\theenumiii.\Alph*}

\AtBeginEnvironment{tikzcd}{\tikzexternaldisable}
\AtEndEnvironment{tikzcd}{\tikzexternalenable}
\newcommand{\drawgenerator}[8]{%
\xymatrix@!0{%
& #8 \ar@{-}[ld]\ar@{.}[dd] \ar@{-}[rr] & & #7 \ar@{-}[ld]  \\%
#1 \ar@{-}[rr] \ar@{-}[dd] &  & #2 \ar@{-}[dd] &            \\%
& #6 \ar@{.}[ld] &  & #5 \ar@{-}[uu] \ar@{.}[ll]       \\%
#3 \ar@{-}[rr] &  & #4 \ar@{-}[ru]                       %
}%
}

\newcommand{\plaquette}[4]{
\xymatrix@!0{%
#1 \ar@{-}[r] \ar@{-}[d]  & #2 \ar@{-}[d] 
\\
#3 \ar@{-}[r]  & #4
}}

\renewcommand{\cal}[1]{\mathcal{#1}}

\definecolor{zijian}{RGB}{0,0,127}

\definecolor{dom}{RGB}{0,127,127}

\definecolor{arpit}{RGB}{127,0,0}

\usepackage{floatrow}
\usepackage[caption=false,label font={bf,normalsize}]{subfig}
\begin{document}
 \title{
Topological defect network representations of fracton stabilizer codes}
\author{Zijian Song}
\email{zjsong@ucdavis.edu}
\affiliation{Department of Physics and Astronomy, University of California, Davis, California 95616, USA}

\author{Arpit Dua}
\affiliation{Department of Physics and Institute for Quantum Information and Matter, California Institute of Technology, Pasadena, California 91125, USA}

\author{Wilbur Shirley}
\affiliation{School of Natural Sciences, Institute for Advanced Study, Princeton, NJ 08540, USA}
\affiliation{Department of Physics and Institute for Quantum Information and Matter, California Institute of Technology, Pasadena, California 91125, USA}

\author{Dominic~J. Williamson}
\thanks{Current Address: Centre for Engineered Quantum Systems, School of Physics,
University of Sydney, Sydney, New South Wales 2006, Australia}
\affiliation{Stanford Institute for Theoretical Physics, Stanford University, Stanford, CA 94305, USA}

\begin{abstract}
\noindent
A topological defect network (TDN) is formed by a network of topological defects embedded within a topological quantum field theory (TQFT). TDNs were introduced recently for the purpose of describing fracton topological phases of matter using the framework of defect TQFT. 
Their effectiveness has been demonstrated through numerous examples, yet a systematic construction was lacking. 
Here we solve this problem by formulating a method to construct TDNs for a wide range of lattice Hamiltonians. 
Our method takes a lattice Hamiltonian as input, applies an ungauging procedure, then creates a refined lattice within each unit cell, followed by regauging the system to produce a TDN as output. 
For topological Calderbank-Shor-Steane (CSS) Pauli stabilizer models, this procedure is guaranteed to produce a phase equivalent TDN. 
This provides TDN representations of canonical fracton models for which no such construction was previously known, including Haah's cubic code and Yoshida's infinite family of fractal spin liquid models. 
We demonstrate the applicability of our method beyond CSS stabilizer models by constructing TDNs for non-CSS models including Chamon's model and the semionic X-cube model. 
\end{abstract}

  \maketitle

% \tableofcontents\thispagestyle{fancy}
\tableofcontents

% \noindent\rule{\textwidth}{1pt}
% \vspace{10pt}
% \newpage

\section{Introduction}

The classification of all phases of matter is central to condensed matter theory. Topological phases~\cite{wegner1971duality,Anderson1973,PhysRevLett.50.1395,PhysRevB.40.7387,WEN1990} form a particularly interesting class of strongly correlated zero temperature quantum phases of matter that have drawn much attention in recent decades due in part to their applications as quantum error correcting codes~\cite{qdouble}. 
In two spatial dimensions, the classification problem for topological phases is elegantly solved by the framework of TQFT~\cite{Witten1988,atiyah1988topological}. 
For a particular class of topological stabilizer code Hamiltonians, a full constructive classification has been found~\cite{Haah2018a}. 
In three spatial dimensions, fracton topological phases~\cite{chamon2005quantum,haah2011local,kim20123d,Chamon_quantum_glassiness,yoshida2013exotic,new_TO_vijay,vijay2016fracton,Williamson_cubic_code} have raised a challenge to this classification program as they are topologically ordered and yet do not yield to a conventional description via TQFT due to an essential interplay of topology and geometry~\cite{shirley2017fracton}. 
These fracton phases have attracted much interest again due to their applications in quantum error correction~\cite{bravyi2011topological,PhysRevLett.107.150504,bravyi2013quantum,Brown2019}, as well as their relation to other areas of condensed matter physics~\cite{PhysRevLett.116.027202,PhysRevB.95.155133,PhysRevLett.120.195301,gromov2017fractional,Doshi2020}, and the unusual quantum field theories that they lead to~\cite{PhysRevB.95.115139,PhysRevB.96.195139,Gromov2018,Slagle2018foliated,Seiberg2019,Seiberg2020a,Seiberg2020,Slagle2021}. 

Recently, an approach based on a network of topological defects within a TQFT, i.e. a TDN, was proposed~\cite{Aasen2020} to extend the descriptive power of TQFT from conventional topological phases to also include fracton topological phases, thus uncovering a possible path towards their full classification. 
The central conjecture in Ref.~\cite{Aasen2020} stipulates that TDNs are capable of describing all gapped phases of matter. 
Here we address a slightly more limited version of this conjecture for gapped phases with superselection sectors that have finite order under fusion, in particular we expect this to cover all topological commuting projector models based on finite dimensional qudit degrees of freedom. We remark that interesting examples violating the above condition have been proposed in systems based on infinite dimensional degrees of freedom~\cite{Ma2020}. 

\subsection{Main results and ideas}
Our main result is a procedure to construct a phase equivalent TDN given an arbitrary topological CSS Pauli stabilizer model. 
This construction confirms the conjecture made in Ref.~\cite{Aasen2020} for the large class of fracton topological orders that are based on topological CSS~\cite{PhysRevA.54.1098,Steane2551} stabilizer codes~\cite{gottesman1997stabilizer}. Hamiltonians based on stabilizer codes form the largest class of known fracton models~\cite{Dua_Classification_2019} and the one that has received the most extensive study due to applications as quantum memories and the ease of relevant calculations~\cite{brown2014quantum}. 
In particular our construction produces a TDN that is equivalent to Haah's cubic code~\cite{haah2011local}, the canonical example of a fracton code with no topological string operators. 
In addition, the construction provides TDNs for Yoshida's infinite family of fractal spin liquid models~\cite{yoshida2013exotic}, beyond the special case considered in Ref.~\cite{Aasen2020}. 
To the best of our knowledge no TDN representations were previously known for these models. We attribute this to the lack of a general prescription for constructing a TDN representation given a topological lattice model, a problem that is solved in this work for topological CSS stabilizer models.

Previously TDNs were constructed on an ad hoc basis by utilizing defects to mimic the elementary neutral topological charge clusters of a desired model. 
In the constructions of Ref.~\cite{Aasen2020} cubes of gauge theory with flux condensing boundaries were used as they support pointlike topological charges. 
These cubes were coupled together via topological defect lines that allowed condensation of elementary neutral clusters of topological charges isomorphic to the elementary charge clusters of target fracton models. 
Here, we extend the idea behind this previous construction into a systematic recipe that produces a TDN in the same quantum phase of matter as an arbitrary topological CSS stabilizer Hamiltonian. 

Our construction procedes as follows: first we utilize the generalized gauging and ungauging procedure for subsystem symmetries to relate a fracton CSS stabilizer model to an unconventional Ising model, following Refs.~\cite{vijay2016fracton,Williamson_cubic_code}. Each spin in the unconventional Ising model is then encoded into a large block of an auxiliary conventional Ising model, i.e. a repetition code. 
This step is reminiscent of reversing Kadanoff’s block-spin renormalization procedure~\cite{Kadanoff1966}. 
We then use a phase equivalence relation to rewrite the original unconventional Ising Hamiltonian interactions in terms of local interactions on the new fine-grained lattice. 
Finally, we (re)gauge  the subsystem symmetry of the fine-grained unconventional Ising model thus produced to arrive at a TDN with the desired locality properties. 
We show that this sequence of ungauging, fine-graining, and regauging provides a phase equivalence between the original fracton topological CSS code Hamiltonian and the new TDN Hamiltonian. 
We also show that the TDN is stable provided that the original fracton topological Hamiltonian has been sufficiently coarse-grained. 

TDNs allow the microscopic lattice scale to be decoupled from the lattice scale of the defect network itself. 
In particular, if the fine-graining step is iterated indefinitely, the microscopic scale can be taken to zero allowing a TDN construction of any CSS fracton Hamiltonian in terms of continuum TQFT containing a network of defects. 
It is an interesting problem to additionally consider the continuum limit of the defect network, which we leave to future work. 

The idea underlying our TDN procedure extends to any fracton topological order that is constructed by gauging a subsystem symmetry acting on layers of TQFTs. To the best of our knowledge such a construction recovers almost all known fracton topological orders with finite order excitations. From this point of view, the case of fracton CSS stabilizer codes we focus on in this work essentially reduces to considering the trivial TQFT being acted upon by subsystem symmetries. 
We also provide a more complicated example where a TDN for the earliest known fracton model, due to Chamon~\cite{chamon2005quantum}, is constructed by gauging a nontrivial subsystem symmetry-protected order. 
While our approach applies quite generally, we leave the question of finding a proof that the TDNs resulting from our procedure applied to more general initial models are in fact phase equivalent to them to future work. 

\subsection{Outline}

The paper is organized as follows. 
In Sec~\ref{background}, we introduce background on topological defect networks and gauging global and subsystem symmetries in spin models, including examples. 
In Sec~\ref{sec:XcubeExample}, we review the TDN construction for the X-cube model~\cite{vijay2016fracton} before rederiving it following our ungauging approach. 
In Sec~\ref{TDNHaah}, we provide a TDN construction for Haah's code (i.e. cubic code A~\cite{haah2014bifurcation}) by following our ungauging approach. 
In Sec~\ref{generalapproach}, we discuss our general approach for constructing TDNs for topological CSS stabilizer models. 
In Sec~\ref{nonCSSTDN}, we demonstrate that this approach extends beyond CSS stabilizer models by constructing a TDN for Chamon's fracon model. 
In Sec~\ref{sec:Conclusion}, we conclude with a discussion about constructing TDNs beyond stabilizer models and in higher dimensions.

\section{Background} \label{Gauging1}
\label{background}

In this section we introduce TDNs~\cite{Aasen2020} and the generalized gauging procedure for subsystem symmetries~\cite{vijay2016fracton,Williamson_cubic_code}. 

\subsection{Topological defect networks}

TDNs were recently introduced~\cite{Aasen2020} to extend the framework of defect TQFT~\cite{carqueville20163} to describe lattice models with fracton topological order (see also Refs.~\cite{Wen2020cellular,Wang2020cellular}). This framework effectively separates the microscopic lattice scale from a potentially larger lattice scale of the defect network itself. In particular, a continuum limit of the microscopic lattice can be taken with the defect network fixed to arrive at a description in terms of defects in a TQFT where the microscopic lattice has been abstracted away. 
It was conjectured~\cite{Aasen2020} that the TDN construction is sufficiently general to describe all gapped phases of matter. Here we add the proviso that the gapped phases being described must have finite order under fusion. 
In particular, the TDN framework straightforwardly contains the classification of 2D gapped quantum phases via TQFT. 
This provides an approach to the seemingly intractable problem of constructing and classifying all topological phases of matter in higher dimension in terms of the more familiar problem of understanding TQFTs and the topological defects they support. 

In order to understand TDNs, we first describe a topological defect. In a $D$+1 dimensional TQFT, one can introduce new degrees of freedom and local interactions on a $d$+1 dimensional sub-manifold with $d<D$ without closing the energy gap~\cite{bombin2010topological,beigi2011quantum,Kitaev2012}. One then asks how the topological excitations of the bulk topological order behave near this altered sub-manifold i.e. the defect. The defect is topologically nontrivial if topological excitations are either condensed or permuted when they pass through the defect.  

\begin{figure}
    \centering
    \includegraphics[width=8cm]{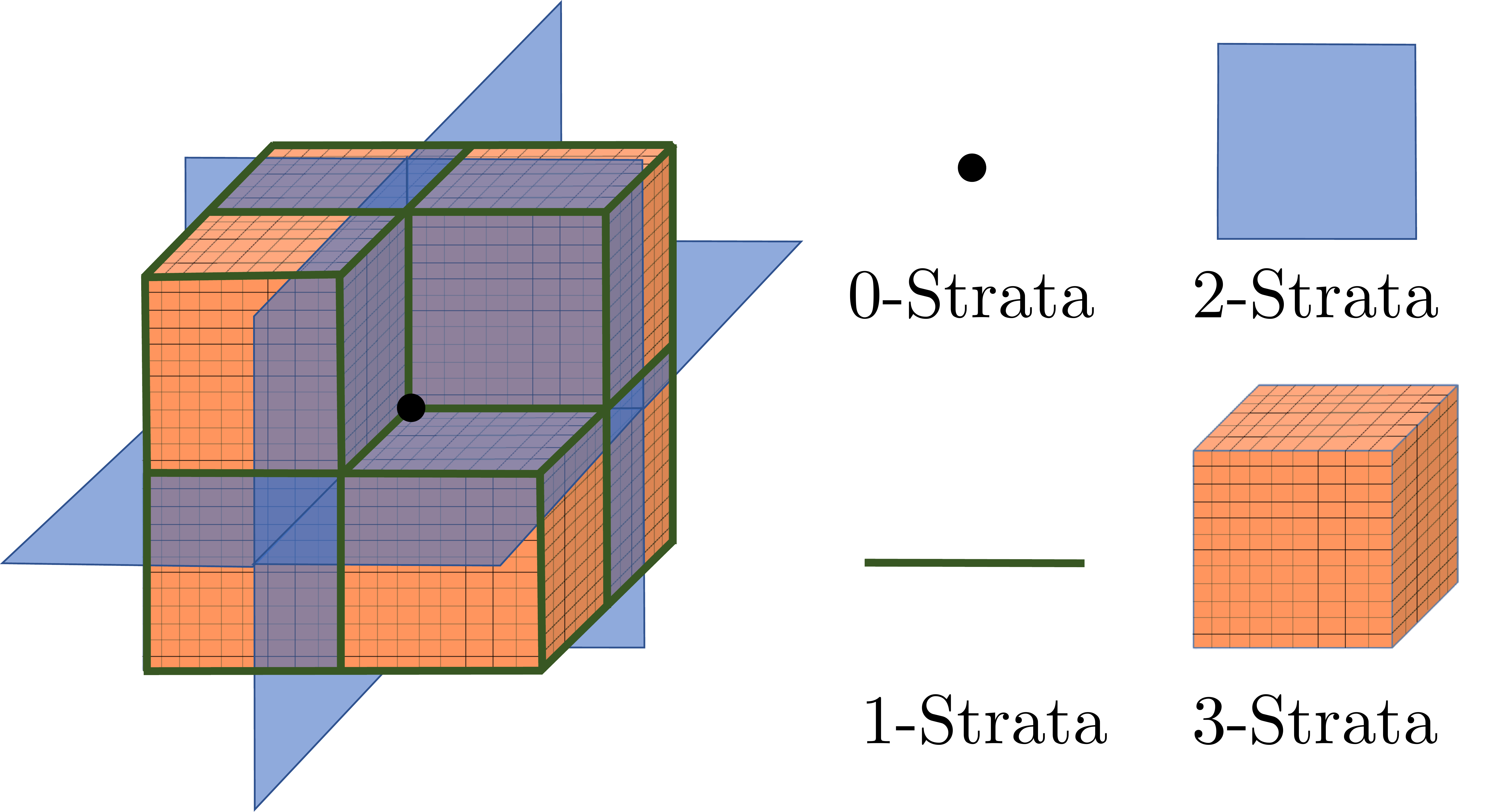}
    \caption{A stratification of 3 dimensional space. The space is stratified into 3-strata (orange blocks), 2-strata (blue planes), 1-strata (magenta lines) and 0-strata (black dots). The black thin lines show the finer lattice inside the 3-strata. 3+1D TQFT lives on each 3-stratum. 3-strata are coupled together by lower dimensional defects (2-strata, 1-strata, 0-strata).
    }
    \label{Strata_2}
\end{figure}

In this work, we consider TDNs in three dimensions. In order to define a general TDN in $\mathbb{R}^3$, we consider a stratification of the 3-dimensional space into $j$-dimensional submanifolds called $j$-strata where $j\leq 3$. 
We illustrate such a stratification in Fig.\ref{Strata_2} for the case of cubic lattice in $\mathbb{R}^3$. 
Now, we put an arbitrary 3+1D TQFT on each of the 3-strata. Then, we couple together these TQFTs via topological defects on the $j$-strata where $j\leq 2$. The local interactions on the topological defects sitting on the $2$-strata allow certain excitations of the 3+1D TQFTs neighboring it to condense. 
When a subset of the 3+1D TQFT condenses on the 2-strata, the excitations that braid nontrivially with this subset are confined as they cannot pass through the 2-strata to the neighboring 3-strata. Similarly, one can introduce local interactions on the 1-strata and 0-strata corresponding to topological defects there. 

As a consequence of the condensation and confinement of topological excitations in the vicinity of topological defects, excitations and their composites can be restricted to move only in certain submanifolds on the scale of the defect network. 
In fact, one can write down TDNs in the same phase as gapped fracton models~\cite{Aasen2020}. 
A simple example is given by a TDN that realizes a phase equivalent to that of the X-Cube model. The TDN for X-cube, defined on a cubic lattice in $\mathbb{R}^3$ uses 3+1D toric codes with flux-condensing gapped boundaries on the 2-strata. 
An appropriate choice of 1-strata imposes mobility constraints on the topological excitations of the 3+1D toric code such that the $e$ particle on a 3-strata maps to the fracton of the X-cube model and an arc of the $m$-loop maps to the lineon of the X-cube model, see Section~\ref{sec:XcubeExample} for further details.

As mentioned in the introduction, our construction of TDNs for gapped fracton phases starts from a defect network for an Ising paramagnet with certain subsystem symmetries and gauges it. Hence, we provide some background on gauging spin models below. 

\subsection{Gauging spin models}
\label{GaugingSpinModels}

In this subsection we review the construction of generalized Ising models with subsystem symmetries, and the generalized gauging procedure for these models. 
\\

\noindent\textit{Subsystem symmetry. } We first define three sets $C$, $M$ and $E$ which correspond to a set of constraint labels, matter qubits, and excitations, respectively. We denote the elements of these sets by $c \in C$, $m \in M$ and $e \in E$. We define $P(M)$ to be the Pauli group on $M$, which is given by all products of Pauli $X$ and Pauli $Z$ operators acting on the matter qubits $M$, working in a basis where
\begin{align}
    X= \left(\begin{array}{cc}
         0 & 1\\
         1 & 0
    \end{array}\right) \qquad Z=\left(\begin{array}{cc}
         1 & 0\\
         0 & -1
    \end{array}\right) . 
\end{align} 
A generating set for $P(M)$ is
\begin{align}
    P_{g}(M) = \{X \otimes \mathbb{I}..., \mathbb{I} \otimes X..., Z \otimes \mathbb{I}..., \mathbb{I} \otimes Z...\}.
\end{align}
Each generator has a non-trivial Pauli $X$ or $Z$ operator on one of the qubits. The constraints are defined via the map
\begin{align}
    \text{con}&: C \rightarrow P(M) \label{con},
\end{align}
i.e. con is a map from the labels $C$ to a set of Pauli terms that generate a constraint algebra. 
Similarly, the set of excitations is defined by a map
\begin{align}
     \text{exc}&: P(M) \rightarrow E, \label{exc}
\end{align}
where exc maps a Pauli operator in $P(M)$ to the labels of the constraint terms that it anti-commutes with. 
Here, we consider $X$-type global (subsystem) symmetries $S^{X}_{i} \in X(M)$ and $Z$-type constraints $\text{con}(c) \in Z(M)$.\footnote{
This covers all Pauli symmetries that are nonanomalous and can hence be gauged~\cite{vijay2016fracton,Williamson_cubic_code} To see this consider the restriction of Pauli symmetries to single sites; if all such single site restricted symmetry actions commute we can change on-site basis to bring these actions into the form of $X$ operators. If some actions do not commute this indicates an anomaly of the symmetry combined with the translation group. In this case we assume coarse graining can be performed to restore on-site commutativity on a coarser lattice scale, otherwise the anomaly can not be removed by explicitly breaking the translation group to a subgroup.}. 
We use $X(M)$ to denote the $X$-sector of $P(M)$, i.e. operators that are generated by all $X$-type generators in $P_g(M)$, and similarly for $Z(M)$. 
In our approach the constraint algebra is used to define the symmetry via the condition that all constraints must commute with the full symmetry group 
\begin{align}
    \left[ \text{con}(c) , S^{X}_{i} \right] = 0. \label{symmetry}
\end{align}
Hence $\text{exc}$ maps $S^{X}_{i}$ to the empty set. 
Consequently, the kernel of the exc map from $X(M)$ to $C$ coincides with the symmetry group. 
In particular, conventional global and subsystem symmetries can be contained in the kernel of exc map. 
\begin{align}
    S^{X}_{i} \in X(\text{ker} \ \text{exc}(m)) \label{globalsymmetry1}
\end{align}

In general, the con map can have a non-trivial kernel also. The elements in the kernel of con define relations. 
The relations of the constraint map satisfy
\begin{align}
    \prod_{c \in r} Z(\text{con}(c)) = \mathbb{I}_M, \label{zrelation1} 
\end{align}
in which $r \in \mathrm{ker}\ \mathrm{con}$. 
We further introduce a map rel, from a new space $R$ to $C$, whose image is the part of the kernel that is generated by elements of finite order (as the system size diverges).  These relations are described by a chain complex,
\begin{align}
     R \xrightarrow{\quad \text{rel} \quad} C \xrightarrow{\quad \text{con} \quad} Z(M). \label{chaincomplex}
\end{align}
The image of the $Z$-relation map are in the kernel of the con map.

It is convenient to introduce a further chain complex, resolving the part of the kernel of exc that corresponds to finite weight symmetry operators. We use $L$ to denote a set of labels for generators of the finite weight symmetry operators (in the limit of infinite system size). We then have the following exact sequence
\begin{align}
    L  \xrightarrow{\quad \text{gen} \quad} X(M) \xrightarrow{\quad \text{exc} \quad} C.
\end{align}
For most of the examples we consider there are no nontrivial symmetry generators of finite weight, rather all symmetry operators have a weight that grows with the size of the system. 
\\

\noindent\textit{Generalized Ising Hamiltonians. }
We now introduce a family of generalized Ising Hamiltonians on the matter qubits, with local terms given by Pauli $X$ operators on single qubits and $Z$-type constraint generators over multiple qubits. 
\begin{align}
    H(J) = - \sum_{c \in C} Z(\mathrm{con}(c)) -J \sum_{m} X(m) .
\end{align}
This model realizes a trivial disordered phase in the limit $J \rightarrow \infty$, and a nontrivial ordered phase in the limit $J \rightarrow 0$ (provided the constraint terms are nontrivial). 
The ordered phase may be classical i.e. a diagonal Hamiltonian, or it may be quantum in which case an algebra of Pauli $X$ terms are generated at finite order in perturbation theory for $J \ll 1$~\cite{Williamson2020a}. 
The Hamiltonian $H(J \ll 1)$ lies in the same topological phase as a Hamiltonian including the constraint terms, along with a generating set of the finite weight symmetry operators
\begin{align}
    H(J \ll 1) \sim - \sum_{c \in C} Z(\mathrm{con}(c)) - \sum_{l \in L} X(\mathrm{gen}(l)) \, ,
\end{align}
where the energies have been rescaled and only a generating set of terms included in the Hamiltonian, both of which are gapped zero temperature quantum phase equivalences~\cite{Haah2013}. 
In the case of a classical Hamiltonian in the $J\rightarrow 0$ limit, assuming the constraint algebra defines a nontrivial global symmetry group, the resulting phase spontaneously breaks the symmetry group as there exist single $Z$ operators (which are charged under the symmetry) that act nontrivially within the ground space. 
Conversely, it is possible to obtain topological phases where generators for the algebra of finite order symmetry operators are present in the Hamiltonian, and prevent the existence of any low weight $Z$ operators that can act nontrivially within the ground space. In general there may be a combination of both topological and spontaneous symmetry broken order. 
\\

\noindent\textit{Generalized gauging map. } 
Gauging is a bijective, isometric duality map from wavefunctions with global symmetries to wavefunctions with gauge (local) symmetries~\cite{vijay2016fracton,Williamson_cubic_code,kubica2018ungauging,shirley2018FoliatedFracton}\footnote{Here, we consider infinite systems to bypass subtle boundary issues, and note that the duality only holds between symmetric subspaces.}. 
We define a new set $G$ that labels gauge qubits. These auxiliary qubits are introduced as part of the gauging prescription. 
The gauging map takes the matter Hilbert space to a gauge invariant subspace of a gauge-matter Hilbert space
\begin{align}
    \cal{G}: P(M) \rightarrow \Pi \big( P(M) \otimes P(G) \big) , 
\end{align}
where $\Pi$ is a projection onto the gauge invariant subspace, see below. 

To gauge the model, we introduce one gauge qubit per constraint. The minimally coupled constraint terms are then given by 
\begin{align}
    \cal{G} \left(Z(\text{con} (c)) \right) = Z(\text{con} (c)) \otimes Z(c). \label{gaugc}
\end{align}
We define a set of local $X$-type gauge symmetry operators that satisfy the gauged version of the commutator in Eq.~(\ref{symmetry}), i.e. 
\begin{align}
    \left[ Z(\text{con} (c)) \otimes Z(c),\ S^{X}_{gauge}\right] = 0. \label{gaugecommu}
\end{align}
By enforcing this commutation relation, the gauge symmetry generators are simply
\begin{align}
    S^{X}_{gauge}(m) = X(m) \otimes X(\text{exc}\ X(m)). \label{gaugs}
\end{align}
The projection onto the gauge invariant subspace is then given by 
\begin{align}
    \Pi := \prod_{m \in M} \frac{1}{2} \big( \mathbbm{1} +  S^{X}_{gauge}(m) \big)
\end{align}
Hence after gauging, the global (subsystem) symmetries are enforced locally. 
These terms are also referred to as generalized Gauss's law terms, as they implement a relation between charges on the matter qubits and a generalized notion of electric flux lines on the surrounding gauge qubits. Throughout the remainder of the text we simply refer to the generalized Gauss's laws as Gauss's laws. 

The $X$ terms in the original Hamiltonian $C_Q$ remain unchanged after gauging. Hence, the gauging map simply acts on them as
\begin{align}
    \cal{G}(X(m)) = X(m).  \label{gaugi}
\end{align} 
Now we see the gauge symmetry operators Eq.~(\ref{gaugs}) commute with all other terms in the gauged Hamiltonian, so the gauged Hilbert space is indeed invariant under the gauge symmetry.

On the other hand, the single body $X$ terms in Eq.~(\ref{gaugi}) still don't commute with the $Z$ terms in Eq.~(\ref{gaugc}). 
This leads to a nontrivial gauged Hamiltonian
\begin{align}
    H(J)_{\text{gauged}}= &- \sum_{c\in C} Z(\text{con} (c)) \otimes Z(c) - J \sum_{m} X(M) 
    \nonumber \\
    &-\Delta \sum_{m} S^{X}_{gauge}(m) \, ,
\end{align}
where the $\Delta$ energetically enforces the gauge constraint, becoming strict as $\Delta\rightarrow \infty$. 
If we take the limit of the coupling strength $J\rightarrow \infty$, all $Z$ operators on matter qubits are expelled from the finite energy subspace. Thus only finite products of gauged $Z$ terms that coincide with the kernel of con survive, which are given by (\ref{zrelation1}). 
Thus, after gauging, finite order terms in the kernel of con map to finite order pure gauge terms in the gauged Hamiltonian, which are generated by 
\begin{align}
    B(r): = \prod_{c \in r} Z(\text{con}(c)) \otimes Z(c) = \prod_{c \in r} Z(c) . \label{puregauge}
\end{align}
As $J\rightarrow \infty$, the matter qubits in $S^{X}_{gauge}(m)$ are fixed, resulting in the pure gauge operator 
\begin{align}
    A(m) : = X(\text{exc}\ X(m)). \label{gaugestarterm}
\end{align}
Hence, in the limit $J\rightarrow \infty$ the gauged Hamiltonian is equivalent to 
\begin{align}
    H(\infty)_{\text{gauged}} \sim - \sum_m A(m) - \sum_{r} B(r) ,
\end{align}
where we have only included a generating set of terms and rescaled the energies as part of the equivalence relation.

In this limit of the pure gauge Hamiltonian, the chain complex from Eq.~(\ref{chaincomplex}) now describes the $X$ and $Z$ type commuting Hamiltonian terms
\begin{align}
     C_Z \xrightarrow{\quad \text{$Z$-relations} \quad} G \xrightarrow{\quad \cal{G}(\text{exc}) \quad} C_X. \label{chaincomplex_2}
\end{align}
$G$ stands for the space of gauge Hamiltonian which is isomorphic to $C$ in Eq.~(\ref{chaincomplex}). $C_Z$ is the space of $Z$-type stabilizers, which is isomorphic to $R$. $C_X$ is the space of $X$-type stabilizers, which is isomorphic to $M$. The image of the $Z$-relation map satisfies Eq.~(\ref{zrelation1}) and Eq.~(\ref{puregauge}). The kernel of the gaug(exc) map satisfies Eq.~(\ref{gaugecommu}) and gives the terms in Eq.~(\ref{gaugs}). In the strong coupling limit, these terms become Eq.~(\ref{gaugestarterm}). Comparing with Eq.~(\ref{chaincomplex}) we can see that gauging corresponds to moving the position of the qubits in the chain complex one step to the left. 

\subsection{Stabilizer formalism} \label{Polynomial}
In this subsection we give a brief introduction to the stabilizer formalism~\cite{gottesman1997stabilizer} and describe the gauging procedure in this language~\cite{Williamson_cubic_code}. We use the computational basis spanned by the eigenstates of $Z$, $|0\rangle$ and $|1\rangle$, which are given by
\begin{align}
    |0\rangle = \left(\begin{array}{c}
         1\\
        0
    \end{array}\right), \quad |1\rangle = \left(\begin{array}{c}
         0\\
        1
    \end{array}\right). 
\end{align}
The actions of Pauli $X$ and $Z$ operators on these states are 
\begin{align}
    &Z |0\rangle = |0\rangle, \quad Z |1\rangle = -|1\rangle, \notag \\ 
    &X |0\rangle = |1\rangle, \quad X |1\rangle = |0\rangle.
\end{align}
Furthermore, we introduce Clifford gates and stabilizer states. The Clifford gates are CNOT gate, Hadamard gate and phase gate module modulo a global $U(1)$ phase on qubits.
\begin{align}
    \mathcal{C} = \{\text{CNOT}_{ij}, H_i, P_i\}/U(1),
\end{align}
in which
\begin{align}
   &\text{CNOT} = \left(\begin{array}{cccc}
         1 & 0 & 0 & 0\\
         0 & 1 & 0 & 0\\
         0 & 0 & 0 & 1\\
         0 & 0 & 1 & 0
    \end{array}\right),\nonumber\\
    &H = \frac{1}{\sqrt{2}} \left(\begin{array}{cc}
         1 & 1\\
         1 & -1
    \end{array}\right),
    \
    P = \left(\begin{array}{cc}
         1 & 0\\
         0 & i
    \end{array}\right). \label{clifford}
\end{align}
A tensor product of Pauli operators $\mathcal{O}_1$ maps to another tensor product of Pauli operators $\mathcal{O}_2$ by conjugation of Clifford gates.
\begin{align}
    \mathcal{C} \mathcal{O}_1 \mathcal{C}^{\dagger} = \mathcal{O}_2.
\end{align}
For example, CNOT gate has the following properties,
\begin{align}
    \mathrm{CNOT}(a,b) X(b) \mathrm{CNOT}^{\dagger} &= X(a) X(b) \notag\\
    \mathrm{CNOT}(a,b) Z(a) \mathrm{CNOT}^{\dagger} &= Z(a) Z(b), \label{cnot2}
\end{align}
in which $a$ is the label of the target qubits and $b$ is the label of the control qubits.

The stabilizer states $|\psi_{S}\rangle$ are generated by a sequence of Clifford gates from a computation basis state. For each stabilizer state, there is an Abelian group $\mathcal{S}$ whose elements $s \in \mathcal{S}$ are the tensor product of Pauli operators satisfying
\begin{align}
    s |\psi_{S}\rangle = |\psi_{S}\rangle. 
\end{align}
$s$ are called stabilizers and $\mathcal{S}$ is called stabilizer group. The generators of the stabilizer group can be used to define the terms in a Hamiltonian. 
Thus the corresponding stabilizer Hamiltonian is given by
\begin{align}
    H = - \sum_{i} s^{g}_{i} - \text{h.c.} 
\end{align}
for the remained of this work we leave the inclusion of Hermitian conjugate terms in Hamiltonians implicit. 
$\mathcal{S}^{g}$ is a generating set of $\mathcal{S}$ and the sum is over operators $s_i^g \in \mathcal{S}^{g}$.

To describe the gauging process more precisely, we now choose $\mathbb{Z}_2$ matrix representations for con and exc and call them $\sigma_{c}$ and $\epsilon_{c}$. The entries of these matrices are $\mathbb{Z}_2$ valued. The dimension of $\sigma_c$ is $2q \times N$, in which $q = |M|$ is the number of matter qubits and $N$ is the number of constraints. Each column of $\sigma_c$ represents for one constraint. The first $q$ rows of each column represent for the $X$-sector of that term. When a term have a Pauli $X$ operator acting on the $i$-th qubit, the $i$-th row of this column will be 1. The last $q$ rows represent for the $Z$-sector. When a Pauli $Z$ operator acts on the $i$-th qubit, the $(q+i)$-th row of this column will be 1. Otherwise the entry is 0. We introduce the symplectic form $\lambda_q$,
\begin{align}
    \lambda_q = \left(\begin{array}{cc}
         0 & \mathbb{I}_q \\
         -\mathbb{I}_q & 0
    \end{array}\right).
\end{align}
$\mathbb{I}_q$ is a $q \times q$ identity matrix. The symplectic form $\lambda_q$ transforms $X$-sector to $Z$-sector, and vice versa. For example, consider a column
\begin{align}
    \sigma = \left(\begin{array}{c}
         1 \\
         0 \\
         0 \\
         1
    \end{array}\right)
\end{align}
This column represents a term on a two-qubit system, which is applying a Pauli $X$ operator on the first qubit and Pauli $Z$ operator on the second qubit. After applying the symplectic form, we get
\begin{align}
    \lambda_{q}: \left(\begin{array}{c}
         1 \\
         0 \\
         0 \\
         1
    \end{array}\right) \mapsto \left(\begin{array}{c}
         0 \\
         1 \\
         1 \\
         0
    \end{array}\right)
\end{align}
Now this term becomes Pauli $Z$ operator applying on the first qubit and Pauli $X$ operator applying on the second qubit. The $X$ and $Z$ sector exchange. 

For a CSS stabilizer Hamiltonian, the matrix representation is schematically given by
\begin{align}
    \sigma = \left(\begin{array}{cc}
         \sigma_X & 0\\
         0 & \sigma_Z
    \end{array}\right) \label{CSS}
\end{align}
Terms of the Hamiltonian are divided into $X$ and $Z$-types. We can gauge this kind of Hamiltonian by applying the maps in Eq.~(\ref{gaugc}), Eq.~(\ref{gaugs}) and Eq.~(\ref{gaugi}). The minimal coupled constraints are given by
\begin{align}
    \cal{G} \left(\sigma_{Z} c\right) = Z(\sigma_Z c) \oplus Z(c) \label{gaugeZ}
\end{align}
The commutation relation Eq.~(\ref{gaugecommu}) can be written as 
\begin{align}
    \left[ Z(\sigma_Z c) \oplus Z(c),\ S^{X}_{gauge}\right] = 0. \label{gaugecommu2}
\end{align}
By enforcing the relation above, we get the Gauss's law terms
\begin{align}
    S_{gauge}^{X} = X(m) \oplus X(\sigma_{Z}^{\dagger}m). \label{gaugesymmetry}
\end{align}
$X(m)$ here is the local representation of global (subsystem) symmetries that acts on qubit $m \in M$. The gauged $X$ terms are given by, 
\begin{align}
    \cal{G} \left(\sigma_X\right) = \sigma_X. \label{gaugeX}
\end{align}
If the kernel of the constraint map is not empty, according to Eq.~(\ref{zrelation1}) and Eq.~(\ref{puregauge}), the pure gauge terms are given by
\begin{align}
    B(r) = \prod_{c\in r} Z(c), \label{loop}
\end{align}
in which $\prod_{c \in r} c$ is the local generating set of $\mathrm{ker} \ \sigma_Z$.

\subsection{Introducing locality} \label{locality}

In the following sections, we focus on local Hamiltonians. We introduce a cubic lattice and place the matter qubits $m \in M$ on the vertices of the lattice. We assume the lattice has been sufficiently coarse-grained such that all local symmetry generators and Hamiltonian terms are supported within the unit cell. 
We focus on generalized Ising matter Hamiltonians defined by single $X$ terms and a generating set of constraint $Z$ terms. Since these terms do not commute, the Hamiltonian supports a nontrivial phase diagram with a tunable parameter $J$ associated to the $X$-type terms. This phase diagram supports an ordered phase in the limit $J\rightarrow 0$ and a disordered trivial phase in the opposite limit $J\rightarrow \infty$ as introduced above. 

The Hamiltonian is again described by
\begin{align}
    H(J) = - J \sum_{m} X(m) - \sum_{c} Z(\sigma_Z c),
\end{align}
Applying Eq.~(\ref{gaugeZ}), Eq.~(\ref{gaugesymmetry}) and Eq.~(\ref{gaugeX}), as above, the gauged Hamiltonian is 
\begin{align}
    H(J)_{\text{gauged}} = &- \Delta \sum_{m}  X(m) 
    X(\sigma_{Z}^{\dagger} m) - \sum_{c} Z(\sigma_{Z} c) 
    Z(c) \notag
    \\&- J \sum_{m} X(m) \label{gaugeham}
\end{align}
The global (subsystem) symmetries of this lattice are contained in $\mathrm{ker}\ \epsilon_c$. We know that $\mathrm{im}\ \sigma_c$ commutes with $\mathrm{ker}\ \epsilon_c$. So the global or subsystem symmetries are given by all symmetries modulo local symmetry generators, which are given by
\begin{align}
    S^{X} = X(\mathrm{ker} \ \sigma_Z^{\dagger}).  \label{globalsym}
\end{align}
In the limit $\Delta \to \infty$ and $J \to \infty$, the ungauged model is deep in the symmetric trivial phase. Thus we can project the value of $X(m)$ to 1. The first term of Eq.~(\ref{gaugeham}) becomes 
\begin{align}
    A_{m} = X(\sigma_{Z}^{\dagger} m),
\end{align}
which is a pure gauge term. Violation of this term gives us electric excitations. The second term of Eq.~(\ref{gaugeham}) is highly suppressed when $J \to \infty$. But if the kernel of $\sigma_Z$ is not empty, pure gauge flux terms can emergence, which is given by Eq.~(\ref{loop}). Violation of this term gives magnetic excitations (fluxes). The pure gauge Hamiltonian is given by 
\begin{align}
    H(\infty)_{\text{gauged}} \sim - \sum_m A(m) - \sum_{r} B(r) \label{generalhamiltonian}.
\end{align}

For non-trivial phases one cannot simply project out the matter qubits with single site $X$ fields after gauging, but the gauging method itself remains valid.
We present a subsystem symmetry-protected topological (SSPT) cluster states model example later and discuss its gauging process. In that case, the gauge Hamiltonian terms are still given by $X(m) X(\sigma_{Z}^{\dagger} m)$ and $Z(\sigma_{Z}) Z(c)$.

In the discussion below, we focus primarily on translation invariant Hamiltonians on the cubic lattice. Translation invariance implies the Hamiltonian stays the same when we translate all the terms by a lattice constant. Because of this symmetry, we can simplify the matrix representation. Instead of writing all terms in the matrix, we only write the generating set of the Hamiltonian. All other terms can be generated by translation. 

We use $\{x, y, z, ...\}$ to represent the spatial directions on the lattice and $\{\bar{x}, \bar{y}, \bar{z},...\}$ to represent the opposite directions. We choose a vertex on the lattice and denote its position by 1. The neighbouring vertex on $x$ direction is denoted by $x$ and the next-neighbouring vertex is denoted by $x^2$, etc. And same for other directions. Consider we have $t$ qubits, and $N'$ terms in the stabilizer generating set, per vertex. The dimension of the matrix representation is $2t \times N'$. The entries of the matrix become polynomials following Refs.~\cite{haah2013commuting,Haah2013,haah2016algebraic,yoshida2013exotic}. 

In addition,  without loss of generality, we only deal with local Hamiltonians where all the entries in the corresponding matrix representation are functions that only depend on the zeroth and first order powers of $\{x,y,z,...\}$. If a Hamiltonian contains next-neighbour, or longer, range terms, we can always perform coarse-graining to make such terms only depend on first order variables. We call such terms {\it ultra-local}. 

Throughout this work, we repeatedly make use the following operations on stabilizer lattice models: Adding decoupled trivial qubits, applying CNOT gates, changing the basis of stabilizer generators and translating Hamiltonian terms or qubits. In Haah's polynomial formalism, adding decoupled qubits prepared in the $Z$-basis to a CSS stabilizer Hamiltonian is represented by the operation
\begin{align}
    \sigma_X \to \sigma_X \oplus \mathbb{0}_{n \times n}, \quad \sigma_X \to \sigma_Z \oplus \mathbbm{1}_{n \times n}, \label{trivial}
\end{align}
in which $\mathbb{0}_{n \times n}$ is an $n \times n$ null matrix and $\mathbbm{1}_{n \times n}$ is an $n \times n$ identity matrix. 
A CNOT gate controlled on qubit $a$, with target qubit $b$, separated by a lattice vector described by a polynomial $f$, acts on the $X$ and $Z$-sector in the following way
\begin{align}
    \mathrm{CNOT}(a, b, f) &: R^{X}_{a} \mapsto R^{X}_{a} + f(x, y, z) R^{X}_b, \notag\\
    \mathrm{CNOT}(a, b, f) &: R^{Z}_{b} \mapsto R^{Z}_{b} + f(\overline{x}, \overline{y}, \overline{z}) R^{Z}_{a} \label{cnotz},
\end{align}
in which $R$ is the row of the stabilizer matrix representation, and $a$, $b$ match those defined in Eq.~(\ref{cnot2}). 
Changing the basis of stabilizer generators is implemented by a column operation on the stabilizer matrix
\begin{align}
    \mathrm{Col} (a, b, f) : C_{a} \mapsto C_{a} + f (x, y, z) C_{b} \, , \label{column}
\end{align}
where $C$ is the column of the stabilizer matrix. Here, $a,b$ refer to stabilizer generators and $f$ again describes a lattice vector between them. 
Translating stabilizer generator terms in the Hamiltonian is implemented by a column operation
\begin{align}
    \mathrm{C}_{a} \mapsto f(x, y, z) \mathrm{C}_{a}.
\end{align}
where $C$ is a column of the stabilizer matrix, $a$ specifies the stabilizer generator and $f$ the vector by which it is translated. 
Similarly, shifting qubits correspond to row operations 
\begin{align}
    \mathrm{R}_{a} \mapsto f(x, y, z) \mathrm{R}_{a} \, ,
\end{align}
where $R$ is a row of the stabilizer matrix, $a$ specifies a qubit, and $f$ the vector by which it is translated.

Now we briefly summarize the relation between equivalence operations that can be performed on the ungauged and the gauged models in the stabilizer matrix formalism. 
\begin{itemize}
    \item Adding trivial qubits prepared in $Z$-basis on the ungauged model is equivalent to adding identity blocks in the ungauged constraint matrix $\sigma_Z$, as we show in Eq.~(\ref{trivial}). After gauging, gauge qubits are coupled to these trivial qubits, as we show in Eq.~(\ref{gaugc}), thus the stabilizer matrix of the gauged model is also enlarged by identity blocks.
    \item CNOT gates are implemented by row operations on the ungauged model~\cite{Haah2013}. They correspond to changing basis on the gauged model, which are column operations.
    \item Adding redundant constraints to the ungauged model is equivalent to adding redundant columns to its stabilizer matrix description.  
    The redundant constraints enlarge the spaces $C$ and $R$ in Eq.~(\ref{chaincomplex}). 
    This corresponds to adding new gauge qubits to the gauged model, which is implemented by adding new rows to its stabilizer matrix description.
    This operation also requires the introduction of further relations between the constraints in the ungauged model. 
    After gauging, this results in the inclusion of further terms in the strongly coupled Hamiltonian. 
    In the strong coupling limit, these gauge qubits couple to each other according to Eq.~(\ref{puregauge}), which results in the addition of new pure gauge terms to the Hamiltonian. 
    This corresponds to adding columns to the stabilizer matrix representation of the gauged model. 
    \item Changing basis on the ungauged model is a column operation, as shown in Eq.~(\ref{column}). On the gauged model, it corresponds to rearrangement of gauge qubits, which is a row operation.
\end{itemize}
All of the above operations correspond to local unitary operators, the addition of auxiliary qubits, or codespace preserving local redefinitions of the Hamiltonian, which are phase preserving. We present examples of gauging below and discuss the exact forms of the above operations in the following sections.

\subsection{Example I: 2D Ising model and toric code} \label{toriccode}

The Hamiltonian of the 2D Ising model is given by,
\begin{align}
    H_{Ising} &= -J \sum_{v} \adjincludegraphics[width=1.3cm,valign=c]{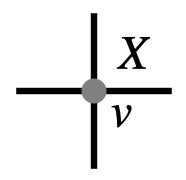} - \sum_{l} \adjincludegraphics[width=1.6cm,valign=c]{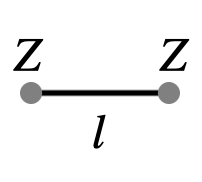}
\end{align}
in which $v \in M$ is the label of matter qubits and $l \in G$ is the label of edges. The matrix representation is given by
\begin{align}
    \sigma = \left(\begin{array}{ccc}
         1  & 0  & 0\\
         0 & 1 + x & 1 + y
    \end{array}\right)
\end{align}
By applying Eq.~(\ref{gaugeZ}), Eq.~(\ref{gaugesymmetry}) and Eq.~(\ref{gaugeX}), the gauged Hamiltonian is given by
\begin{align}
    H^g = &-J \sum_{v} \adjincludegraphics[width=1.3cm,valign=c]{Figures/Ising1.png} - \sum_{l} \adjincludegraphics[width=1.4cm,valign=c]{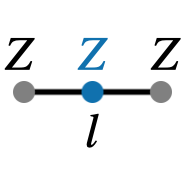}\notag
    \\&-\sum_{v} \adjincludegraphics[width=2.2cm,valign=c]{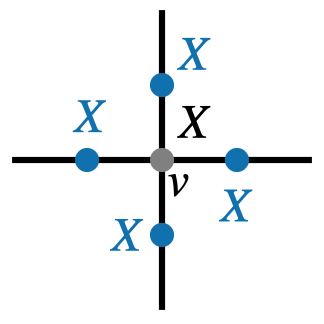}
\end{align}
The gray dots represent for matter qubits and the blue dots represent for the gauge qubits. In the strong coupling limit, $J \rightarrow \infty$, we get the Hamiltonian in the form of Eq.~(\ref{generalhamiltonian}), which is given by
\begin{align}
    H_{TC} &= -\sum_{v} \adjincludegraphics[width=2cm,valign=c]{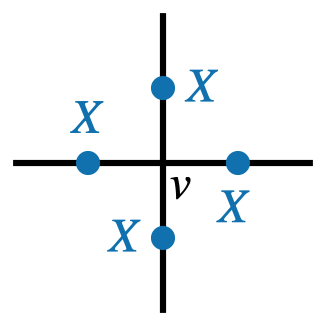} - \sum_{p} \adjincludegraphics[width=2cm,valign=c]{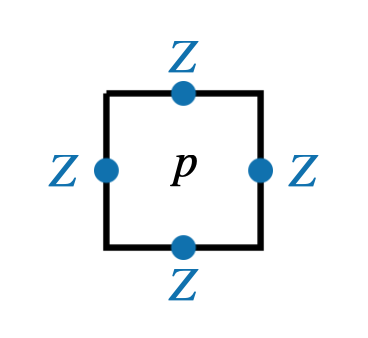},
\end{align}
in which $v$ is the label of vertex and $p$ is the label of plaquette. This is the toric code model~\cite{qdouble}. 

Following  Eq.~(\ref{globalsym}) the global symmetry before gauging is given by
\begin{align}
    S_{global}^{X} = X(\mathrm{ker} \ \sigma_Z^{\dagger}) = \prod_{v \in M} X(v),
\end{align}
which is the product of $X$ operators on all matter qubits, as expected. After gauging, this symmetry maps to identity.

In general, we can define toric code of level $k \in \{1, ..., D-1\}$ in dimension $D$. Matter qubits are placed at $(k-1)$-simplex and gauge qubits are placed at $k$-simplex. As an example, we can write down the Hamiltonian of the 3D toric code, 
\begin{align}
    H_{TC}^3 &= - \sum_{p} \adjincludegraphics[width=2cm,valign=c]{Figures/TC2.png} - \sum_{v} \adjincludegraphics[width=2.2cm,valign=c]{Figures/3dIsing1.png}
\end{align}
$v$ is the label of vertices and $p$ is the label of plaquettes. There are two kinds of excitations of 3D toric code denoted by $e$ and $m$. The electric charge $e$ corresponds to the violation of the $X$-type stabilizer generators. When an $X$-type operator located at vertex $v$ gets an $-1$ eigenvalue, there is an $e$ excitation existing at $v$. 
When a string of Pauli $Z$ operators are applied to the ground state, a pair of $e$ excitations are created at the endpoints of the string. 
The magnetic flux excitation $m$ corresponds to the violation of the $Z$-type stabilizer generators. 
When there is a membrane of Pauli $X$ operators applied to the ground state, the magnetic flux loops are created on the boundary plaquettes. Hence in 3D, there are point-like excitation $e$ and membrane-like excitation $m$. In higher dimensions, this generalizes directly to 0-dimensional and codimension-1 topological excitations with $-1$ braiding. There are also further generalizations of the toric code  in $d\geq 2$ spatial dimensions realizing $k\leq d-2$ and $d-k-2$ dimensional excitations with a mutual $-1$ braiding. 
The 3D toric code in particular plays an important role in the construction of the TDNs that we discuss in the sections below.

\subsection{Example II: 2D cluster state with linear subsystem symmetries}

Now we discuss a 2D linear subsystem symmetry protected topological (SSPT) model~\cite{you2018subsystem,subsystemphaserel}. We consider a square lattice and its dual lattice and label them as $\alpha, \beta$. We place one qubit per vertex. The Hamiltonian can be written as
\begin{align}
    H &= -\sum_{v \in \alpha} \adjincludegraphics[width=2cm,valign=c]{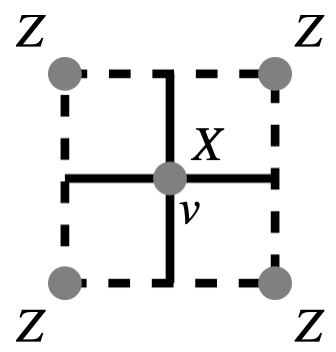} - \sum_{v' \in \beta} \adjincludegraphics[width=2cm,valign=c]{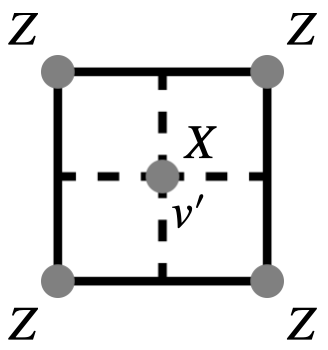}
\end{align}
in which $v, v'$ is the label of vertices on lattice $\alpha$ and $\beta$, depicted by solid lines and dashed lines, respectively. 
While this model is non-CSS, on each sublattice the Hamiltonian terms can be divided into $X$ and $Z$-types. 
Thus the methods we introduced above remain applicable. After applying Eq.~(\ref{gaugeZ}), Eq.~(\ref{gaugesymmetry}) and Eq.~(\ref{gaugeX}), the gauge symmetry operators of each sublattice are given by,
\begin{align}
    A_{v}= \adjincludegraphics[width=2cm,valign=c]{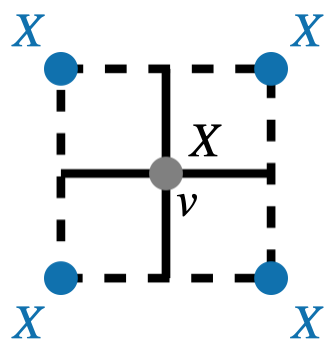}, \quad A_{v'} = \adjincludegraphics[width=2cm,valign=c]{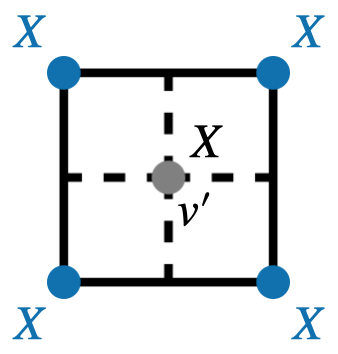}
\end{align}
The minimal coupled constraints are given by
\begin{align}
    B_{v} &= \cal{G} \left(Z(\sigma_{Z} v) X(v)\right) \notag\\
    &= Z(\sigma_{Z} v) X(v) %\oplus 
    Z(v)\\
    B_{v'} &= \cal{G} \left(Z(\sigma_{Z} v') X(v')\right) \notag\\
    &= Z(\sigma_{Z} v') X(v') %\oplus 
    Z(v').
\end{align}
We arrive at 
\begin{align}
    B_{v} = \adjincludegraphics[width=2cm,valign=c]{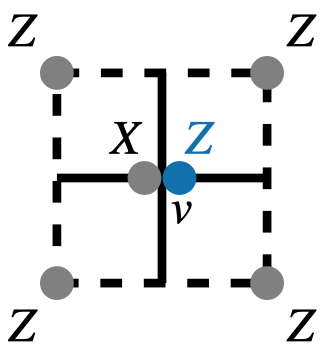},\quad B_{v'} = \adjincludegraphics[width=2cm,valign=c]{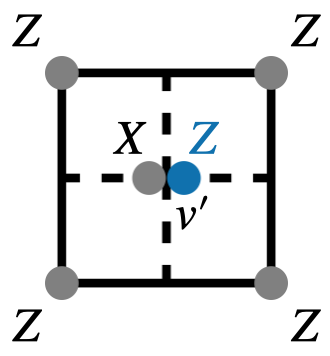} \, .
\end{align}
The kernel of the constraint map for this model is trivial for this Hamiltonian, hence we find no pure gauge terms. Thus the gauged Hamiltonian is given by
\begin{align}
    H = &-\sum_{v \in \alpha} \adjincludegraphics[width=2cm,valign=c]{Figures/Cluster3.png} - \sum_{v' \in \beta} \notag \adjincludegraphics[width=2cm,valign=c]{Figures/Cluster4.png} \\&-\sum_{v \in \alpha} \adjincludegraphics[width=2cm,valign=c]{Figures/Cluster5.png} - \sum_{v' \in \beta} \adjincludegraphics[width=2cm,valign=c]{Figures/Cluster6.png} \, .
\end{align}
The subsystem symmetries are given by ${S_{\text{subsystem}}^{X} = X(\mathrm{ker}\ \epsilon_c)}$, which are,
\begin{align}
    S^{X}_{m} = \prod_{v \in L_m} X(v)
\end{align}
in which $L_m$ can be any row or column of the $\alpha$ or $\beta$ lattice. 

\subsection{Example III: 3D plaquette Ising model and X-cube} \label{example3}

Before reviewing the TDN construction of the X-cube model in the next subsection, we first introduce its ungauged variant, the plaquette Ising model~\cite{vijay2016fracton} and discuss gauging it. The Hamiltonian of the plaquette Ising model is given by
\begin{align}
    H_{PI} = - J \sum_{v} \adjincludegraphics[width=1.3cm,valign=c]{Figures/Ising1.png} - \sum_{p} \adjincludegraphics[width=1.6cm,valign=c]{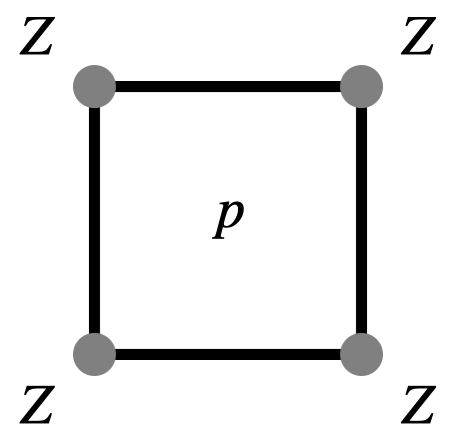}, \label{HamPI}
\end{align}
where $v$ labels vertices and $p$ labels plaquettes of a 3D cubic lattice. 
The constraint terms in this Ising model lead to planar subsystem symmetries generated by spin flips applied to all spins within a plane of the cubic lattice~\cite{vijay2016fracton}. 
These subsystem symmetries are given by ${S_{\text{subsystem}}^{X} = X(\mathrm{ker}\ \epsilon_c)}$, which are,
\begin{align}
    S^{X}_{n} = \prod_{v \in P_n} X(v) \, ,
\end{align}
where $P_{n}$ denotes the set of qubits in a plane that is indexed by $n$.

After gauging these subsystem symmetries, following Eqs.~(\ref{gaugeZ}),~(\ref{gaugesymmetry}), and~(\ref{gaugeX}), we obtain gauge symmetry operators given by
\begin{align}
    A_{v}= \adjincludegraphics[width=2.3cm,valign=c]{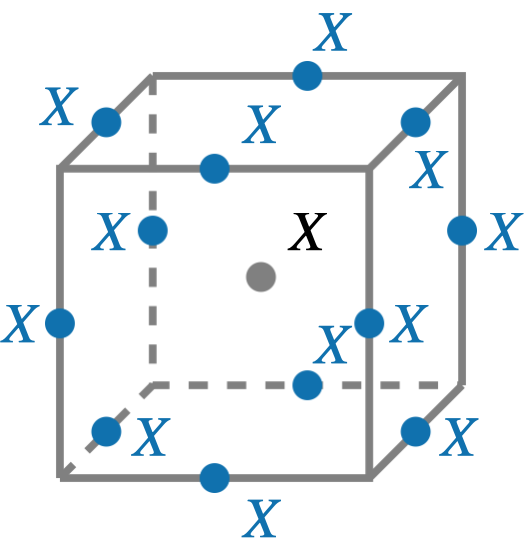}, 
\end{align}
where the gauge qubits live on the edges of the dual lattice. 
The gauged constraint terms are 
\begin{align}
    B_{v} &= \cal{G} \left(Z(\sigma_{Z} v)\right) = Z(\sigma_{Z} v) %\oplus 
    Z(v) = \adjincludegraphics[width=1.6cm,valign=c]{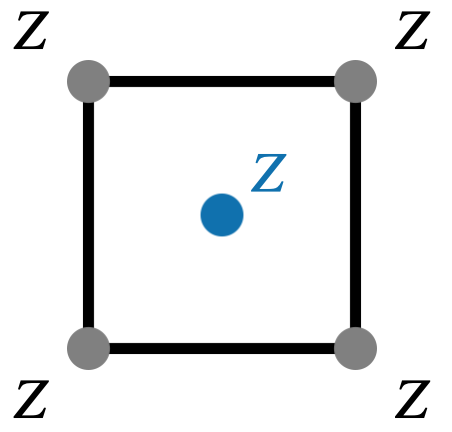}.
\end{align}
The kernel of the constraint map for the plaquette Ising model turns out to be non-empty. The local relations are generated by products of four plaquette constraints adjacent to a single cube. Hence in the strong coupling limit of the gauged model, there are pure gauge flux terms, see Eq.~(\ref{loop}). These correspond to four body star terms on edges adjacent to a vertex within a single plane on the dual lattice. 
The gauged Hamiltonian in the strong coupling limit turns out to be the X-cube model 
\begin{align}
    H_{XC} = &-\sum_{c} \adjincludegraphics[width=2.3cm,valign=c]{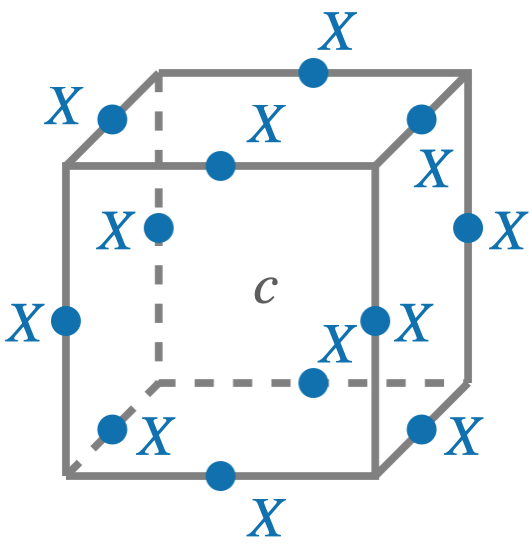} -\sum_{v} \adjincludegraphics[width=2cm,valign=c]{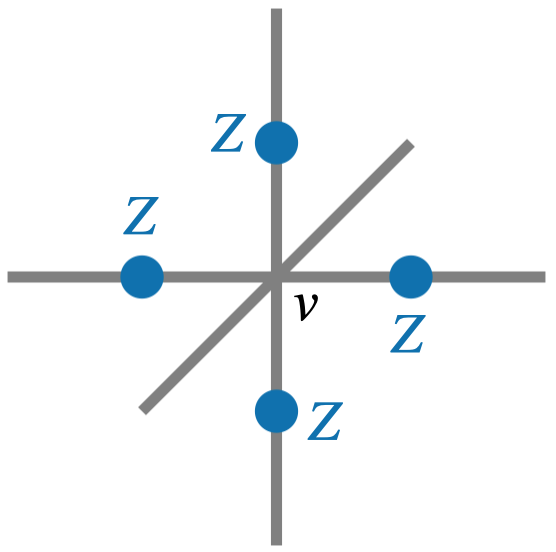} \notag\\&-\sum_{v} \adjincludegraphics[width=2cm,valign=c]{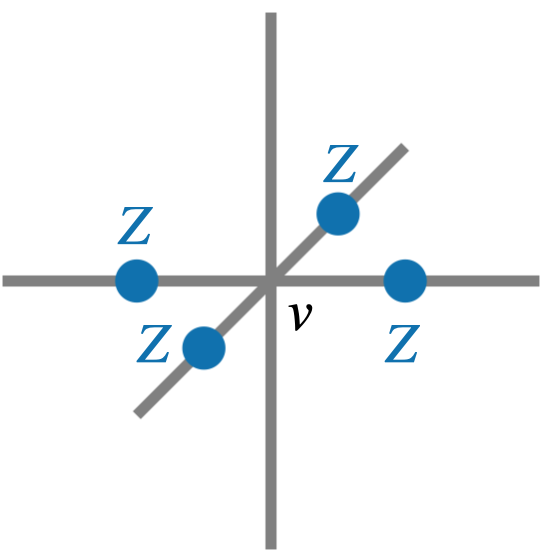} -\sum_{v} \adjincludegraphics[width=2cm,valign=c]{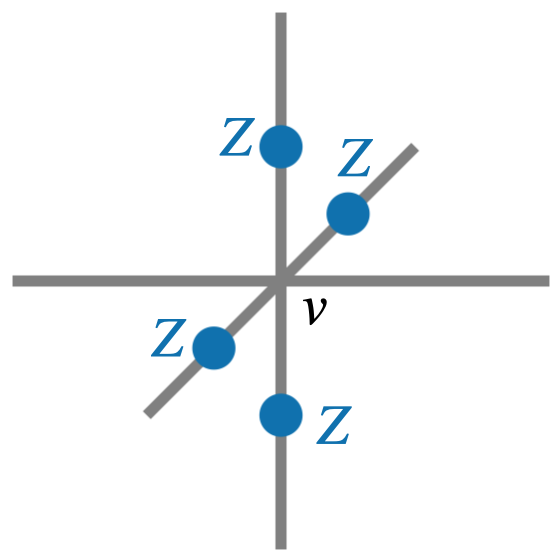}. \label{Hamxcube}
\end{align}
The planar subsystem symmetries of the plaquette Ising model are mapped to a relation involving the product of cube terms over a dual plane after the gauging procedure.

There are two types of topological excitations that generate all others in the X-cube model, fractons and lineons. 
Fractons correspond to violations of the $X$-type stabilizer generator. 
A cluster of four fractons, at the corners of a dual plaquette, is created by the application of a $Z$ operator to an edge of the lattice. 
A single fracton is immobile unless an energy penalty is paid to create further excitations as it is moved. More precisely, a fracton can move one step on the lattice by creating a neighboring pair of fractons. This implies that a pair of adjacent fractons can be moved within the plane orthogonal to the edge separating them without any energy penalty. Hence such a composite excitation is known as a planon. 

Lineons correspond to violations of the $Z$-type stabilizer generators. 
The application of an $X$ operator to an edge of the lattice creates a pair of lineons on the adjacent vertices. By applying further $X$ operators these lineons can move along a straight line. However, when a lineon turns a corner another lineon with mobility in the perpendicular direction is created. Hence single lineons have their mobility restricted to a single line. On the other hand, a pair of adjacent lineons of the same type have mobility within a lattice plane orthogonal to the edge separating them (for one choice this composite excitation is trivial). Again such a composite excitation is known as a planon. 

The X-cube model is the most extensively studied  type-I fracton model, in the terminology of Ref.~\cite{vijay2016fracton}, due to the simplicity it retains while still supporting topological excitations with a range of fracton, lineon, and planon, restricted mobilities.  
In constrast, models which support only fracton topological excitations, known as type-II fracton models~\cite{vijay2016fracton}, are significantly more complicated. 
We discuss TDN constructions of both kinds of fracton model in the sections below.

\section{Revisiting the X-cube TDN}
\label{sec:XcubeExample}

In this section we revisit the construction of a TDN for the X-cube model, first reviewing the construction in Ref.~\cite{Aasen2020}, and then describing a construction following our new approach which arrives at the same TDN. 

\subsection{Review of the X-cube TDN}

The TDN representation of the X-cube model is defined on a cubic stratification of a 3D manifold $\mathcal{M}$ that is isomorphic to $\mathbb{R}^3$ or the 3D torus $\mathbb{T}^3$. 
On that manifold we have 3-strata, 2-strata, 1-strata, and 0-strata that form a cubic lattice as shown in Fig.~\ref{Strata_2}. 
To each 3-stratum we assign a 3D toric code, while topological defects are assigned to the j-strata with $j < 3$. These defects can be specified by a set of topological excitations that condenses on each of the j-strata. 
The condensing excitations determine the set of remaining excitations which cannot pass through each of the strata due to non-trivial braiding relations.  
Schematically, the Hamiltonian of the TDN can be written as the sum of Hamiltonian terms $H_i$ associated to each strata.
\begin{align}
    H = \sum_{j=0}^{3} H_{i}.
\end{align}
The details of this Hamiltonian are discussed  below.

The basic strategy, following Ref.~\cite{Aasen2020}, is to assign a 3D toric code to each 3-stratum and condense excitations from the ambient 3-strata on lower dimensional strata to make the behaviour of the uncondensed excitations equivalent to those in the X-cube model. 
On 2-strata, the condensate is generated by the magnetic excitations from the neighbouring 3-strata, denoted by $\pm$
\begin{align}
    \big\langle m_{+}, m_{-}\big\rangle.
\end{align}
Due to the braiding statistics of the excitations in the 3D toric code, no electric excitations can pass through the 2-strata.

The condensate on each 1-strata is generated by
\begin{align}
    \big\langle e_1 e_2 e_3 e_4,m_1 m_2, m_2 m_3, m_3 m_4\big\rangle. \label{XC1strata}
\end{align}
The subscripts label the four neighbouring 3-strata. 
This condensate implies that an $e_1 e_2 e_3 e_4$ charge pattern can be created by applying string operators in the vicinity of the 1-strata. 
Due to the condensations on the 2-strata, the electric excitations cannot pass through the 2-strata to another 3-strata. The only way to move an $e$ excitation between 3-strata is via a 1-strata, at the cost of generating a pair of additional $e$ excitations in different 3-strata. 
%For example, $e_1$ to a nearby 1-stratum, and because of the $e_1 e_2 e_3 e_4$ condensation there, $e_1$ can only move to a neighbouring 3-stratum by creating another two $e$ excitations, which costs additional energy. 
These $e_1 e_2 e_3 e_4$ condensing excitations are isomorphic to the local neutral clusters of fractons in the X-cube model. 

Demonstrating the existence of lineon excitations in the TDN is somewhat more involved, and is explained in detail in Ref.~\cite{Aasen2020}, we briefly summarize the explanation here. 
The mobility is due to the fact that a flux loop near a 1-stratum can be moved to the corner of the 3-stratum to create the excitation shown on the left of Eq.~\eqref{3loop}. Each disconnected arc of magnetic line excitation corresponds to a lineon along the associated edge direction. Three lineons with different orientations can annihilate at a corner
\begin{align}
    \adjincludegraphics[width=6cm,valign=c]{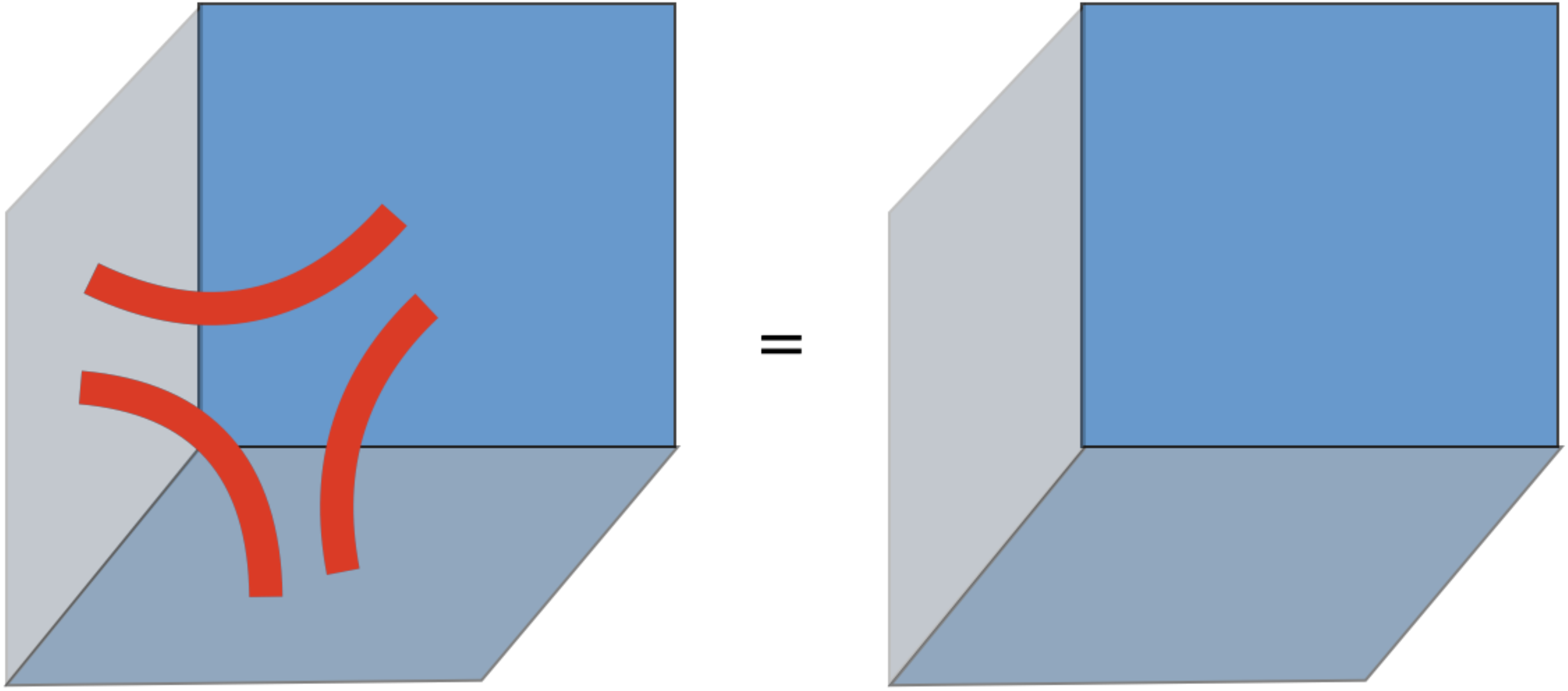} \, . \label{3loop}
\end{align}
Hence a single lineon at a corner is equivalent to a pair of lineons with different orientations. 
By using this property and the condensations of $m$ excitations on 1-strata, we find the following process
\begin{align}
    \adjincludegraphics[width=7cm,valign=c]{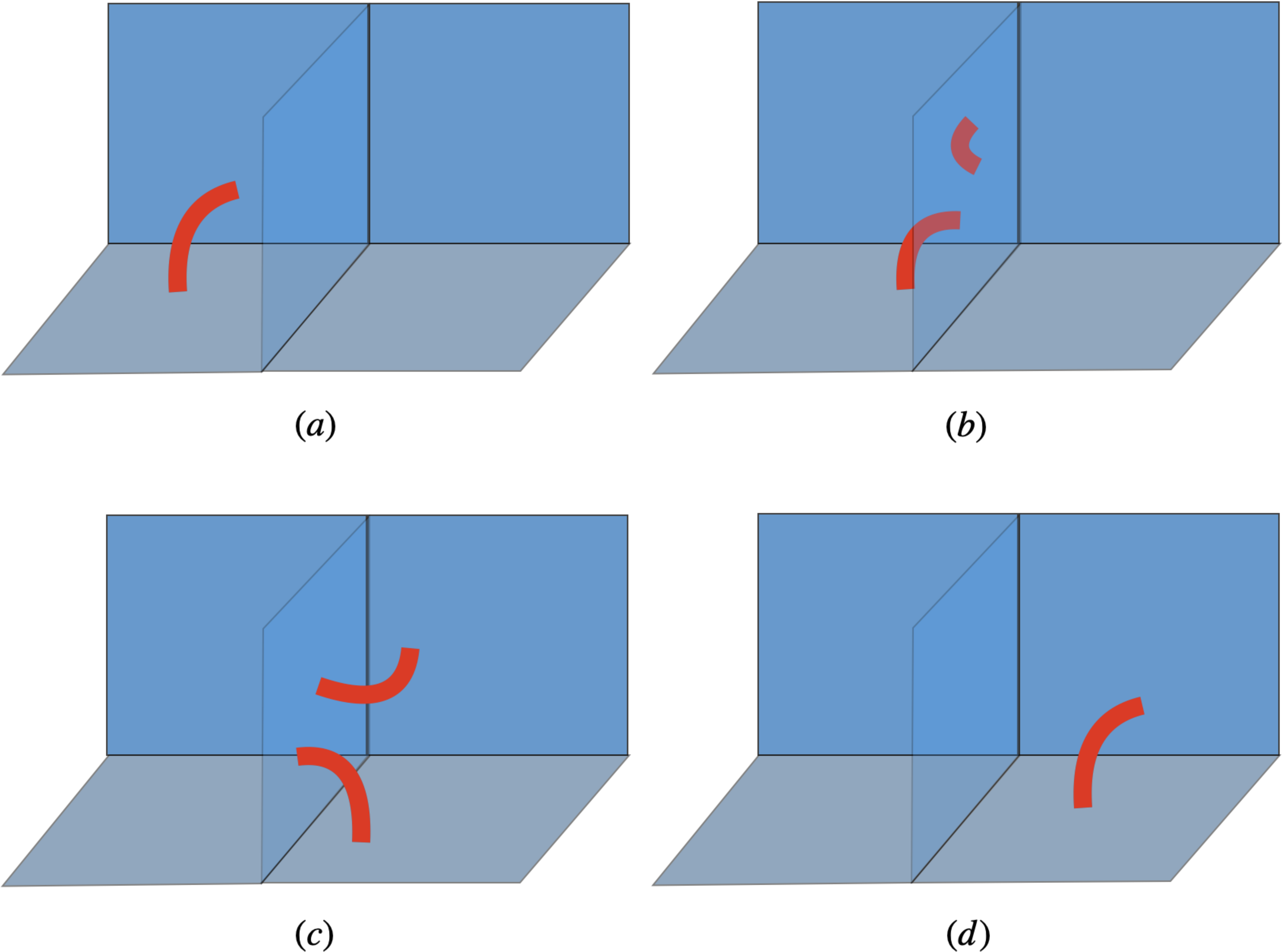} \, . \label{lineons}
\end{align}
Going from (a) to (b) requires an application of Eq.~(\ref{3loop}) to turn one lineon on a 1-stratum at the corner into a pair of lineons on with different orientations. 
To pass from (b) to (c) we make use of $m$ condensations on the 1-strata.
%We multiply them with certain magnetic condensations and get the fluxes on the other sides of 2-strata. 
To go from (c) to (d) we apply Eq.~(\ref{3loop}) again. During this process, No additional excitations are left behind. 
This exactly mimics the behaviour of lineon excitation in the X-cube model. 
The planons of the X-cube model are similarly formed by pairs of fractons or lineons.

In the TDN construction of the X-cube model just described, we use the 3D toric code in 3-strata and topological defects on 1- and 2-strata. No further excitations condense on the 0-strata, to fully specify the defect there we specify the eigenvalue of all topological operators linking the point to be $+1$ as was done in Ref.~\cite{Aasen2020}. 

\subsection{TDN representation of X-cube via ungauging}

In the previous subsection, we reviewed the TDN representation of the X-cube model. 
As explained in Section \ref{example3}, we can also obtain the X-cube model in  Eq.~(\ref{Hamxcube}) by gauging the plaquette Ising model in Eq.~(\ref{HamPI}). 
In this section, we show how to construct the TDN of the X-cube model explicitly from the plaquette Ising model. 

The sketch of the construction is the following. 
We start from the plaquette Ising model, and extend it to a fine grained lattice on a stratified space by adding trivial qubits and applying unitary gates, specifically CNOT gates as well as redefinitions of the stabilizer generators. After the stratification, each qubit of the previous plaquette Ising model becomes a 3-stratum containing a block of 3D Ising model, and the original constraint terms now couple different 3-strata together. 
This is the ungauged defect network. We obtain the TDN of the X-cube model by gauging it.

\subsection{Defect network for the plaquette Ising model} \label{Plaquette}

The Hamiltonian of the plaquette Ising model is given by,
\begin{align}
    H_{PI} = - J \sum_{v'} \adjincludegraphics[width=1.2cm,valign=c]{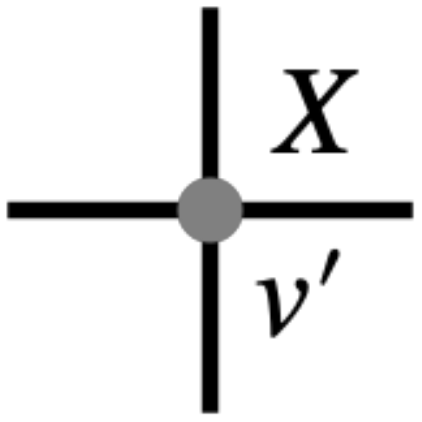} - \sum_{p} \adjincludegraphics[width=1.6cm,valign=c]{Figures/PI1.png}. \label{plaquetteIsing}
\end{align}
where $v'$ is the label of vertices on a cubic lattice and $p$ is the label of plaquettes.

\begin{figure*}
    \centering
    \includegraphics[width=16cm]{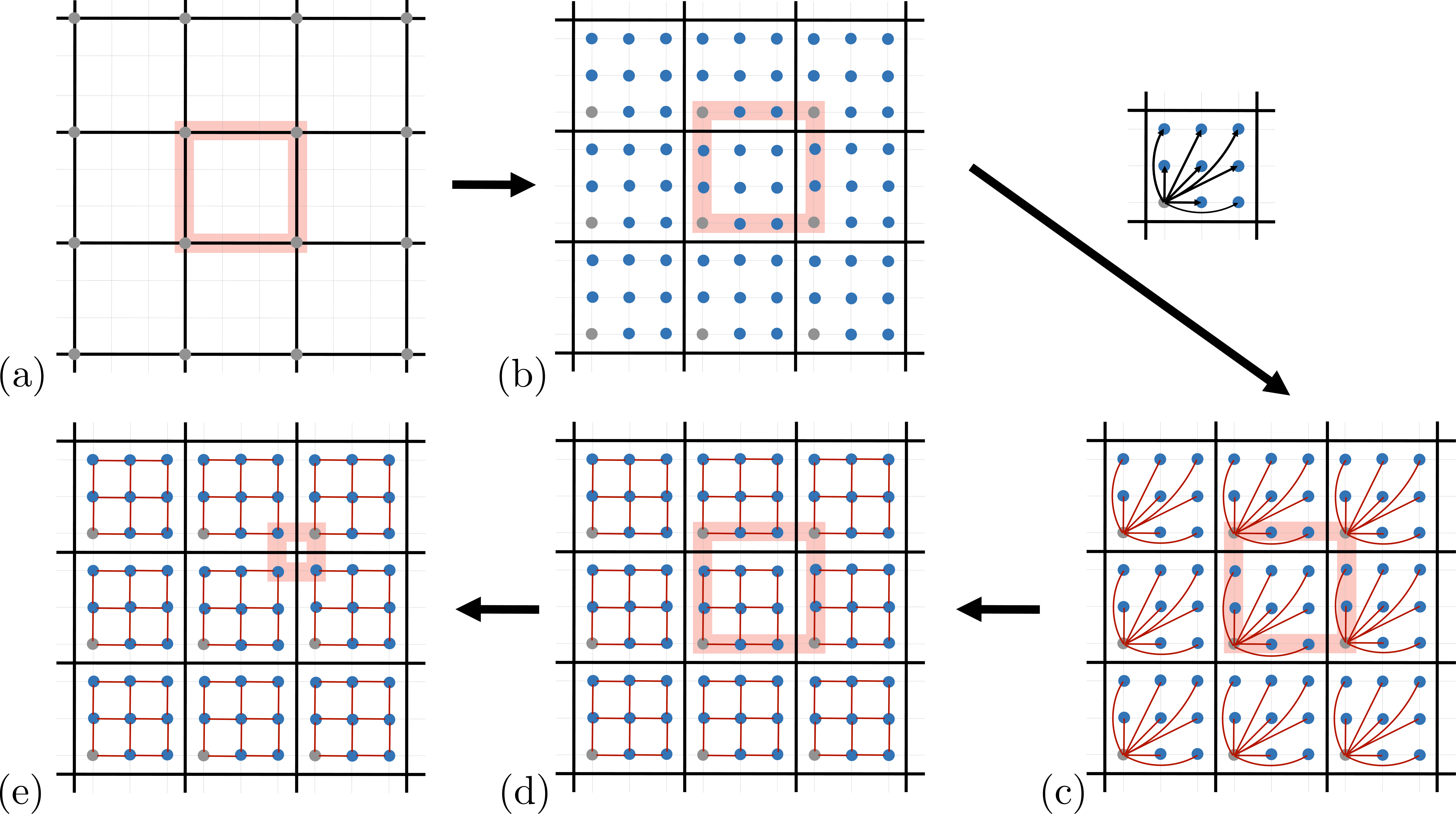}
    \caption{
    An illustration of our defect network construction applied to the 2D plaquette Ising model. 
    (a) Qubits (grey circles) governed by plaquette Ising model constraint terms (transparent orange) are shown on the original coarser lattice (black lines). A finer lattice (grey lines) is introduced for reference. The plaquettes, edges, and vertices of the original lattice define 2-, 1-, and 0-strata for the defect network construction. 
    (b) The finer lattice is shifted by half a lattice spacing in all directions. Decoupled qubits (blue circles) governed by trivial single site constraints are introduced onto the vertices of the refined lattice. 
    (b$\rightarrow$c) CNOT gates (black arrows) are applied from the original (grey) qubit in each 2-strata to the newly introduced (blue) qubits in the same 2-strata.
    (c) The original (grey) qubit in each 2-strata is now coupled to all newly introduced (blue) qubits in the same 2-strata via two-body $ZZ$ Ising constraint terms (red lines). 
    (d) We choose a new, redundant, generating set of two-body Ising constraints (red). These constraints simply correspond to blocks of the Ising model within 2-strata, and are ultra-local. 
    The introduction of additional redundancy to the generating set of constraints requires the addition of further relations on the constraints. An ultra-local generating set of these relations is given by products of constraints around plaquettes that are fully contained within 2-strata (i.e. bounded by red edges). 
    Up to, and including, this step the plaquette Ising model couplings (transparent orange) on the original (grey) qubits remained unchanged. They couple qubits on the scale of the coarser lattice and hence are not yet ultra-local. 
    (e) We make use of the two-body Ising terms within the 2-strata to choose a new ultra-local generating set of plaquette Ising terms (transparent orange) located at 0-strata. 
    In this example there are no Ising terms associated to 1-strata, and no further relations. 
    In the general construction of a defect network for an ungauged topological CSS stabilizer models (that has been sufficiently coarse grained), an ultra-local generating set of relations involving the generalized Ising couplings is chosen at this stage, see Section~\ref{generalapproach}. 
    }
    \label{TDNs}
\end{figure*}

To produce a defect network, we first introduce an auxiliary cubic lattice with lattice spacing much smaller than the original one. This is depicted for a 2D example in Fig.~\ref{TDNs}a. We refer to the newly introduced auxiliary lattice as the \emph{finer lattice} and call the original lattice the \emph{coarser lattice}. When a term only couples qubits within the same unit cell of the finer lattice, we call it \emph{ultra-local}. 
Next, we shift the finer lattice, along with the matter qubits, relative to the coarser lattice by half a finer lattice spacing in each of the spatial directions. 
This step moves each of the matter qubits from a vertex of the coarser lattice to a corner vertex of the finer lattice within a 3-cell. 
The cubes of the coarser lattice define the 3-strata, faces 2-strata, edges 1-strata and vertices 0-strata. 
Due to the half shift of the finer lattice, the 0-strata are associated to cubes of the refined lattice, the 1-strata are associated to plaquettes and cubes of the refined lattice that intersect them, the 2-strata are associated to edges, plaquettes and cubes of the refined lattice that intersect them, and the 3-strata are associated to the vertices, edges, faces and cubes of the refined lattice that they contain strictly. 

Next we add trivial qubits on the vertices of the finer lattice prepared in the ground state of the trivial Hamiltonian
\begin{align}
    H_{trivial} = - \sum_{v} Z(v) - J \sum_{v} X(v) \, ,
\end{align}
where $v$ is the label of the vertices on the finer lattice. 
This process is shown for a 2D example in Fig.~\ref{TDNs}b. 
We then apply CNOT gates between the qubits on the coarser lattice and the qubits on the finer lattice inside the corresponding cubes. Recall, the properties of CNOT gates are given by Eq.~(\ref{cnot2}). By taking the qubits of the original plaquette Ising model to be the control qubits and the newly introduced trivial qubits in the corresponding cubes as target qubits, we can get a complete set of Ising constraint terms within each 3-stratum. This step is shown for a 2D example in Fig.~\ref{TDNs}c.
Next, we change the basis of generators to make all of the $ZZ$ constraint terms nearest neighbor on the finer lattice, and hence ultra-local. 
Then, we add redundant $ZZ$ terms to the 3-strata on every edges of the finer lattice within a cube. The resulting 3-strata are Ising paramagnets. 
This is shown for a 2D example in Fig.~\ref{TDNs}d. 
This step also involves the addition of new relations between the added constraint terms. Furthermore it is possible to choose a set of generating relations that are themselves ultra-local within the cubes.

While the $ZZ$ terms in the 3-strata are ultra-local, the constraint terms of the plaquette Ising model are not. They still couple qubits of the coarser lattice. For a depiction of this in a 2D example see Fig.~\ref{TDNs}d.  
To remedy this, we change the basis of constraint generators by using the freedom to multiply with two-body terms within each 3-strata to bring all the constraints into an ultra-local form in the vicinity of one of the strata. Which strata they are localized on depends on their dimension. 
In particular, $n$-dimensional terms become ultra-local on $(3-n)$-strata. 
Since the plaquette terms in Eq.~(\ref{plaquetteIsing}) are 2-dimensional we make them ultra-local on 1-strata. 
Next We add redundant terms to those strata, along with additional relations, such that the model is homogenous on each strata, and the relation generators are ultra-local. For a depiction of an ungauged defect network in a 2D example, see Fig.~\ref{TDNs}e. 
A justification for why the above process results in a phase equivalent TDN after gauging is provided in Section~\ref{generalapproach}. 

The Hamiltonian of 3-strata is thus given by
\begin{align}
    H_3 &= -J \sum_{v} \adjincludegraphics[width=1.3cm,valign=c]{Figures/Ising1.png} - \sum_{l} \adjincludegraphics[width=1.6cm,valign=c]{Figures/Ising2.png}. \label{Ising3strata}
\end{align}
These are just the terms of the 3D paramagnet and ultra-local on the finer lattice.

As the Hamiltonian of the X-cube model doesn't have 1-dimensional (or 3-dimensional) constraint terms, there are no ultra-local terms on 2-strata (or 0-strata) of the defect network. Therefore, the Hamiltonian on 2-strata (or 0-strata) is trivial.

For 1-strata, there are ultra-local plaquette terms on them. The Hamiltonian on one of the 1-stratum is thus given by
\begin{align}
    H_{1} &= - \sum_{p} \adjincludegraphics[width=1.6cm,valign=c]{Figures/PI1.png} \label{Xcube1}
\end{align}
$p$ is the label of plaquettes on 1-strata and these terms couple four 3-strata together. So far we get the Hamiltonian of the defect network of the plaquette Ising model.

The above prescription can also be described in Haah's polynomial formalism as we introduced in Section \ref{Polynomial}. The polynomial matrix of the plaquette Ising model is given by
\begin{align}
   \footnotesize \left(\begin{array}{cccc}
         1 & 0 & 0 & 0  \\
         0 & 1+x+y+xy & 1+y+z+yz & 1+x+z+xz
    \end{array}\right).\normalsize
\end{align}
We can use one of the plaquette term to show how to build the defect network. Consider we have the term $1+x+y+xy$ and add trivial qubits in one of the cubes, we get
\begin{align}
    \sigma_Z = \left(\begin{array}{ccccc}
            1 + x + y + xy & 0 & 0 & ... & 0\\
           0 & 1 & 0 & ... & 0\\
           0 & 0 & 1 & ... & 0\\
           ... & ... & ... & ... & ...\\
           0 & 0 & 0 & ... & 1
    \end{array}\right).
\end{align}
The first row are the $Z$ constraint term on the coarser lattice. The following rows are the $Z$ sector of the trivial qubits in one of the 3-strata. We apply CNOT gates between the qubits of the original model and the trivial qubits. The CNOT gates on $X$ and $Z$ sectors are given by Eq.~(\ref{cnotz}). After applying CNOT gates, we get
\begin{align}
    \sigma_Z = \left(\begin{array}{ccccc}
            1 + x + y + xy & 1 & 1 & ... & 1\\
           0 & 1 & 0 & ... & 0\\
           0 & 0 & 1 & ... & 0\\
           ... & ... & ... & ... & ...\\
           0 & 0 & 0 & ... & 1
    \end{array}\right).
\end{align}
We see between the trivial qubits there is a complete set of $ZZ$ terms. To make these terms ultra-local, we apply column operations, as shown in Eq.~(\ref{column}). Then we add redundant nearest-neighbor $ZZ$ terms to make them on every edges of the finer lattice. We get
\begin{widetext}
\begin{align}
    &\sigma_Z = \Big(\begin{array}{cccc}
          1 + x + y + xy & [ijk] + [(i+1)jk] & [ijk] + [i(j+1)k] & [ijk] + [ij(k+1)]
    \end{array}\Big).
\end{align}
\end{widetext}
Here we use $[ijk]$ as the coordinates of the finer lattice in one 3-stratum. We use $[ijk] + [(i+1)jk]$ to represent for the nearest neighbor $ZZ$ terms along direction $i$, and same for the other directions $j$ and $k$.

However, this is not the end of the story. As we can see the first term $1+x+y+xy$ is not ultra-local. So we need to apply further column operations Eq.~(\ref{column}) to make it ultra-local on 1-stratum. Finally we add redundant plaquette terms along the 1-stratum and choose a set of local relation generators. After these steps, the construction of the ungauged defect network of X-cube model is complete.

A comment here about the construction of defect networks is that, all the operations we have, including adding trivial qubits, applying CNOT gates and column operations are allowed transformations that keep the phase of matter invariant. They are generalized local unitary transformations. This is discussed in more detail in Section~\ref{generalapproach}. 

\subsection{Gauging}

We know from Section \ref{example3} that the plaquette Ising model is mapped to the X-cube model under gauging. 
So by gauging the defect network of plaquette Ising model, we expect to also find the TDN of the X-cube model. 
We demonstrate this explicitly in this subsection.

By applying Eq.~(\ref{gaugeZ}) and Eq.~(\ref{gaugesymmetry}), and in the strong coupling limit, the gauged Hamiltonian of the 3-strata Eq.~(\ref{Ising3strata}) is given by,
\begin{align}
    H_{3}^g &= - \sum_{p} \adjincludegraphics[width=2cm,valign=c]{Figures/TC2.png} - \sum_{v} \adjincludegraphics[width=2.2cm,valign=c]{Figures/3dIsing1.png}.
\end{align}
$p$ and $v$ are the labels of plaquettes and vertices on the finer lattice. The blue dots are the gauge qubits of the 3D toric code. 

Similarly, the gauged Hamiltonian of the 1-strata is given by
\begin{align}
    H_{1}^g &= - \sum A_1 - \sum B_1. \label{H1X}
\end{align}
$A_1$ are the Gauss's law terms
\begin{align}
    A_1 = \adjincludegraphics[width=2.3cm,valign=c]{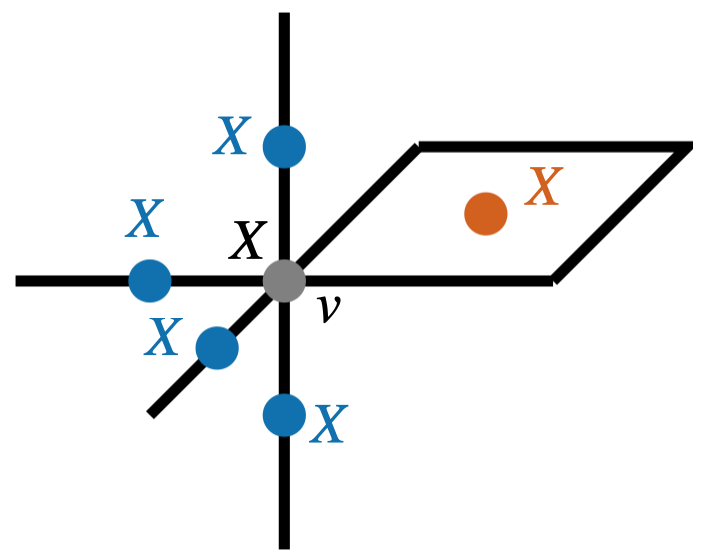}. \label{gaugesymmetryH1X}
\end{align}
The blue dots here are the gauge qubits of 3D toric code. The orange dot is the gauge qubit on the center of plaquettes on 1-strata. These terms are given by following Eq.~(\ref{gaugeZ}), Eq.~(\ref{gaugecommu2}) and Eq.~(\ref{gaugesymmetry}).

$B_1$ are the flux terms. These flux terms lie in the kernel of the constraint map according to Eq.~(\ref{loop}). During the gauging procedure, we couple gauge qubits to constraint terms. 
The gauged plaquette terms are then given by,
\begin{align}
    \adjincludegraphics[width=1.6cm,valign=c]{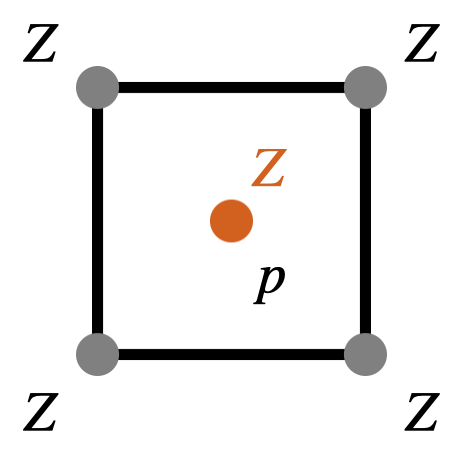}, \label{gaugepla}
\end{align}
in which $p$ is the label of plaquettes on 1-strata.
There are also gauged Ising constraint terms adjacent to the 1-strata. The kernel of the constraint terms on the 1-strata is given by combinations of these two types of constraints of the form
\begin{align}
    \adjincludegraphics[width=2.3cm,valign=c]{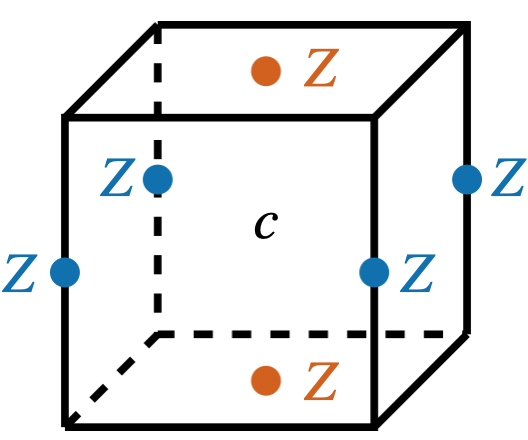}.  \label{H1XFlux}
\end{align}
The orange qubits lie on the plaquettes on the 1-strata. While the blue qubits are on edges adjacent to the 1-strata.

On 0-strata, the kernel of the constraint map is generated by products of four terms of the form shown in Eq.~(\ref{gaugepla}) from different 1-strata. The flux terms on 0-strata are thus given by
\begin{align}
    \adjincludegraphics[width=7cm,valign=c]{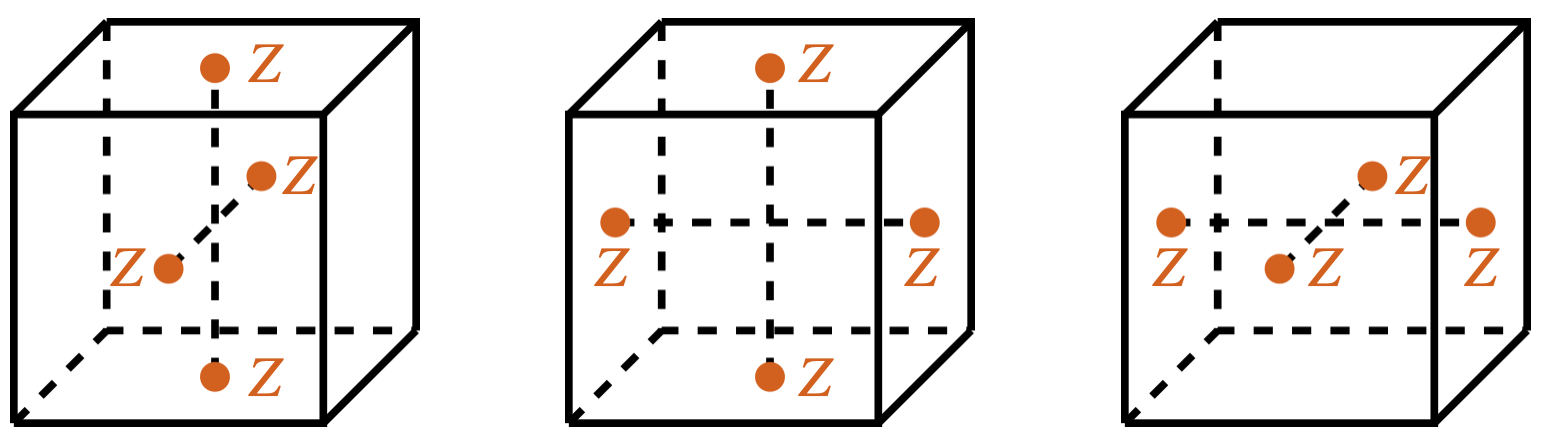},  \label{H0XFlux}
\end{align}
As there is no ultra-local constraint terms on 2-strata, there is no non-trivial flux terms on 2-strata. 
We have now completed the construction of the X-cube TDN Hamiltonian.

\subsection{Condensations and excitations in the X-cube TDN lattice model} \label{sec35}

Due to the appearance of new Gauss's law and flux terms in the vicinity of lower dimensional strata after gauging, the behavior of bulk topological excitations is modified there. 
For instance, some patterns of bulk excitations are condensed on the defects, which means they are identified with the vacuum sector there. 
Additionally,  defects may also permute the topological superselection sectors of bulk excitations that pass through them. 
In this subsection, we show the equivalence of the X-cube model and the TDN constructed in the preceeding subsection by verifying that the condensations and hence the behaviour of excitations is equivalent. 

We first check the electric condensations. These electric condensations correspond to patterns of bulk electric charges that condense on the defects. These patterns can also be created by local operators on the defects. These local operators commute with the Hamiltonian Eq.~(\ref{H1XFlux}) on the defects but anti-commute with the Gauss's law terms Eq.~(\ref{gaugesymmetryH1X}) in the bulk. The minimal coupled local operator that satisfy the above conditions are single $Z$ operators on the orange qubits on the 1-strata. 
By applying a single $Z$ operator on the 1-strata, Gauss's law terms Eq.~(\ref{gaugesymmetryH1X}) in four neighboring 3-strata are violated, which produces the anticipated electric charge pattern $e_1 e_2 e_3 e_4$. 
Here we use the subscripts 1 to 4 to denote the four adjacent 3-strata. 
These charges can each be moved within their corresponding 3-strata via the application of local operators therein. 
These electric excitations become identified with the X-cube fractons.

On 2-strata, the only possible local operators are single $Z$ operators on the boundary of the adjacent 3-strata. These operators create pairs of electric charges in the same 3-strata. 
Such pairs can already locally annihilate within the 3-strata so there are no nontrivial electric condensations on 2-strata. A similar argument applies to the 0-strata, showing there are no further electric condensation there either. 

Next we check the magnetic condensations. Generally there are two ways to find them. One way is to find local operators that commute with the Hamiltonian Eq.~(\ref{H1XFlux}) on the defects, but anti-commute with the flux terms in the neighboring bulks. The second, equivalent, way is to find a generating set of composite magnetic flux excitations that braid trivially with the electric composite excitations that condense on the relevant strata. 
Following the first method for 2-strata, the relevant local operators are single $X$ operators on the adjacent Ising gauge qubits. These operators anti-commute with the plaquette terms in the bulk and we find the magnetic condensations $\big\langle m_{+}, m_{-}\big\rangle$. Where we have used $+$ and $-$ to represent the neighboring 3-strata.
On 1-strata, we find the following local operators
\begin{align}
    \adjincludegraphics[width=7.3cm,valign=c]{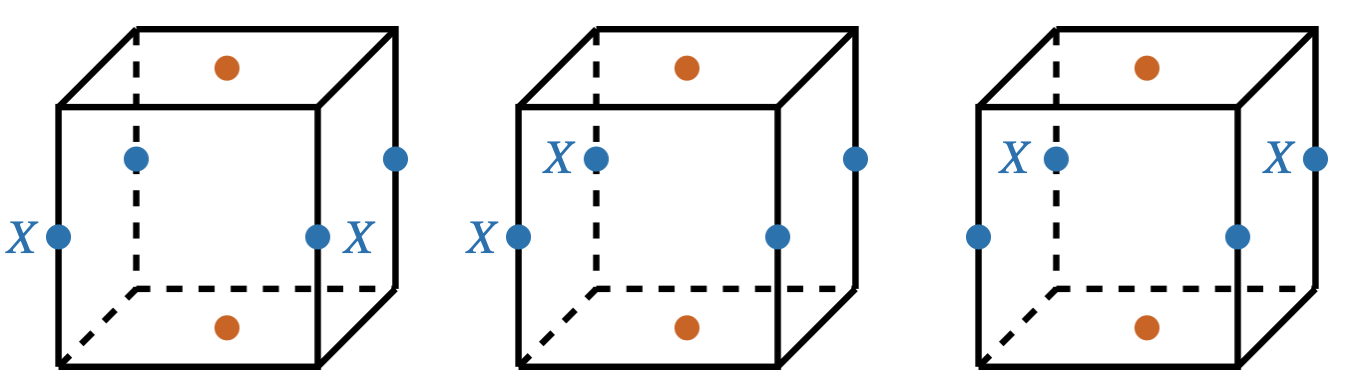}.  \label{X0Strata2}
\end{align}
These operators are independent from one another and commute with Eq.~(\ref{H1XFlux}) but anti-commute with the flux terms in the bulk. The following magnetic condensation can then be read off $\big\langle m_1 m_2, m_2 m_3, m_3 m_4\big\rangle$.

If we instead follow the second method, we start from the electric condensations $\big\langle e_1 e_2 e_3 e_4 \big\rangle$ on 1-strata and the magnetic condensations on the neighboring 2-strata $\big\langle m_1, m_2, m_3, m_4\big\rangle$. 
The magnetic condensations must braid trivially with the electric condensations, such magnetic composite excitations are given by $\big\langle m_1 m_2, m_2 m_3, m_3 m_4\big\rangle$. 
For more complicated defect Hamiltonians, it can be difficult to find all local operators that induce a complete generating set of magnetic condensations. 
However, it remains straightforward to find the electric condensations. 
Thus the second method often turns out to be much simpler than the first in practice.

From Eq.~(\ref{H0XFlux}) we know that there are no Ising gauge qubits in the vicinity of the 0-strata. Thus there is no magnetic condensation on 0-strata. However, we are still able to find local operators that commute with the Hamiltonian on 0-strata, which are
\begin{align}
    \adjincludegraphics[width=7cm,valign=c]{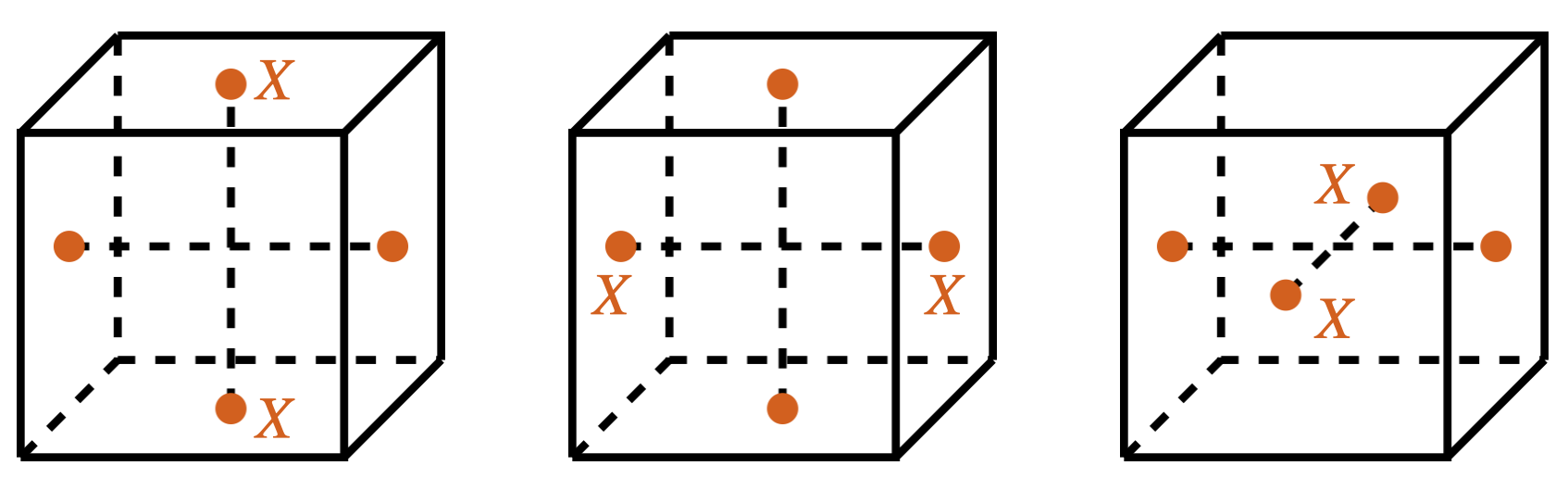}.  \label{X0Strata3}
\end{align}
These operators don't create any excitations in the bulk, but they anti-commute with Eq.~(\ref{H1XFlux}). We find that these operators play an important role when transporting a magnetic flux along 1-strata. 
Recall that the magnetic excitations of the X-cube model are also lineons, which have their mobility restricted to lines unless additional excitations are created. 

There are other local operators in the vicinity of 0-strata that commute with the full Hamiltonian. An example is shown below 
\begin{align}
    \adjincludegraphics[width=3cm,valign=c]{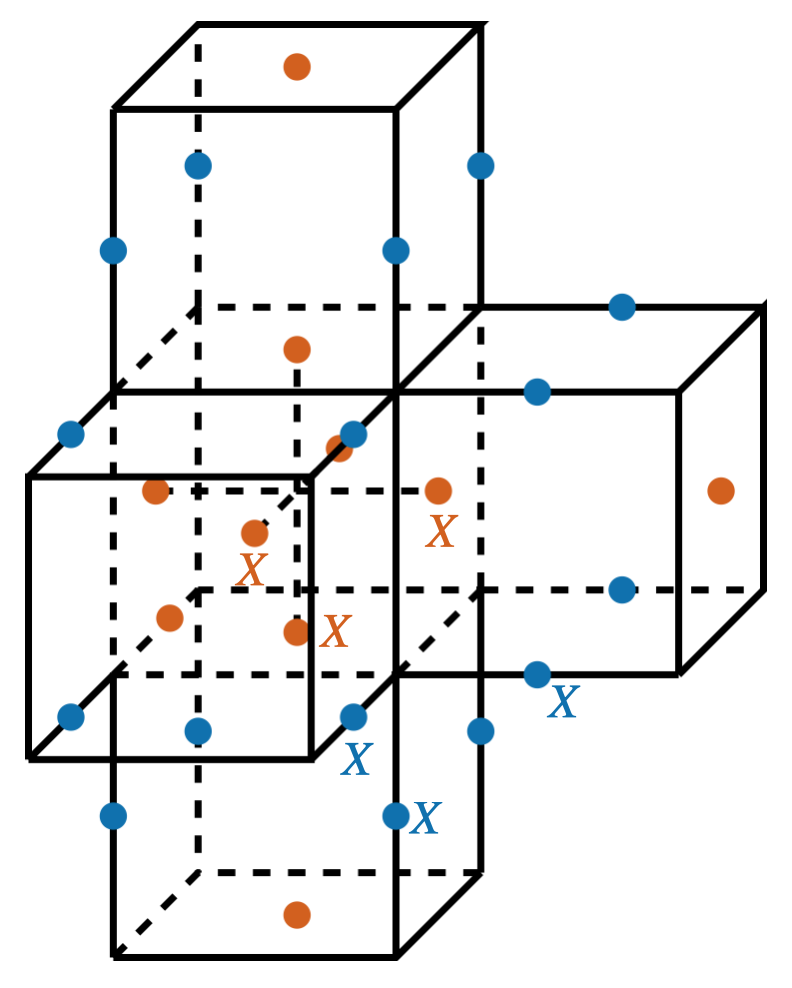}.  \label{X0Strata5}
\end{align}
This operator corresponds to the fact that 3-loops can annihilate at a vertex as we shown in Eq.~(\ref{3loop}).

The behavior of X-cube lineons can be derived from our TDN lattice model, reproducing Eq.~(\ref{lineons}), 
\begin{align}
    \adjincludegraphics[width=7cm,valign=c]{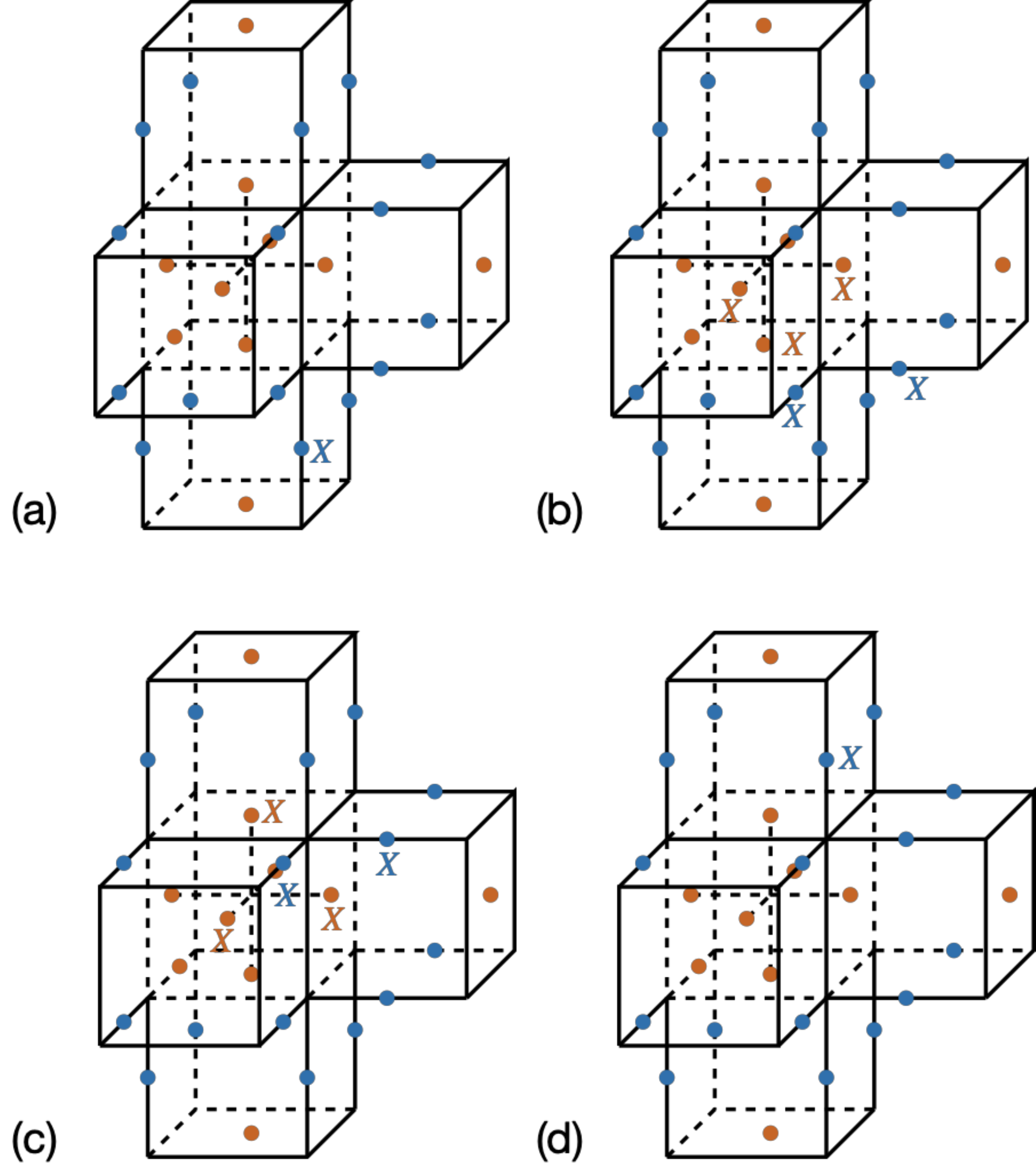} .  \label{X0Strata1}
\end{align}
Starting from (a), there is a magnetic flux on the 1-stratum below. Next we apply Eq.~(\ref{X0Strata5}) to get (b). Then we apply Eq.~(\ref{X0Strata3}) and Eq.~(\ref{X0Strata2}) to get (c). Finally we apply Eq.~(\ref{X0Strata5}) with the opposite orientation to find (d). The operator in (d) creates a magnetic flux on the 1-stratum above. During this process we only use operators that commute with the Hamiltonian. Therefore no additional energy is generated. 
This demonstrates that the magnetic flux arcs are in fact lineons, as they indeed have restricted mobility along lines in the TDN we have constructed. 

In the final part of this subsection we discuss the behavior of electric condensations. The electric condensations on 1-strata result in the mobility constraints of fractons. 
Consider a fracton in a 3-stratum that we want to move to the next 3-stratum. 
It cannot pass through the 2-stratum directly as on the 2-stratum there is a condensate generated by $\big\langle m_{+}, m_{-}\big\rangle$ and the electric charge braids nontrivially with elements of this condensate. 
The only possible route is via the 1-strata. 
We can apply local $Z$ operators to move the electric charge. 
When the charge reaches the 1-stratum, similar to  the $e_4$ in Eq.~(\ref{X0Strata4}), to move it to the adjacent 3-stratum we apply a local $Z$ operator on the plaquette. 
\begin{align}
    \adjincludegraphics[width=7cm,valign=c]{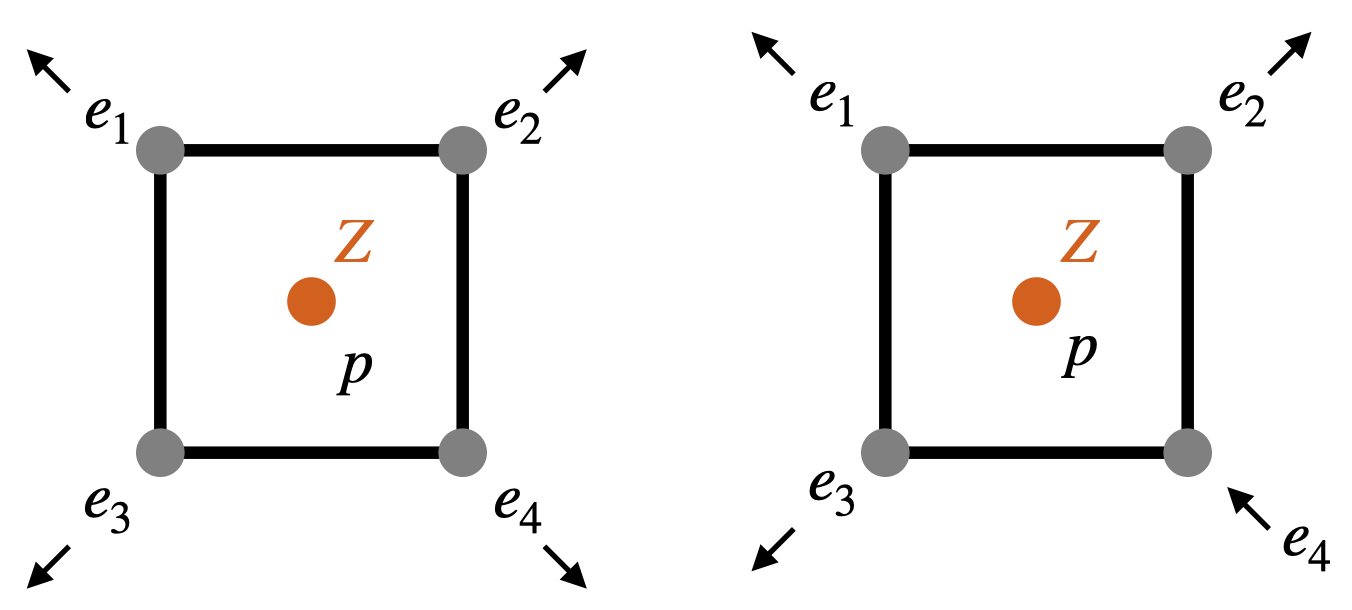}  \label{X0Strata4}
\end{align}
This operator annihilates $e_4$ but creates $e_1, e_2$ and $e_3$ on the neighboring 3-strata. 
So although the electric charge is moved to the neighboring 3-stratum, another pair of electric charges is also created. This process costs additional energy. Thus the electric charges in our TDN are fully immobile on the scale of the 3-strata, and furthermore mimic the mobility of fractons in the X-cube model

In summary, in this section we have provided a lattice construction of a TDN for the X-cube model from its ungauged variant, the plaquette Ising model, by applying phase equivalence transformations. 
The TDN we construct has equivalent excitation structure to the X-cube model. In Section~\ref{generalapproach} we present a general argument that implies the TDN Hamiltonian is in fact in the same phase of matter as the X-cube model.

\section{TDN for Haah's cubic code A}
\label{TDNHaah}

In this section we use our gauging construction to find a TDN representation of Haah's cubic code~\cite{haah2011local}. 
No TDN was previously known for the cubic code, due to the lack of a systematic procedure for converting lattice models into TDNs before the construction introduced here. 

\subsection{Introduction to Haah's cubic code A}
Haah's code, also known as cubic code A is the canonical example of a type-II fracton model~\cite{haah2011local}. It has the following Hamiltonian, 
\begin{align}
    H_{CC1} &= -\sum_{c} \adjincludegraphics[width=2cm,valign=c]{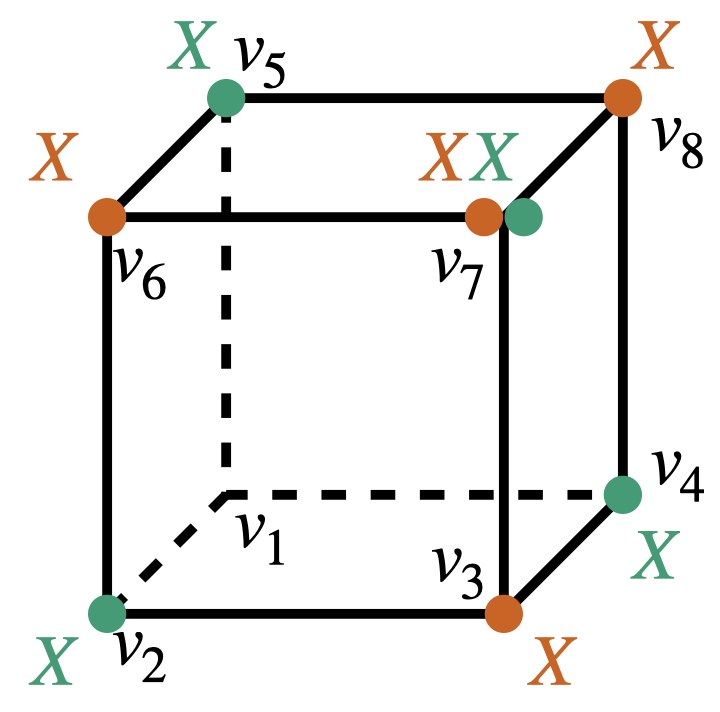} - \sum_{c} \adjincludegraphics[width=2cm,valign=c]{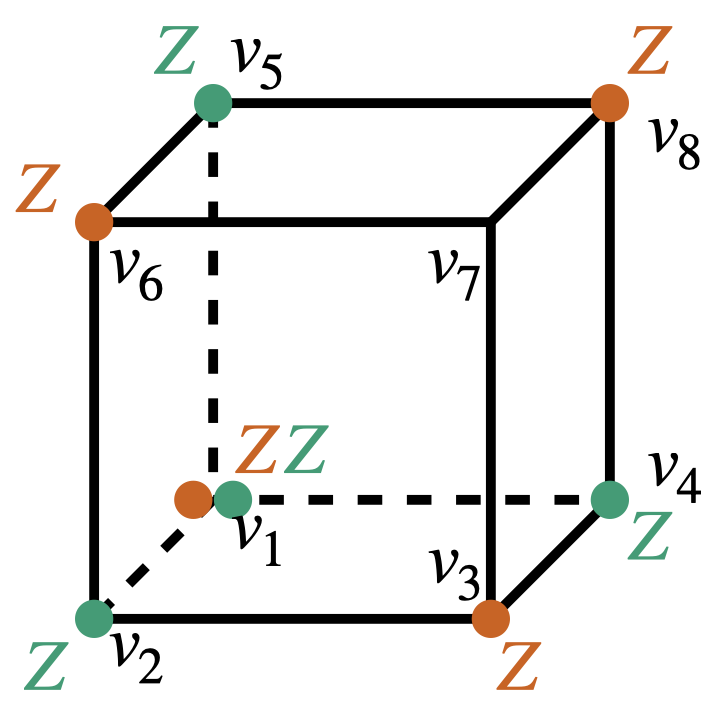}. \label{HaahA}
\end{align}
There are two qubits on each vertex, which are denoted by orange and green dots in Eq.~(\ref{HaahA}). The excitation patterns created by local Pauli $Z$ operators are shown below.
\begin{align}
   \adjincludegraphics[width=6cm,valign=c]{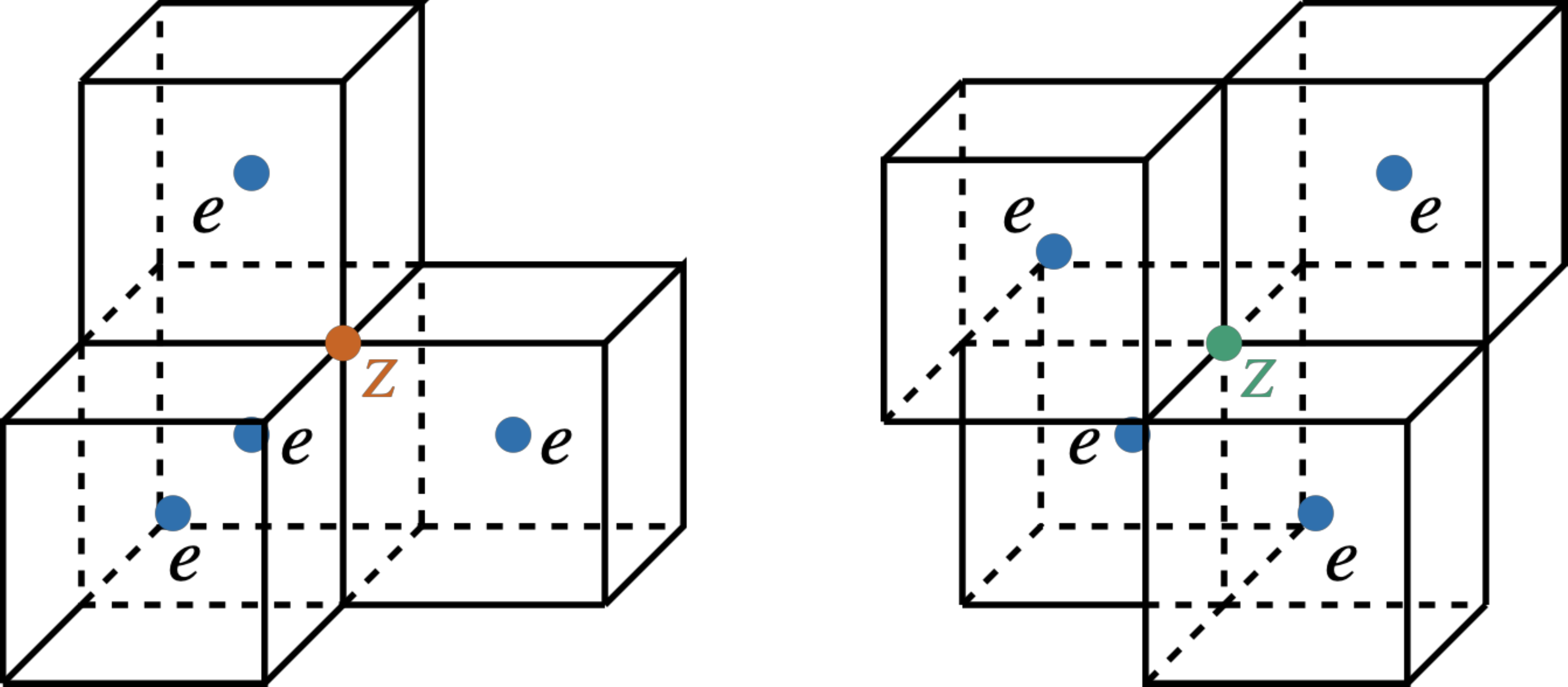}.\notag
\end{align}
The excitation pattern for local $X$ operators are similar. These excitations are called fractons. Each step fractons take on lattice creates two extra excitations, i.e. they cannot move without costing additional energy.

Cubic code A can be achieved by gauging a fractal Ising model~\cite{vijay2016fracton,Williamson_cubic_code} in the $J\rightarrow \infty$ limit, whose Hamiltonian is given by
\begin{align} 
    H = &-J \sum_{v} \adjincludegraphics[width=1.3cm,valign=c]{Figures/Ising1.png}\notag \\&- \sum_{c} \adjincludegraphics[width=2cm,valign=c]{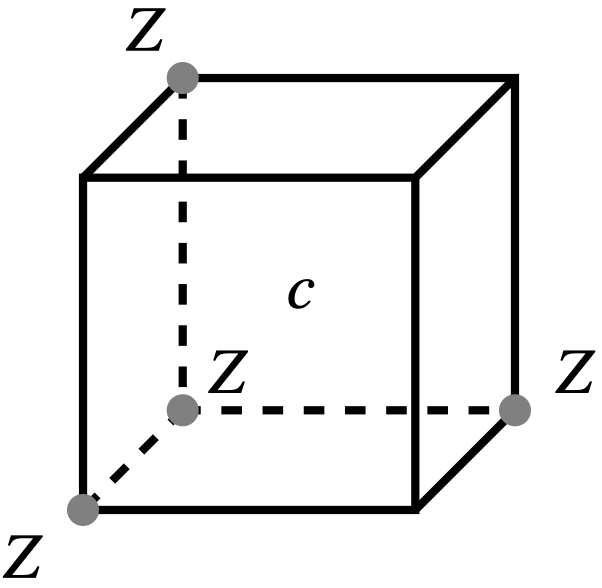} - \sum_{c} \adjincludegraphics[width=2cm,valign=c]{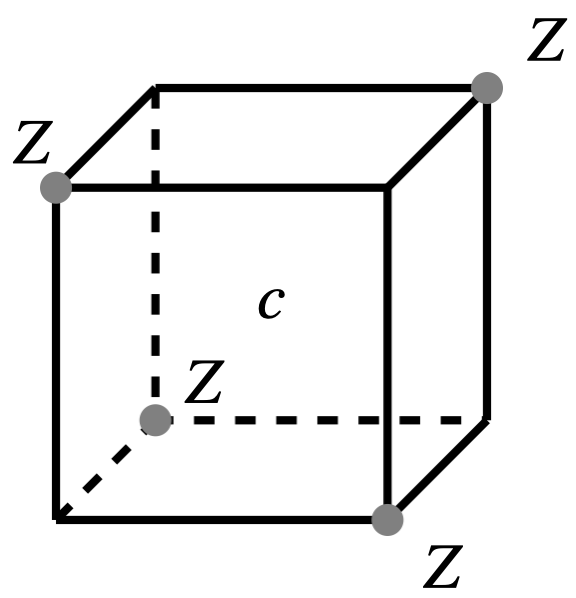}. \label{FractalIsing1}
\end{align}
In the following subsections, we describe a lattice construction of the TDN of Haah's cubic code A from this fractal Ising model.

\subsection{Defect network for the fractal Ising model}
\label{UngaugedCubicCodeTDN}

We now describe the failure of a naive attempt to make a fractal Ising TDN that results in a TDN for the cubic code after gauging. 
We move on to describe how, by first coarse graining the fractal Ising model, we can successfully construct an alternate fractal Ising TDN that results in a valid TDN for cubic code after gauging. 

We start from the fractal Ising model Eq.~(\ref{FractalIsing1}). To convert it into a defect network, we follow the procedure described in Section~\ref{Plaquette}, first introducing a refined cubic lattice. 
Next we add trivial qubits on the vertices of the finer lattice, prepared in the ground state of trivial Hamiltonian
\begin{align}
    H_{trivial} = - \sum_{v} Z(v) - J \sum_{v} X(v),
\end{align}
where $v$ labels the vertices of the finer lattice. 
Again, the cubes of the coarser lattice play the role of 3-strata, faces 2-strata, edges 1-strata and vertices 0-strata. 
We then apply CNOT gates between the qubits on the coarser lattice and their corresponding qubits on the finer lattice. This process is depicted in Fig.~\ref{TDNs}. 
The properties of CNOT gates are described in Eq.~(\ref{cnotz}). 
The CNOT gates introduce a complete set of Ising constraint terms inside each 3-stratum. 
Next, we change the choice of stabilizer generators to make all of the $ZZ$ constraint terms within a 3-stratum nearest neighbor and hence ultra-local. 
This step is implemented by column operations in the stabilizer formalism, see Eq.~(\ref{column}).
We then add redundant Ising $ZZ$ terms to all remaining edges of the finer lattice within each 3-strata. 
This step also involves the introduction of additional relations between these redundant Ising terms, which can be generated by the standard plaquette relations of the Ising model. 
The interactions and relations within the resulting 3-strata are simply those of the cubic lattice Ising model.

At this stage the $ZZ$ terms in the 3-strata are ultra-local. However, the original constraint terms of the fractal Ising model are not ultra-local as they couple qubits of the coarser lattice. 
Thus, we change our choice of the set of stabilizer generators in each 3-stratum. 
This is implemented by column operations in the stabilizer formalism, see Eq.~(\ref{column}). 
In particular, we multiply the fractal Ising constraint terms with the two-body Ising terms introduced in the bulk of the 3-strata to make all the constraints ultra-local on an appropriate strata. 
The strata that each fractal Ising constraint is localized to depends on the dimension of a hypercube containing the term. 
Concretely, $n$-dimensional terms are made ultra-local on $(3-n)$-strata. 
For example, the plaquette terms in Eq.~(\ref{plaquetteIsing}) are 2-dimensional, thus they are made ultra-local on 1-strata. 
Following the same logic, the constraint terms in Eq.~(\ref{FractalIsing1}) can be made ultra-local in the vicinity of a 0-strata, as both such terms are 3-dimensional. 
However the corresponding TDN produced by regauging the resulting Hamiltonian is in fact \textit{not} ultra-local. 
This is caused by the following subtlety, after gauging the flux terms can not be made ultra-local as they necessarily involve \textit{multiple} corners of the same cubic 3-stratum. 
In the ungauged model this manifests in the impossibility to choose a generating of relations that are ultra-local. 

\begin{figure}[t]
    \centering
    (a)\includegraphics[width=7cm]{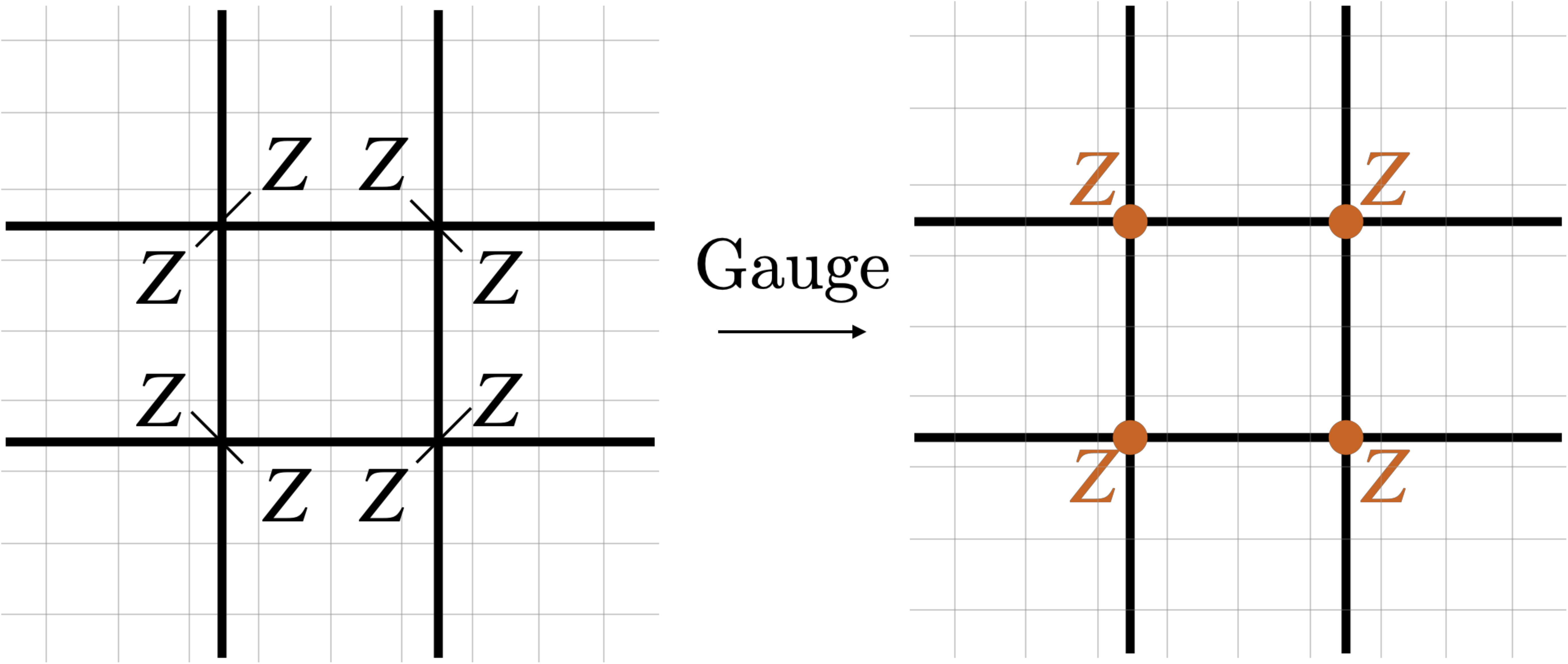}\\ 
    \vspace{10mm}
    (b) \includegraphics[width=7cm]{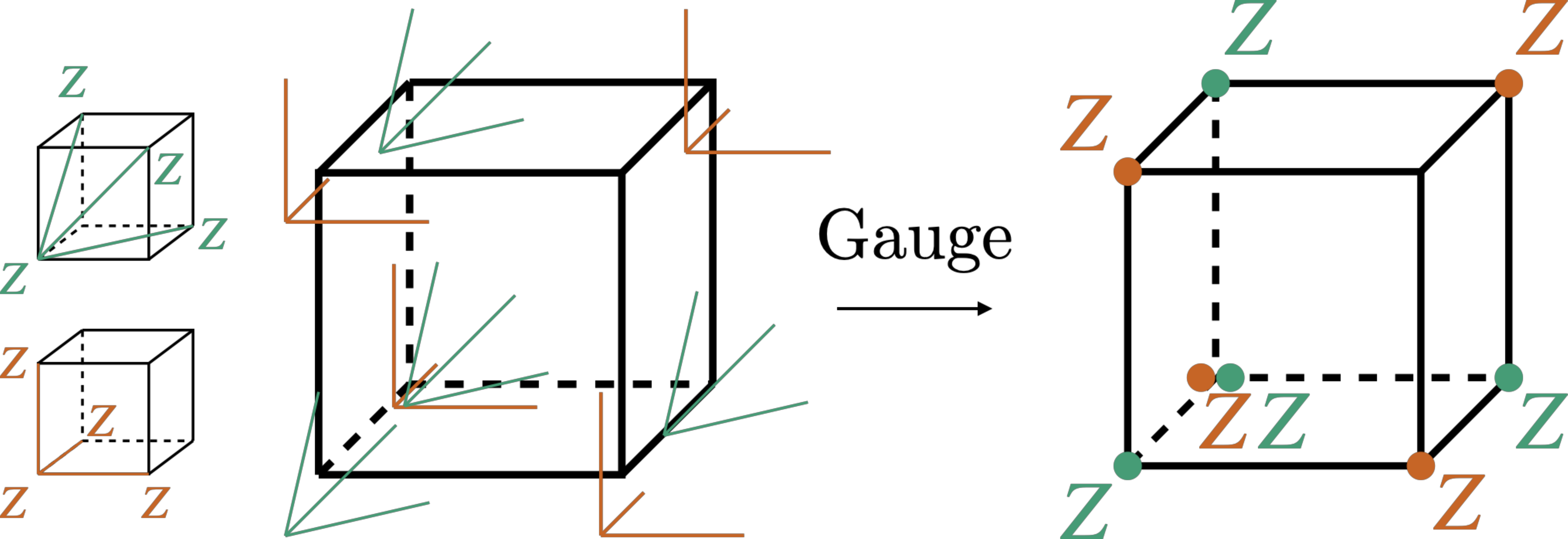} 
    \caption{
    (a) A 2D example that demonstrates how non-ultra-local flux terms may occur even with constraint terms that are ultra-local at 0-strata. The bold lines are the coarser lattice. The plaquettes of the coarser lattice are 2-strata, edges are 1-strata and vertices are 0-strata. The light grey lines form the finer lattice. The defect network for the Ising model in Eq.~\eqref{2Dnonlocal} is depicted on the left. Its matter qubits live on the vertices of the finer lattice. It has ultra-local constraint terms at each vertex. Four such terms indicated on the left, along with $ZZ$ terms in the 2-strata, generate a relation that is not ultra-local. After gauging this leads to a flux term that is not ultra-local, depicted on the right. 
    \\
    (b) A demonstration of why the naive TDN for the cubic code is not ultra-local, similar to the example shown above in (a). While the constraint terms of the ungauged model (left) are ultra-local, the relations involving them are not. After gauging, this leads to a flux term in the cubic code TDN that is not ultra-local, shown on the right. 
    We find that an additional coarse graining step along each axis is sufficient to construct an ultra-local TDN, as explained in the main text. 
    }
    \label{2Dnonlocal1}
\end{figure}

To convey this point, we take a small diversion and note that a similar difficulty can already be seen in the simple 2D example depicted in Fig.~\ref{2Dnonlocal1}a.  
This example is described by the following Ising constraint terms,
\begin{align}
    \sigma_Z = \left(\begin{array}{ccc}
          0 & 0\\
          1 + xy & x + y
    \end{array}\right) \, . 
    \label{2Dnonlocal}
\end{align}
Since both types of constraint terms are two dimensional (i.e. they involve $x$ and $y$ terms in the polynomial), they are made ultra-local on 0-strata as shown in Fig.\ref{2Dnonlocal1}a. 
Following Eq.~(\ref{loop}), the gauge flux terms are determined by the kernel of the constraint map. 
In this case, we need to multiply terms involving four distinct 0-strata to produce a flux term. 
This is not ultra-local as it has a size on the order of the coarser lattice scale, which becomes the lattice scale of the defect network itself. 

Similarly, if we make the constraint terms of Haah's cubic code ultra-local at 0-strata, after gauging, gauge qubits at different 0-strata are coupled together, and hence  the flux terms are not ultra-local. This is depicted in Fig. \ref{2Dnonlocal1}b. 
To demonstrate this more explicitly, we first recall that as part of the gauging procedure each constraint term is coupled to a gauge qubit. 
The gauged constraint terms of the fractal Ising model are given by
\begin{align}
   \adjincludegraphics[width=6cm,valign=c]{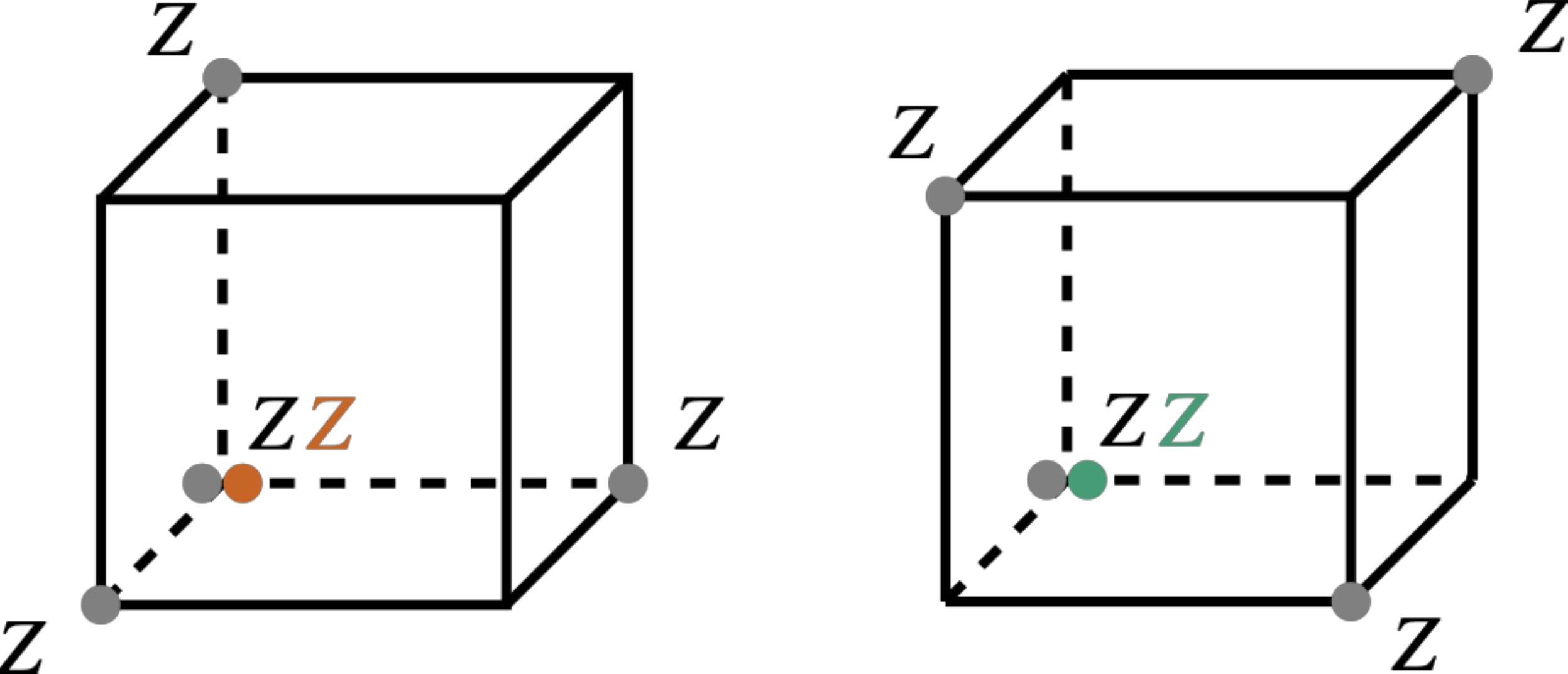} \, , \notag
\end{align}
where the orange and green dots represent distinct gauge qubits. 
Following Eq.~(\ref{puregauge}), the flux terms are given by the kernel of the constraint map. 
A generating set of flux terms is constructed by multiplying the gauged constraint terms on different 0-strata together, as shown in Fig.\ref{2Dnonlocal1}b. 
The size of the 3-strata is considered macroscopic compared to the finer lattice we have introduced.  
Thus the flux terms we find are not ultra-local. 
Fortunately, it is straightforward to avoid such non ultra-local flux terms. To achieve this we simply coarse-grain the fractal Ising model by a factor of two along each axis before gauging it.

The resulting coarse-grained fractal Ising model has 8 qubits per site. 
To construct a defect network, we again introduce a finer cubic lattice inside the cubes. To each vertex on the finer lattice we assign 8 qubits. 
These qubits are prepared in the ground state of the trivial Hamiltonian
\begin{align}
    H_{trivial} = - \sum_{v_i} Z(v_i) - J \sum_{v_i} X(v_i) \, ,
\end{align}
where $v_i$ is the label of vertices on the finer lattice and $i \in \{1,2,...,8\}$ is the label of qubits on each vertex. 
We refer to the qubits with the same label $i$ as being from the same layer. 
Next we apply CNOT gates between the original qubits and the trivial qubits of the same label $i$ to create 8 blocks of the Ising model within the 3-strata, following the same method as above in this and the previous section. 
We note this also involves choosing a new redundant generating set of constraints, along with an associated set of relations. 

In the failed construction above, we made all the fractal constraint terms ultra-local at 0-strata. 
Here, after coarse-graining, the 16 types of fractal Ising constraint terms split up into a pair of dimension 0, three pairs of dimension 1, three pairs of dimension 2, and a pair of dimension 3. 
Of the dimension 1 terms, a pair is associated to each lattice direction, and similarly for the dimension 2 terms. 
We next make the terms ultra-local by making a new choice of generating set for the constraints. This results in a pair of different terms assigned to each 3-strata, 2-strata, 1-strata and 0-strata. 
We further include redundant copies of each type of fractal Ising term in the 1-, 2-, and 3-strata, to make the Hamiltonian homogeneous over each strata. 
This step is accompanied by the introduction of additional relations due to the redundant copies. 
In this case, we find that it is possible to choose a generating set of relations that are ultra-local, similar to the X-cube example above. 
This is because the terms involved in each relation in the original uncoarse-grained fractal Ising model are all mapped into the vicinity of a single vertex by the coarse-graining step. 
Hence, after gauging a generating set of flux terms should also be ultra-local, which we confirm below. 

We now define the Hamiltonian of the coarse-grained fractal Ising defect network. 
On the 3-strata the terms are given by
\begin{align}
    H_3 = &-J \sum_{v_i} \adjincludegraphics[width=1.2cm,valign=c]{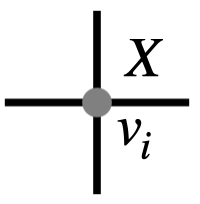} - \sum_{l_i} \adjincludegraphics[width=1.4cm,valign=c]{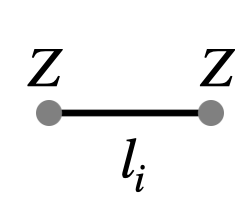} \nonumber\\
    &- \sum_{v} \adjincludegraphics[width=2.5cm,valign=c]{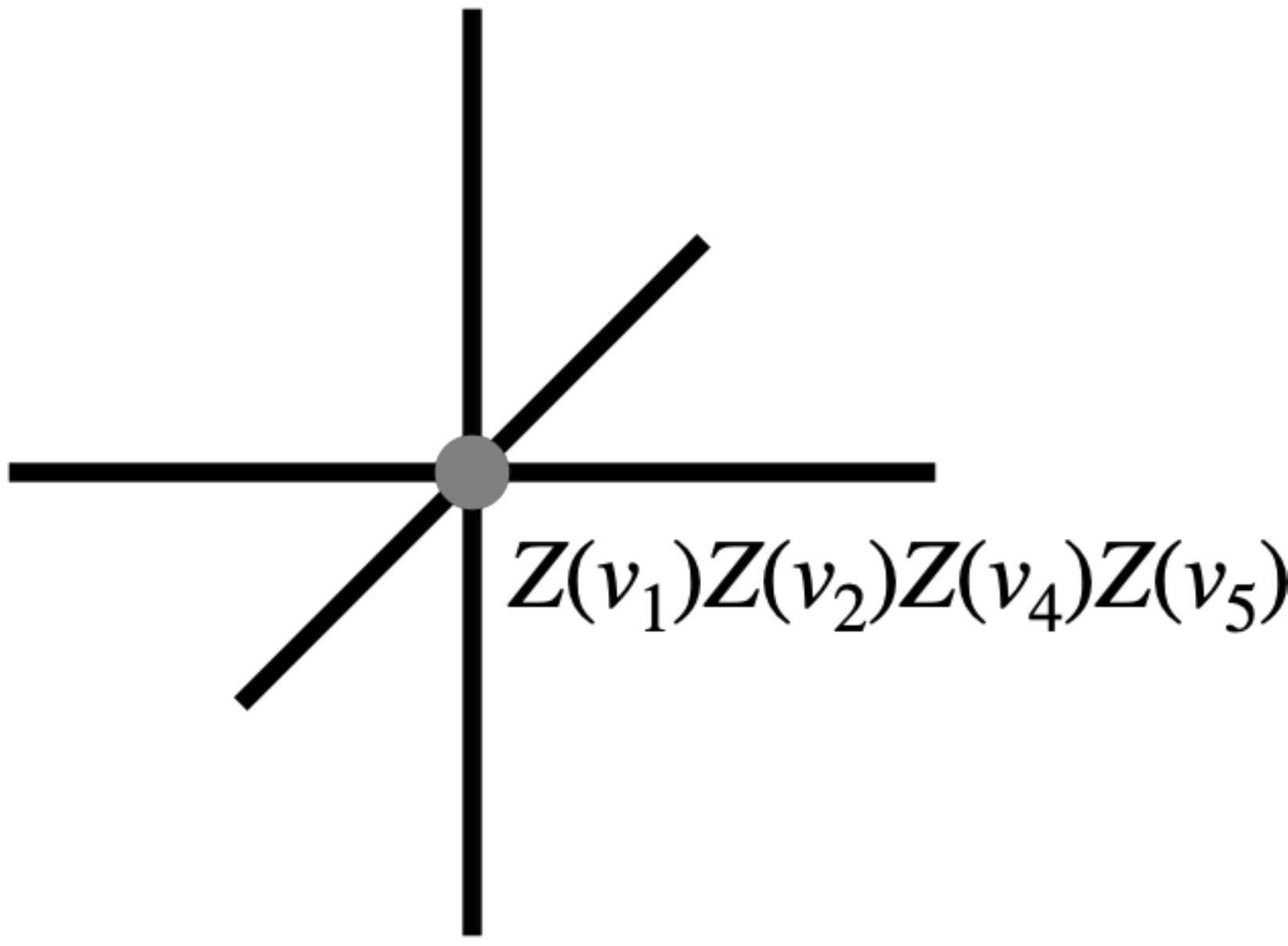} - \sum_{v} \adjincludegraphics[width=2.5cm,valign=c]{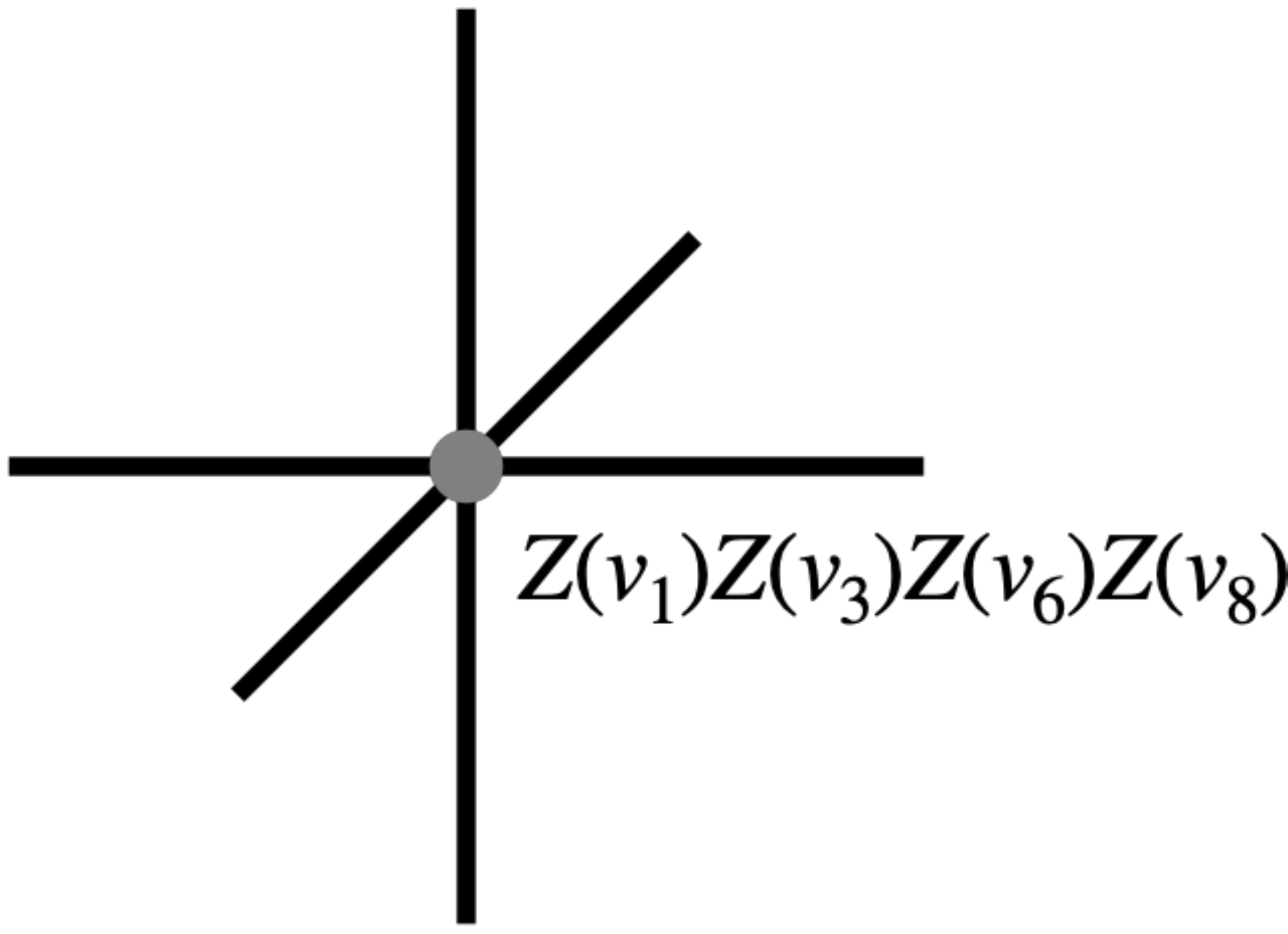}.  \label{Haah3}
\end{align}
The first two terms are Ising terms on each layer. The latter two terms are the 0-dimensional fractal constraints that couple different layers together at the same vertex. Hence, we see that some of the fractal terms become point-like within the 3-strata after coarse graining. 

On 2-strata, as no CNOT gates are applied between pairs of distinct 3-strata, there are no two-body Ising terms between the 3-strata. 
The ultra-local representation of the fractal constraint terms on 2-strata are line-like. They couple different vertices between neighboring 3-strata. 
The Hamiltonian on the $yz$-oriented 2-strata is given by
\begin{align}
    H_2 = &- \sum_{\tilde{l}} \adjincludegraphics[width=2.2cm,valign=c]{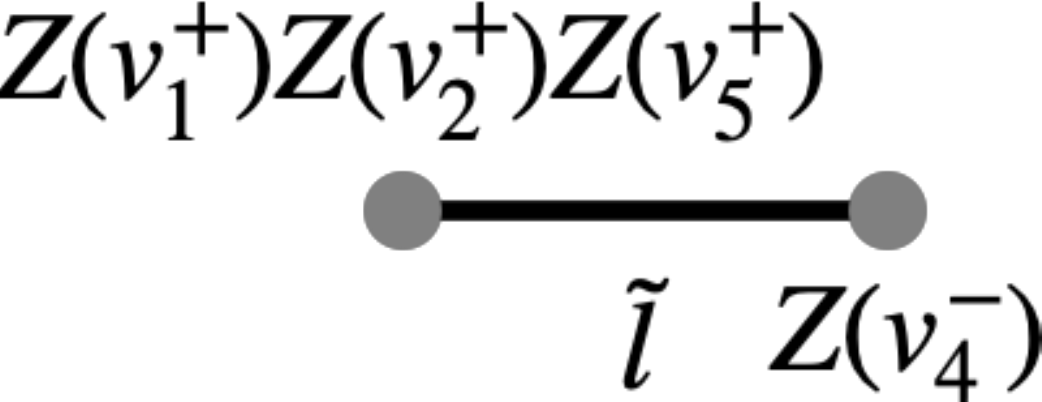} - \sum_{\tilde{l}} \adjincludegraphics[width=2.1cm,valign=c]{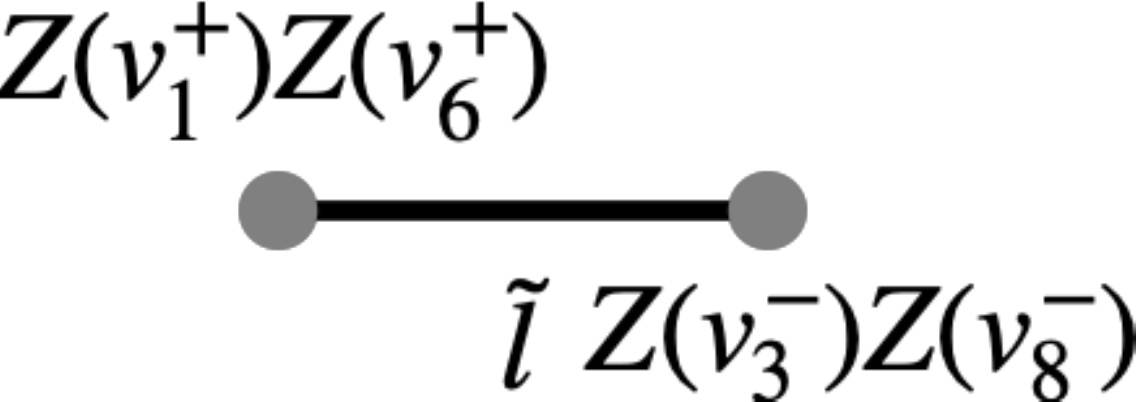}. \label{Haah2}
\end{align}
These two terms are the fractal constraint terms on 2-strata. Here, $\tilde{l}$ is the label of edges connecting two neighboring 3-strata and $+$ and $-$ are the labels of 3-strata.

Similarly, the fractal constraint terms on 1-strata are plaquette-like. They couple four distinct 3-strata. The Hamiltonian on $z$-oriented 1-strata is given by
\begin{align}
    H_1 = &- \sum_{\tilde{p}} \adjincludegraphics[width=2cm,valign=c]{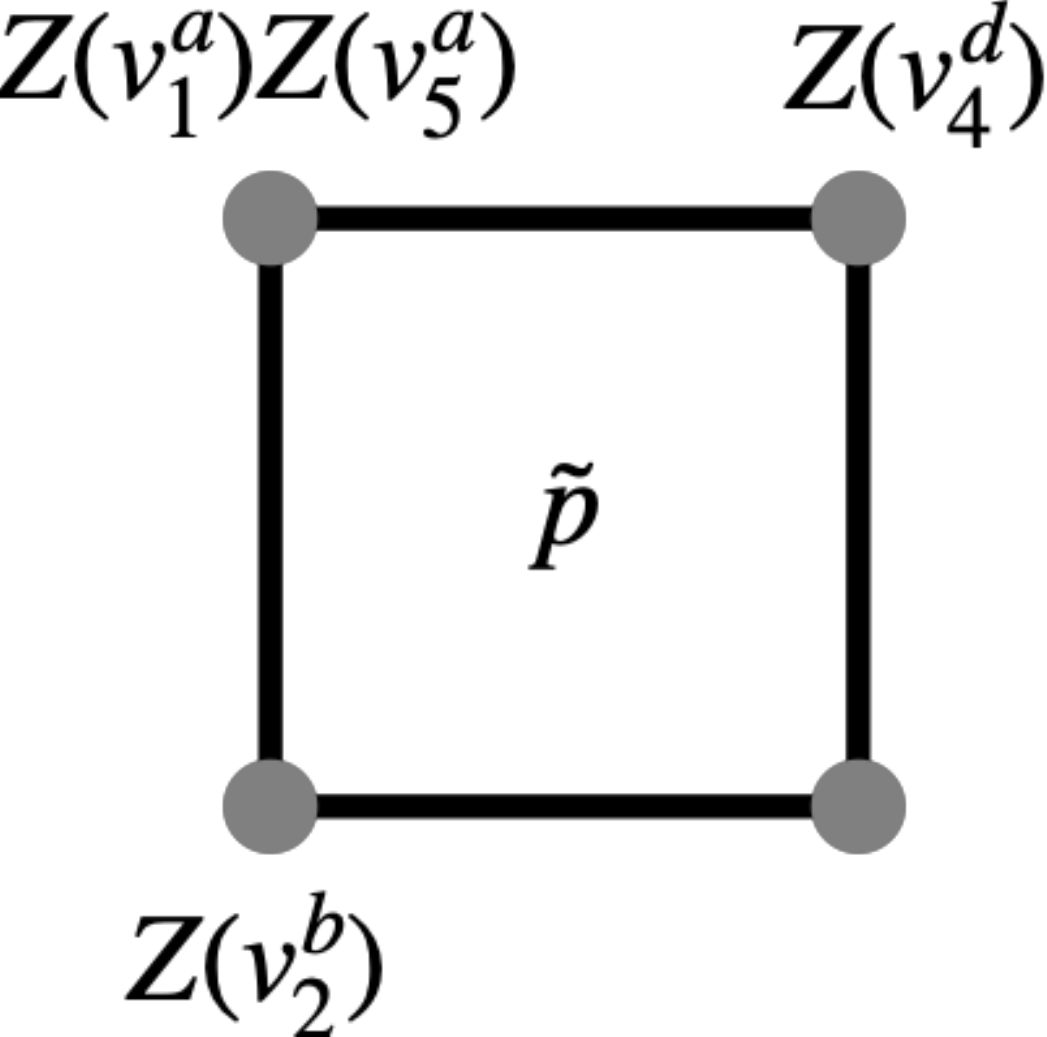} - \sum_{\tilde{p}}\adjincludegraphics[width=2cm,valign=c]{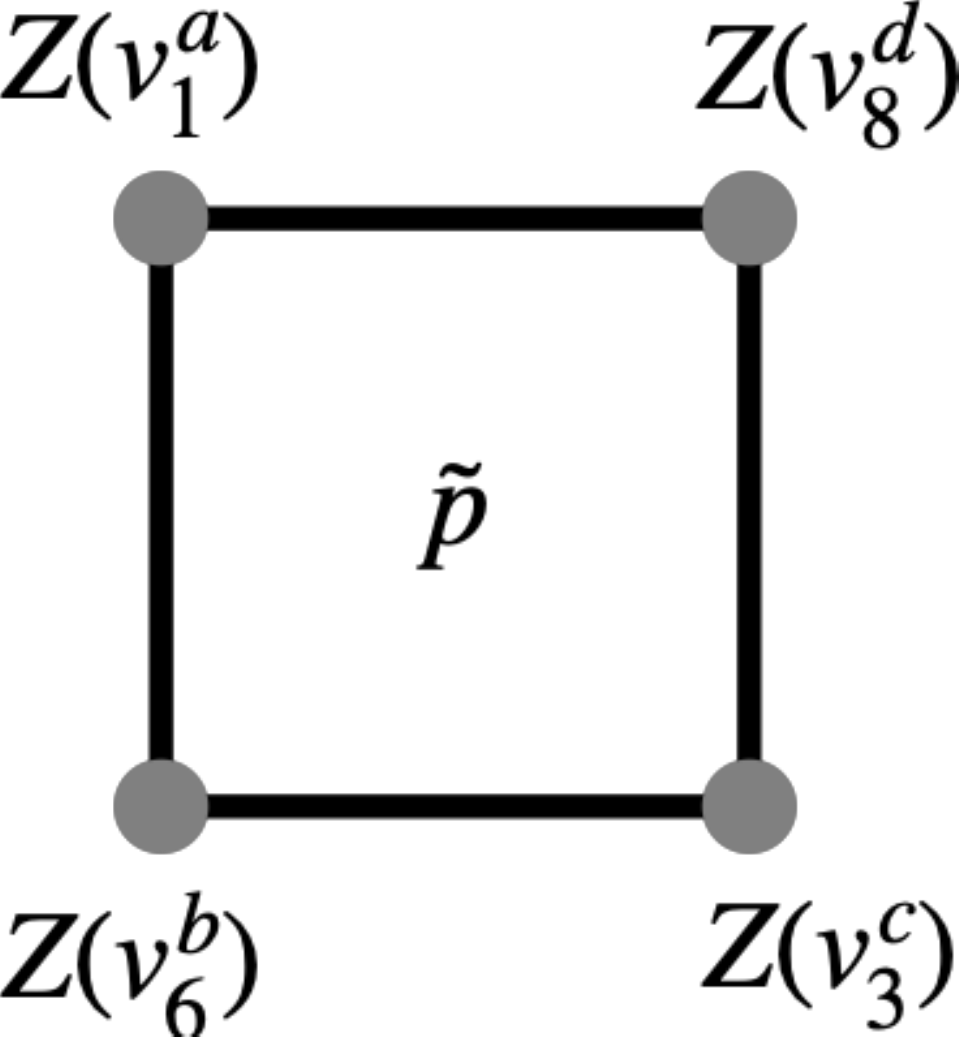} \label{Haah1}
\end{align}
$\tilde{p}$ is the label of plaquettes on 1-strata. $a, b, c, d$ are the labels of neighboring 3-strata. 

There are only two terms on each 0-stratum. The Hamiltonian is given by
\begin{align}
    H_0 =&- \adjincludegraphics[width=2.4cm,valign=c]{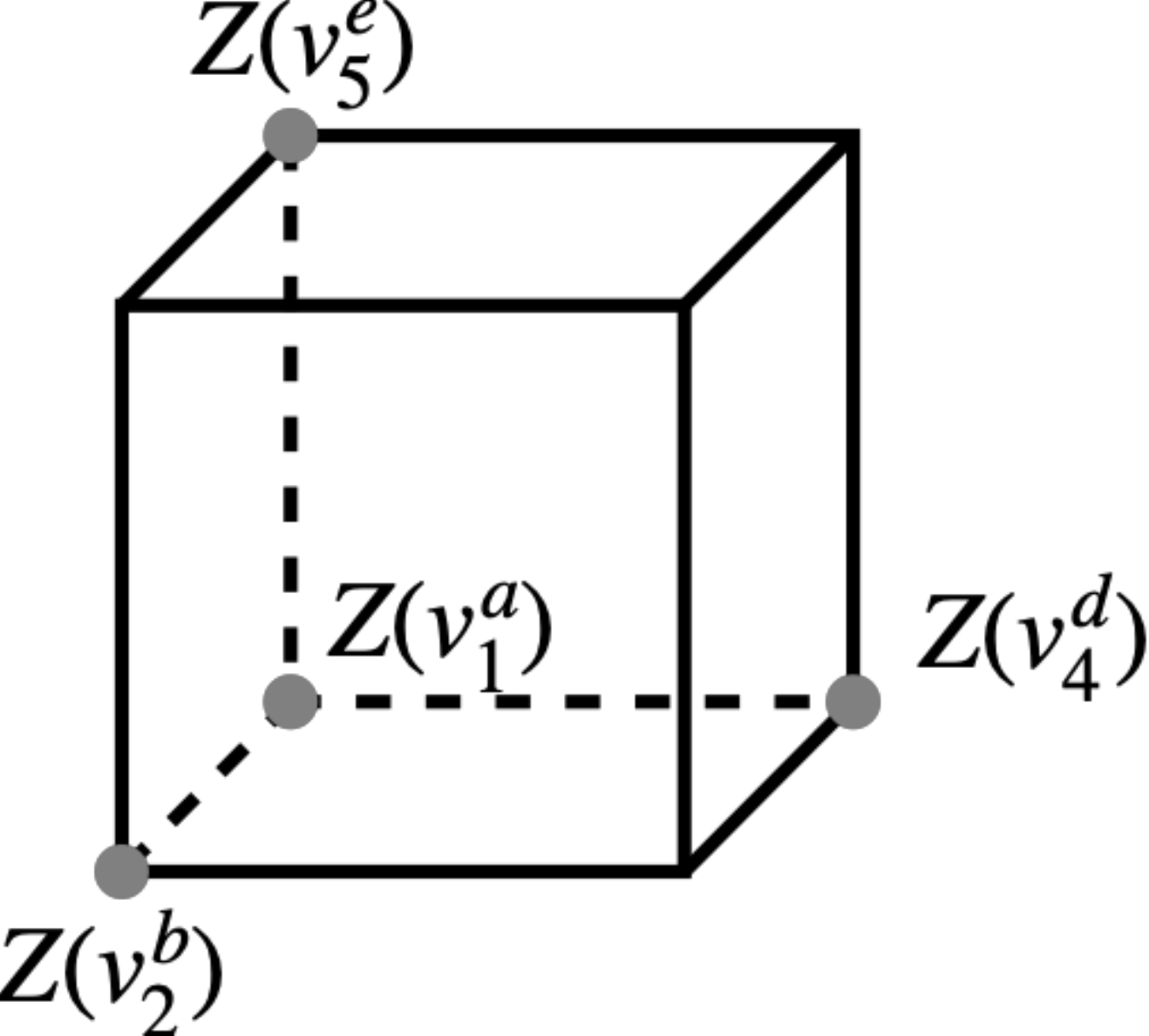} - \adjincludegraphics[width=2.4cm,valign=c]{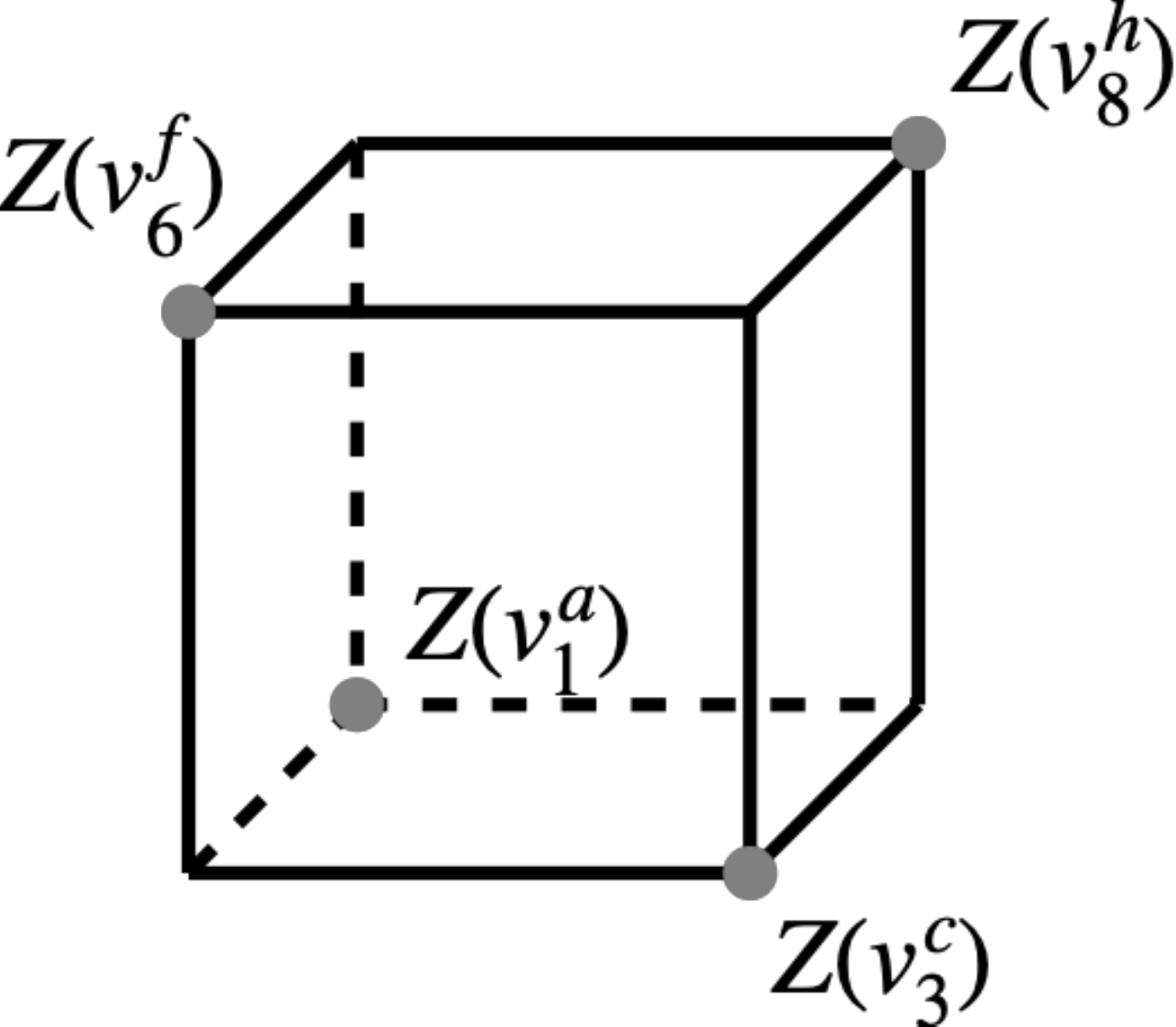} \label{Haah0}
\end{align}
We use $\{a,...,h\}$ to label the eight neighboring 3-strata.

While we have only described the Hamiltonian explicitly for certain orientations, due to the $2 \pi/3$ rotation symmetry of the cubic code about $(1,1,1)$, the Hamiltonians along the remaining orientations of 1- and 2-strata follow by applying the permutation of the axes $x\rightarrow y, y \rightarrow z, z \rightarrow x,$ which correspond to a permutation of labels $2 \to 4, 4 \to 5, 5 \to 2$ and $3 \to 8, 8 \to 6, 6 \to 3$. We use the standard cycle notation $(2 4 5)(3 8 6)$ to denote this permutation.

\subsection{TDN for Haah's cubic code} \label{HaahCC1}

To gauge the coarse-grained fractal Ising model defect network, we first couple gauge qubits to the constraint terms. 
As there are two different types of constraint terms, we introduce two different types of gauge qubits. 
The two gauged constraint terms of the fractal Ising model in the 3-strata, beyond the standard gauged two-body Ising terms, are given by,
\begin{align}
   \adjincludegraphics[width=7cm,valign=c]{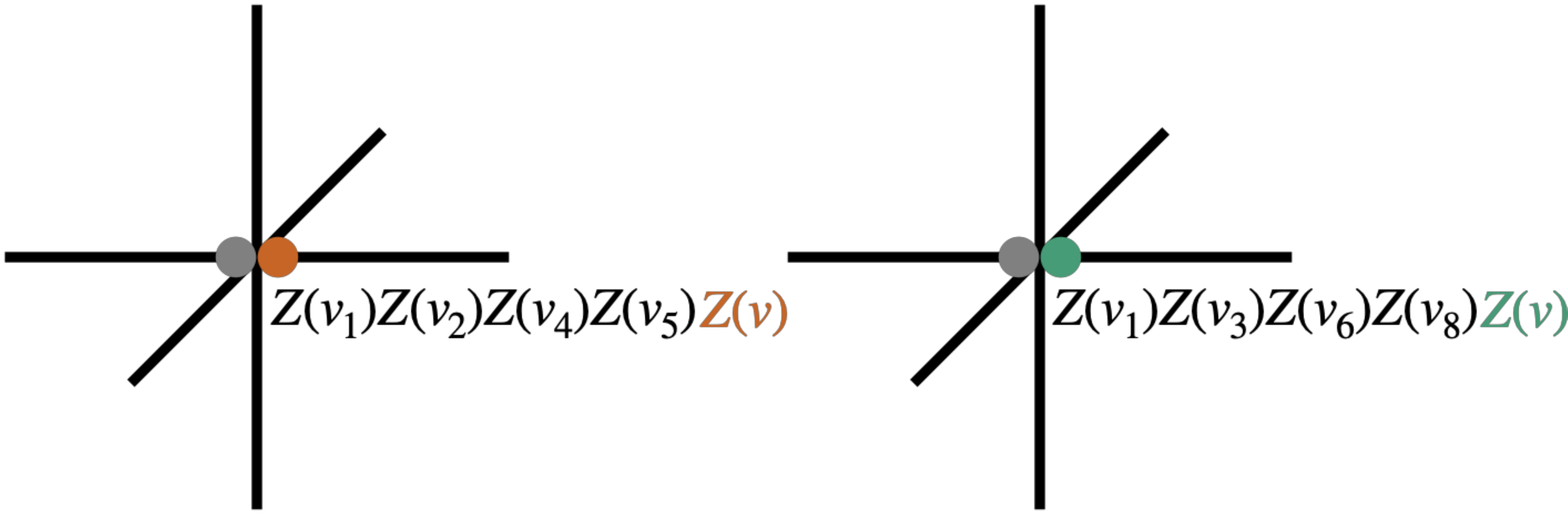}. \notag
\end{align}
The orange and green qubits depict the gauge qubits that associated to the fractal constraint terms. Following Eq.~(\ref{gaugesymmetry}), the Gauss's law operators within the 3-strata are given by,
\begin{align}
 \adjincludegraphics[width=2.3cm,valign=c]{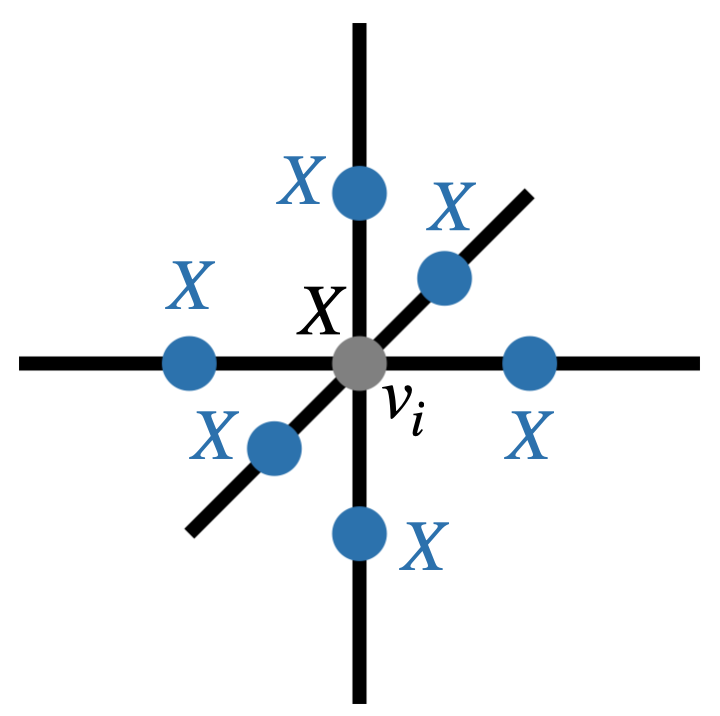}\quad \adjincludegraphics[width=2.3cm,valign=c]{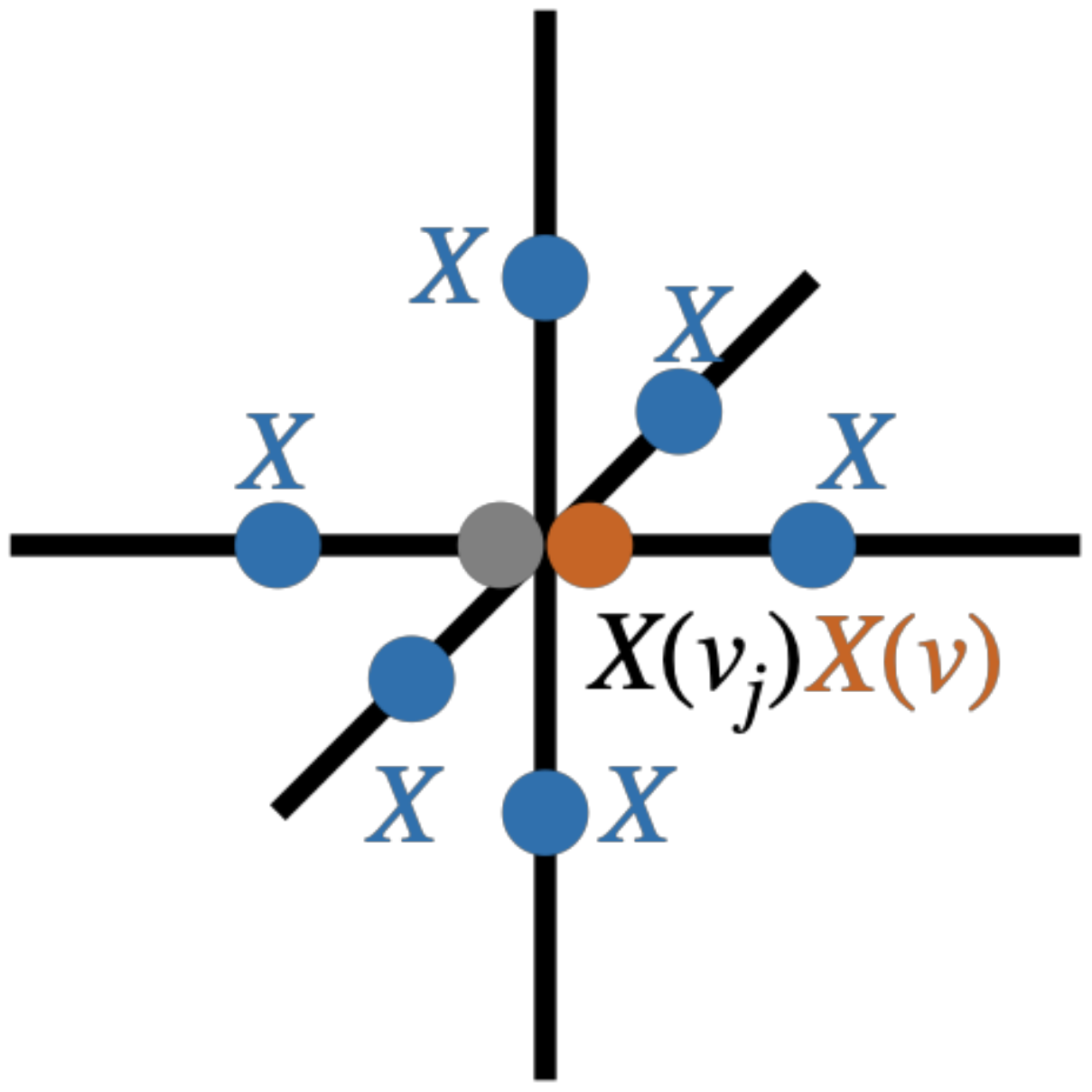}\quad
 \adjincludegraphics[width=2.3cm,valign=c]{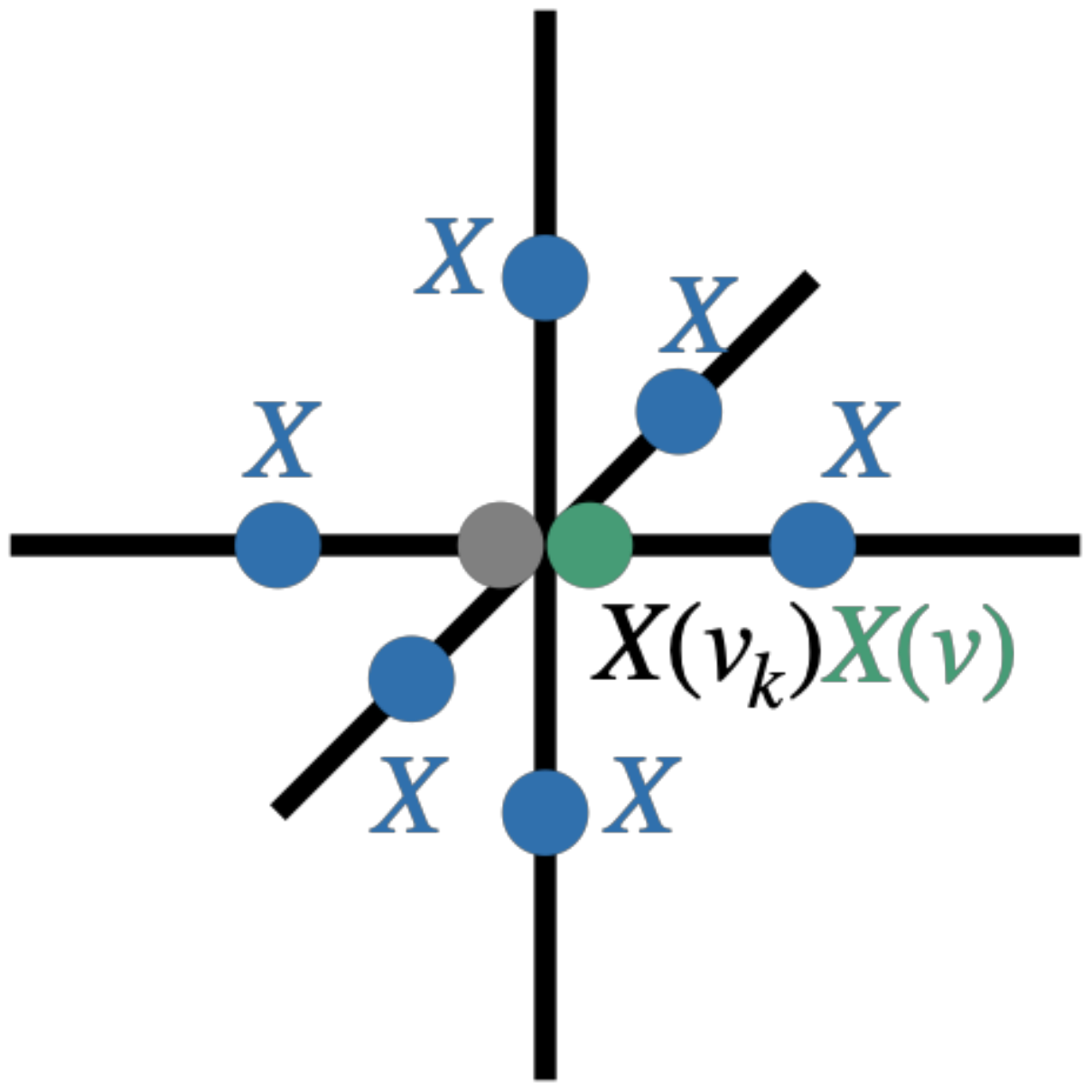}
\end{align}
where $i\in \{1,...,8\}$, $j\in\{1,2,4,5\}$ and $k\in \{1,3,6,8\}$. Thus, on 3-strata, by applying Eq.~(\ref{gaugeZ}) and Eq.~(\ref{loop}) to Eq.~(\ref{Haah3}) in the strong coupling limit, the gauged Hamiltonian is 
\begin{align}
    H_3^g = &- \sum A_3 - \sum B_3 \label{Haahga3} ,
\end{align}
where  $A_3$ are the Gauss's law terms, which are given by
\begin{align}
    \adjincludegraphics[width=7cm,valign=c]{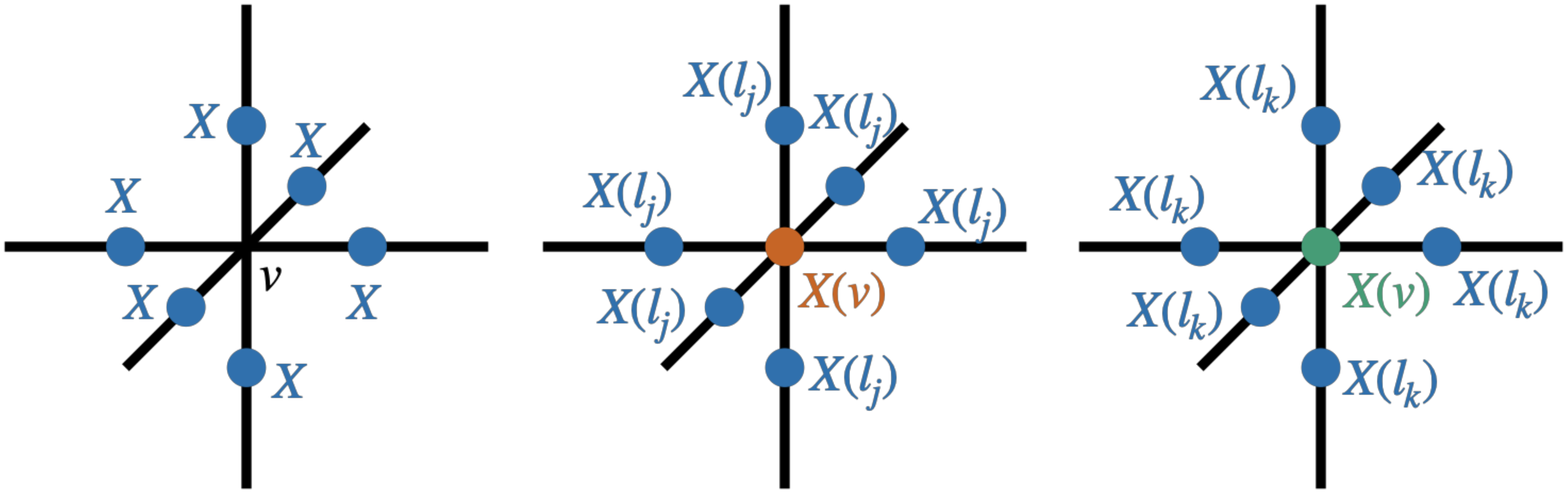} \, , \nonumber
\end{align}
where $j$ and $k$ here are the same as above. The $B_3$ terms are the flux terms, which are given by
\begin{align}
    \adjincludegraphics[width=7cm,valign=c]{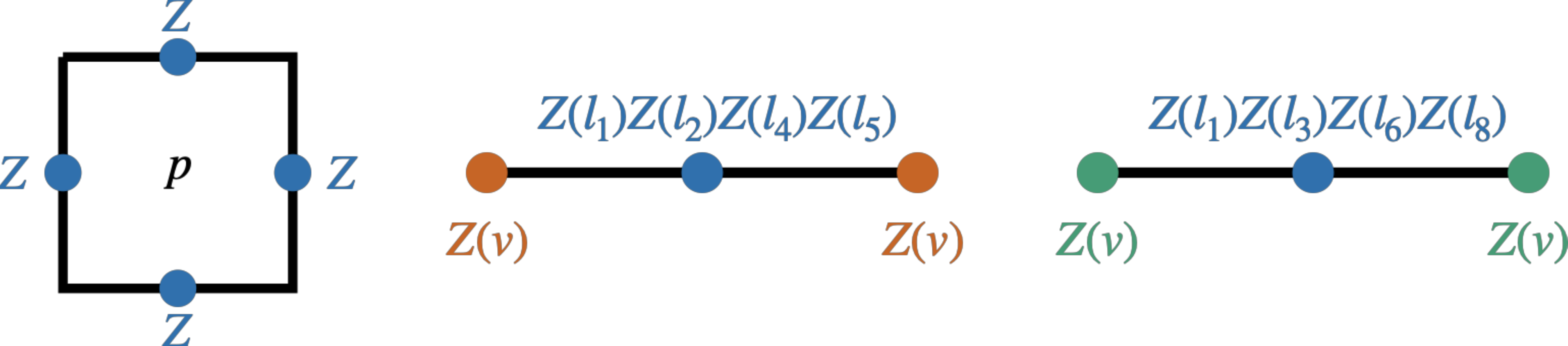}. \notag
\end{align}

The same method applies to Eq.~(\ref{Haah2}), resulting in the following gauged Hamiltonian on $yz$-oriented 2-strata in the strong coupling limit 
\begin{align}
    H_2^g = &- \sum A_2 - \sum B_2 . \label{Haahga2}
\end{align}
Here, the $A_2$ Gauss's law terms are given by
\begin{align}
    \adjincludegraphics[width=8cm,valign=c]{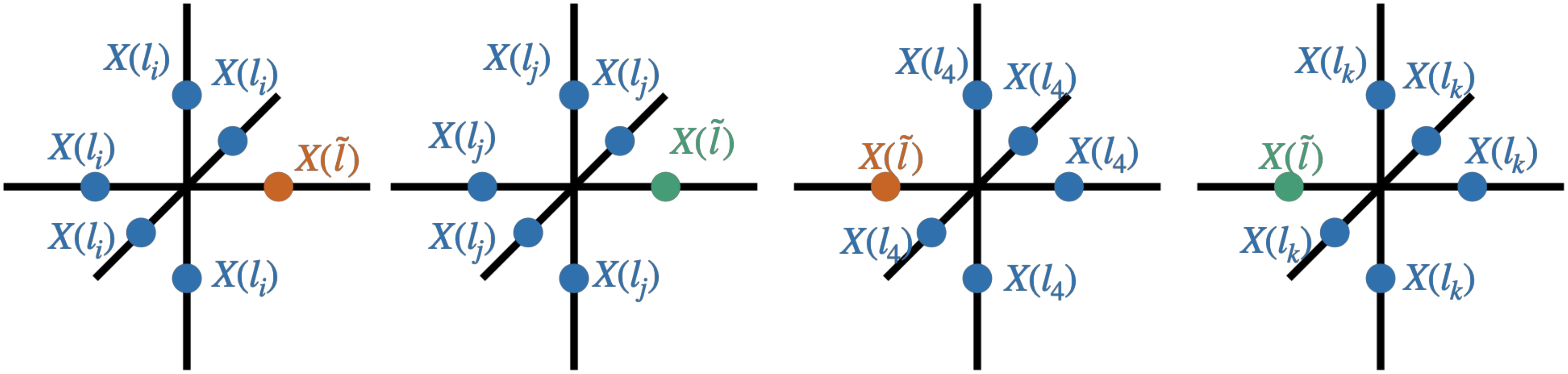} , \notag
\end{align} 
for 
$i \in \{1, 2, 5\}$, $j \in \{1, 6\}$ and $k \in \{3, 8\}$. The $B_2$ flux terms are given by
\begin{align}
    \adjincludegraphics[width=6cm,valign=c]{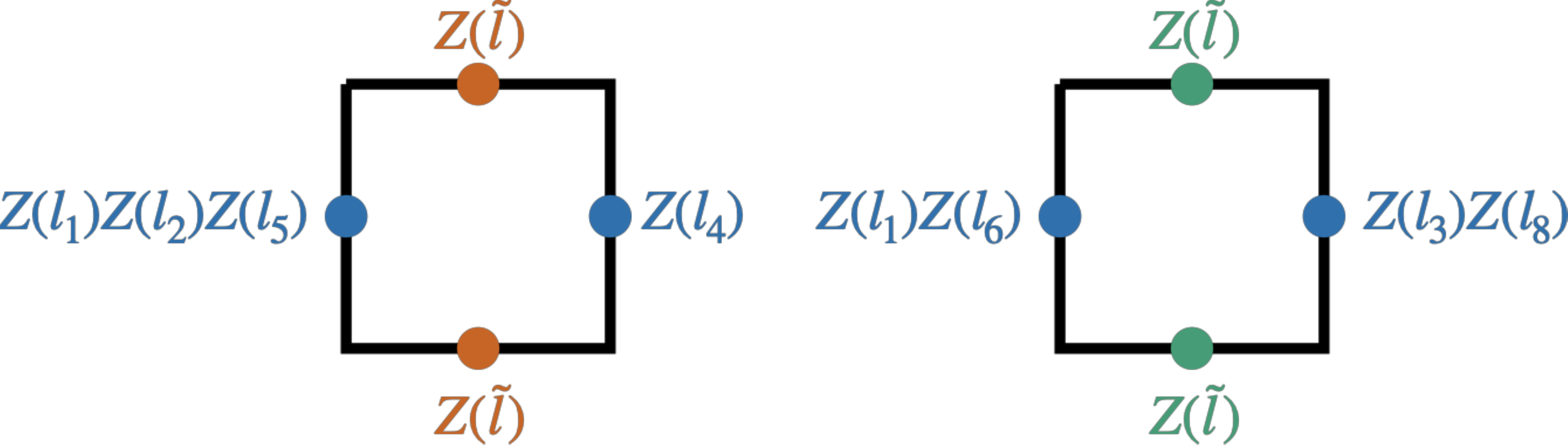} \, , \notag
\end{align}
where $\tilde{l}$ are the edges on the $yz$-oriented 2-strata that connect two different 3-strata.

The gauged Hamiltonian on 1-strata from Eq.~(\ref{Haah1}) is 
\begin{align}
    H_1^g = &- \sum A_1 - \sum B_1, \label{Haahga1}
\end{align}
where the Gauss's law terms $A_1$ are given by
\begin{align}
    \adjincludegraphics[width=8.2cm,valign=c]{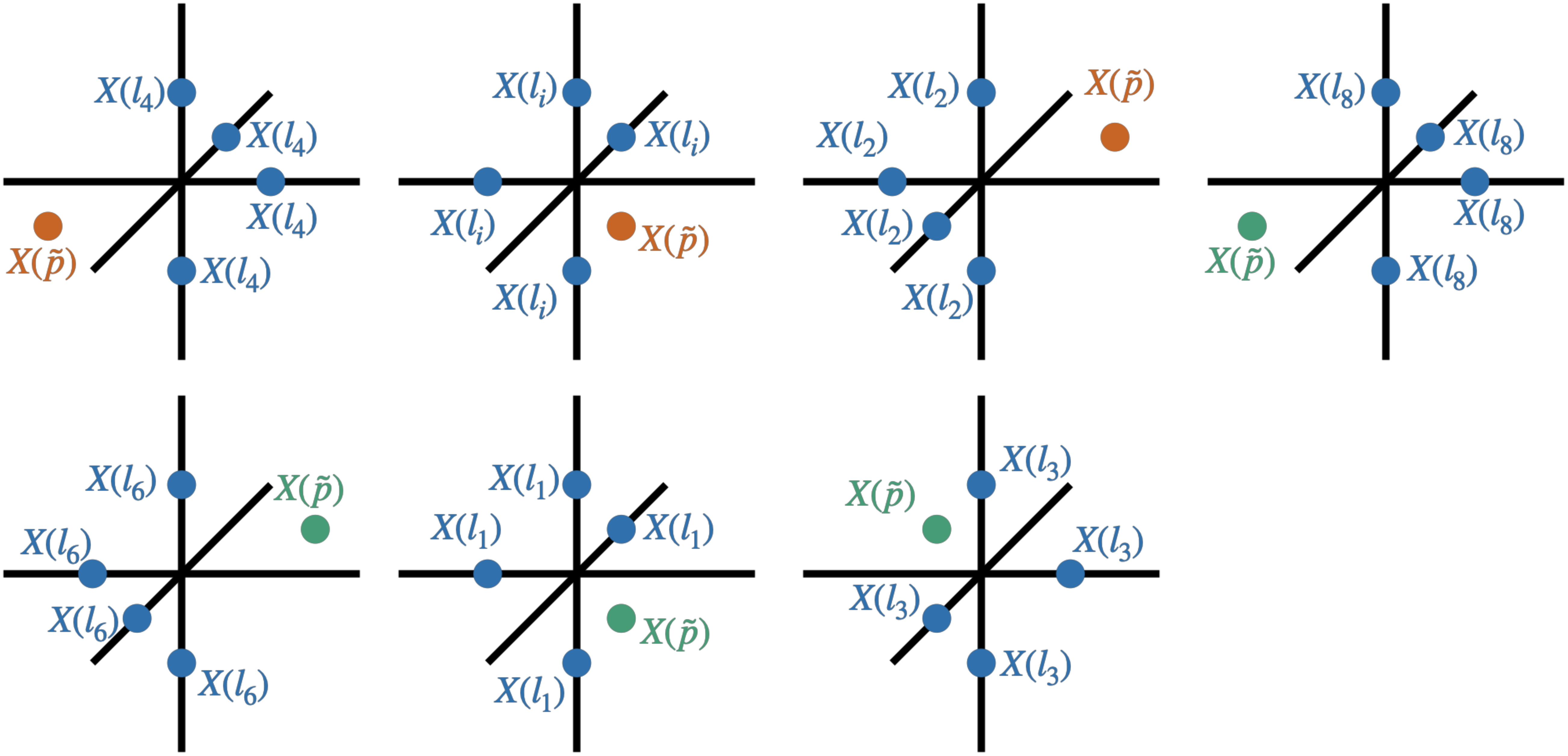}. \notag
\end{align}
While the flux terms $B_1$ are given by
\begin{align}
    \adjincludegraphics[width=5.6cm,valign=c]{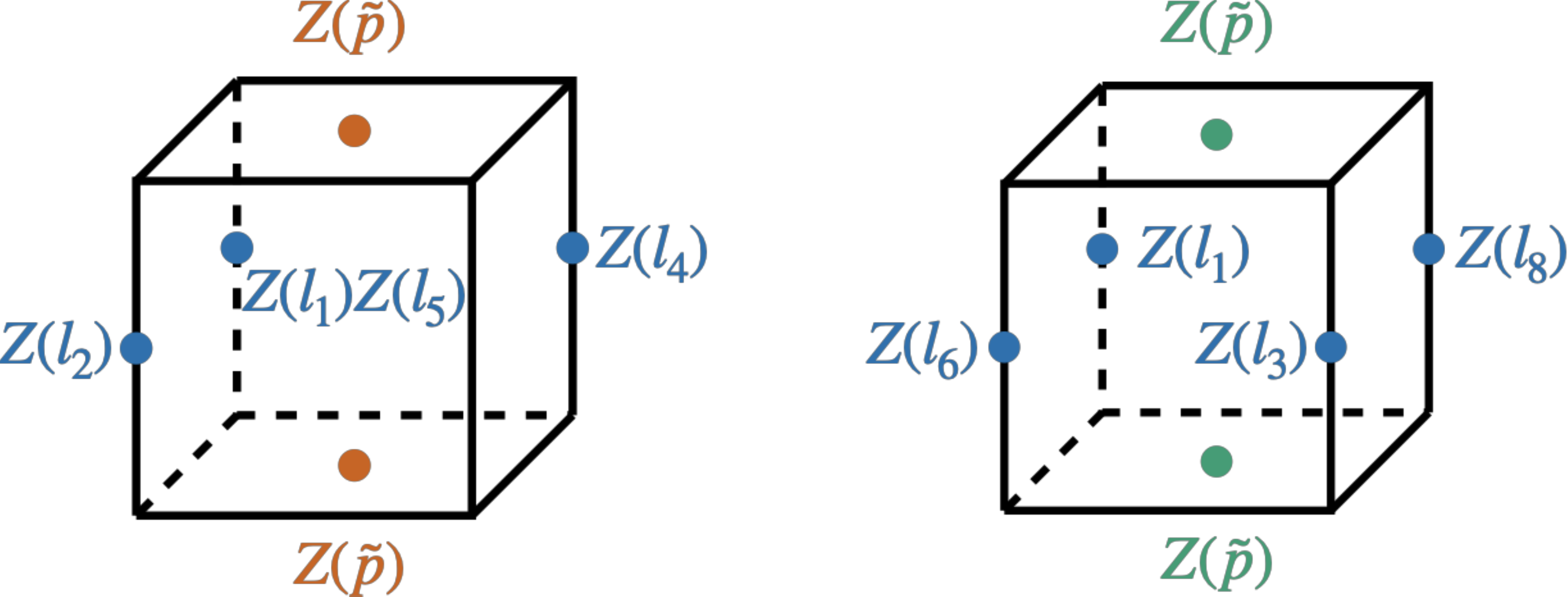}\, , \notag
\end{align}
where $\tilde{p}$ are the plaquettes on the $z$-oriented 1-strata that connect four different 3-strata.

As a result of gauging the Hamiltonian on 0-strata from  Eq.~(\ref{Haah0}), the Gauss's law terms near 0-strata are given by
\begin{align}
    \adjincludegraphics[width=8.2cm,valign=c]{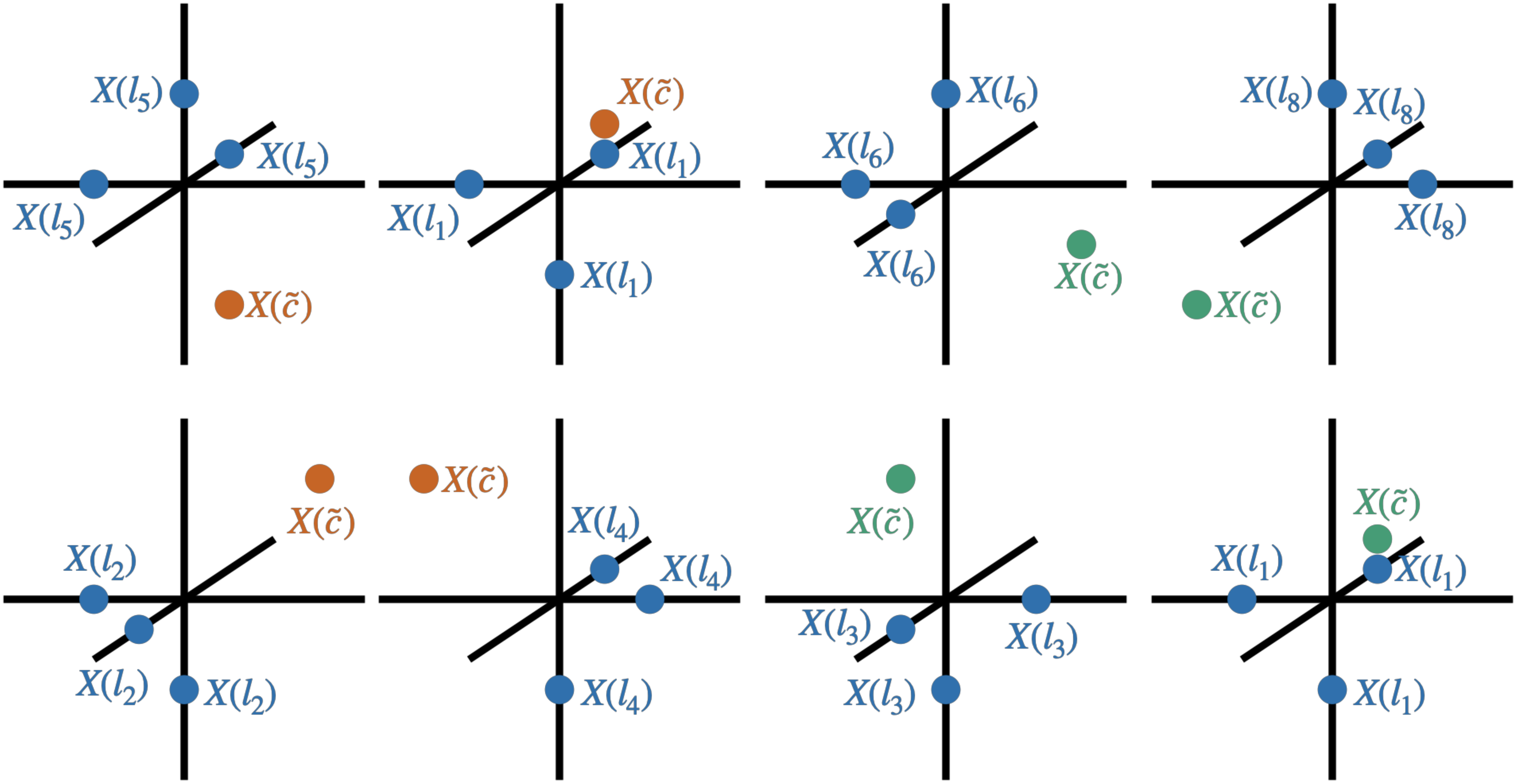}. \notag
\end{align}
As there are only two gauge qubits per 0-stratum, there are no flux terms strictly contained on the 0-strata. 
Rather, there are ultra-local flux terms adjacent to the 0-strata, which are given by
\begin{align}
     \adjincludegraphics[width=2.5cm,valign=c]{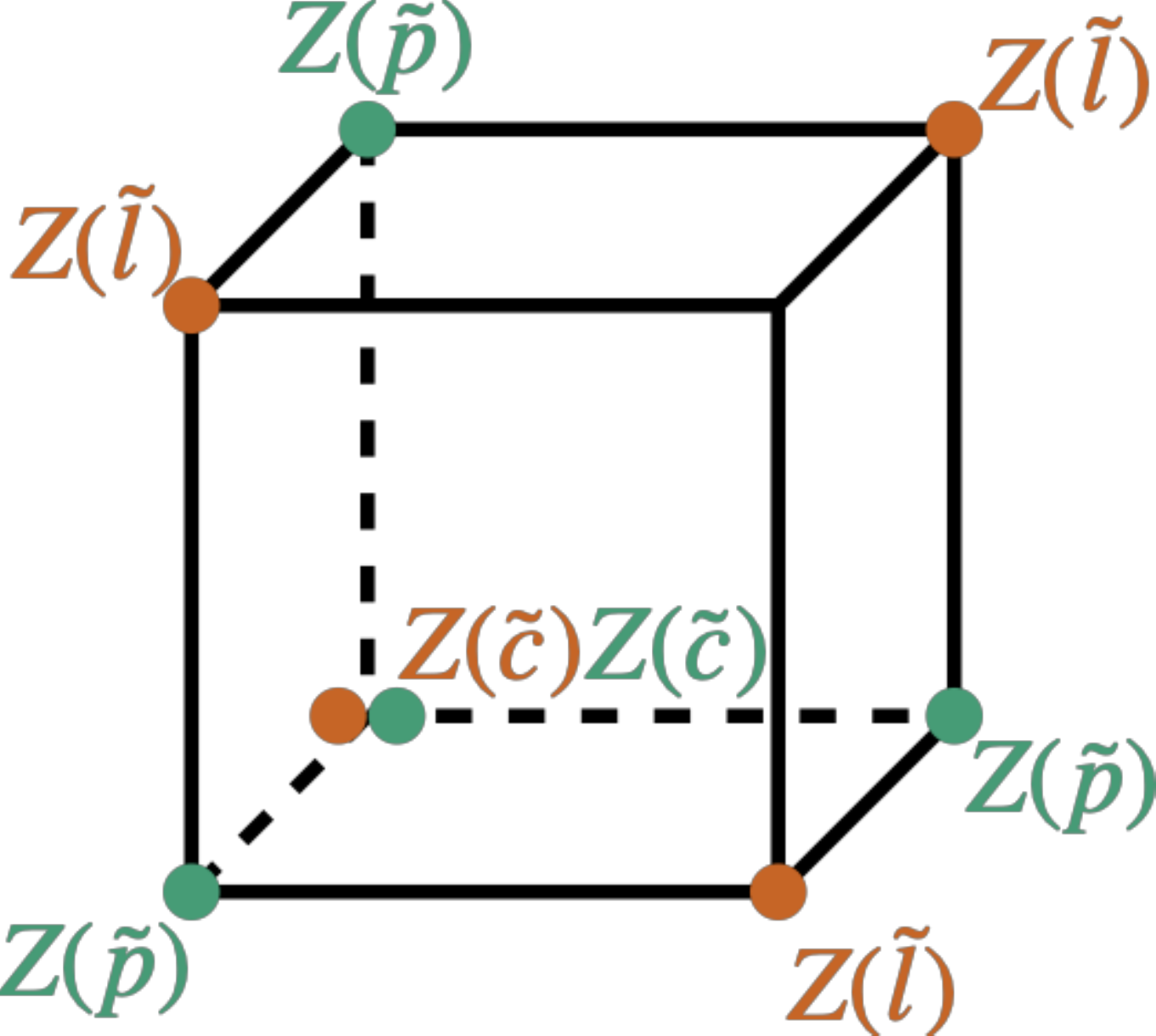} \, ,
     \label{gauged_H0}
\end{align}
where $\tilde{l}$ and $\tilde{p}$ are the edges and plaquettes on the neighbouring 2- and 1-strata. 
This is precisely the flux terms of Haah's cubic code 1 in Eq.~(\ref{HaahA}). Furthermore, this terms is ultra-local, which solves the non-locality problem that occurs in the uncoarse-grained model.

\subsection{Condensations and excitations}
\label{sec:Haahcond}

\begin{figure}[t!]
    \centering
    \includegraphics[scale=0.2]{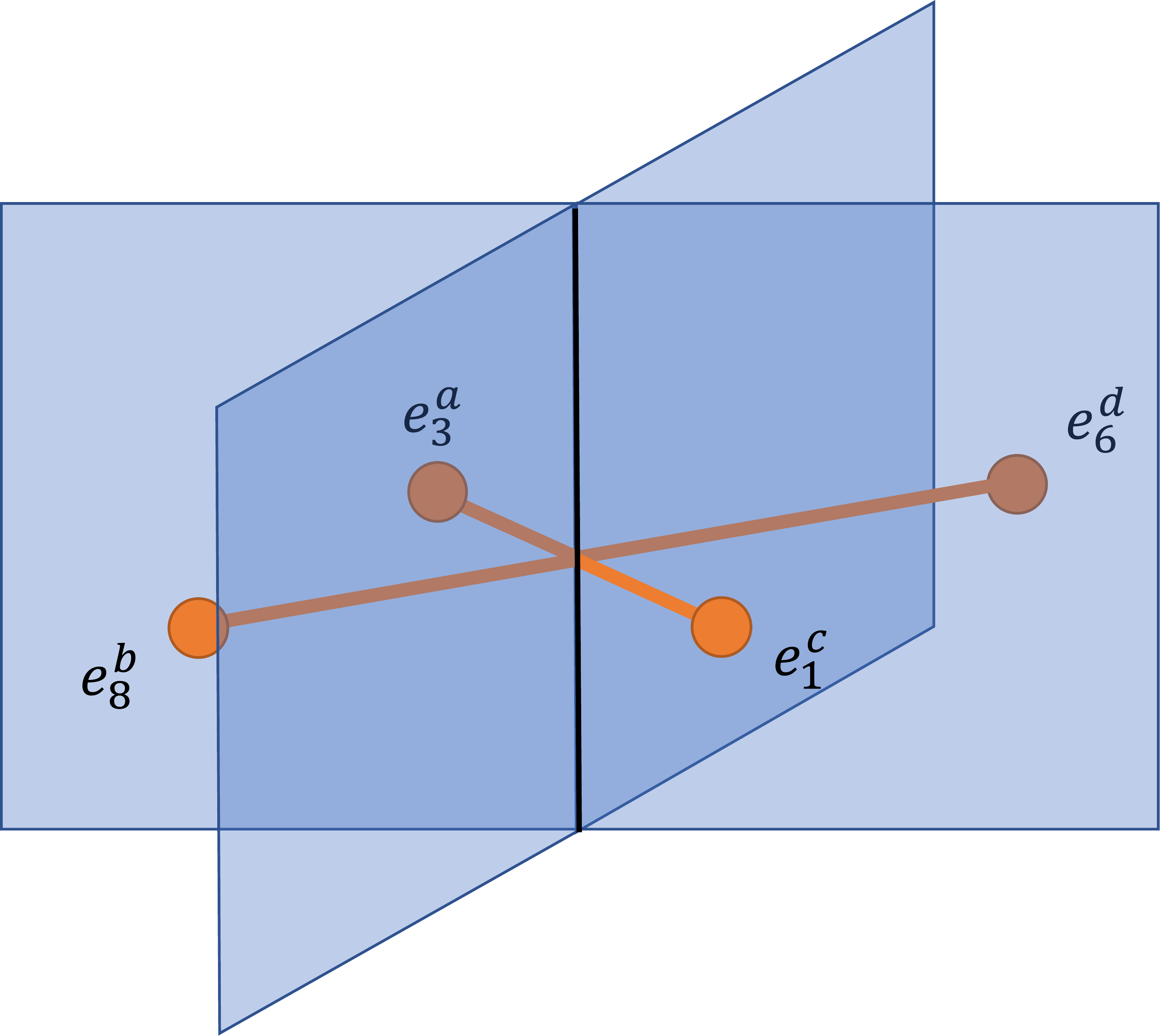}
    \caption{An example of an electric condensation on a 1-stratum. We label the four adjacent 3-strata as $\{a,b,c,d\}$. On each 3-stratum, there are 8 layers of 3D toric code. We label these layers $\{1,2,...,8\}$. 
    Using these labelling conventions, we depict a cluster of electric topological charges  $e_3^a e_7^a e_4^b e_2^d$ that condenses on the 1-stratum.}
    \label{Stratas_1}
\end{figure}

\begin{figure}[t!]
    \centering
    \includegraphics[scale=0.2]{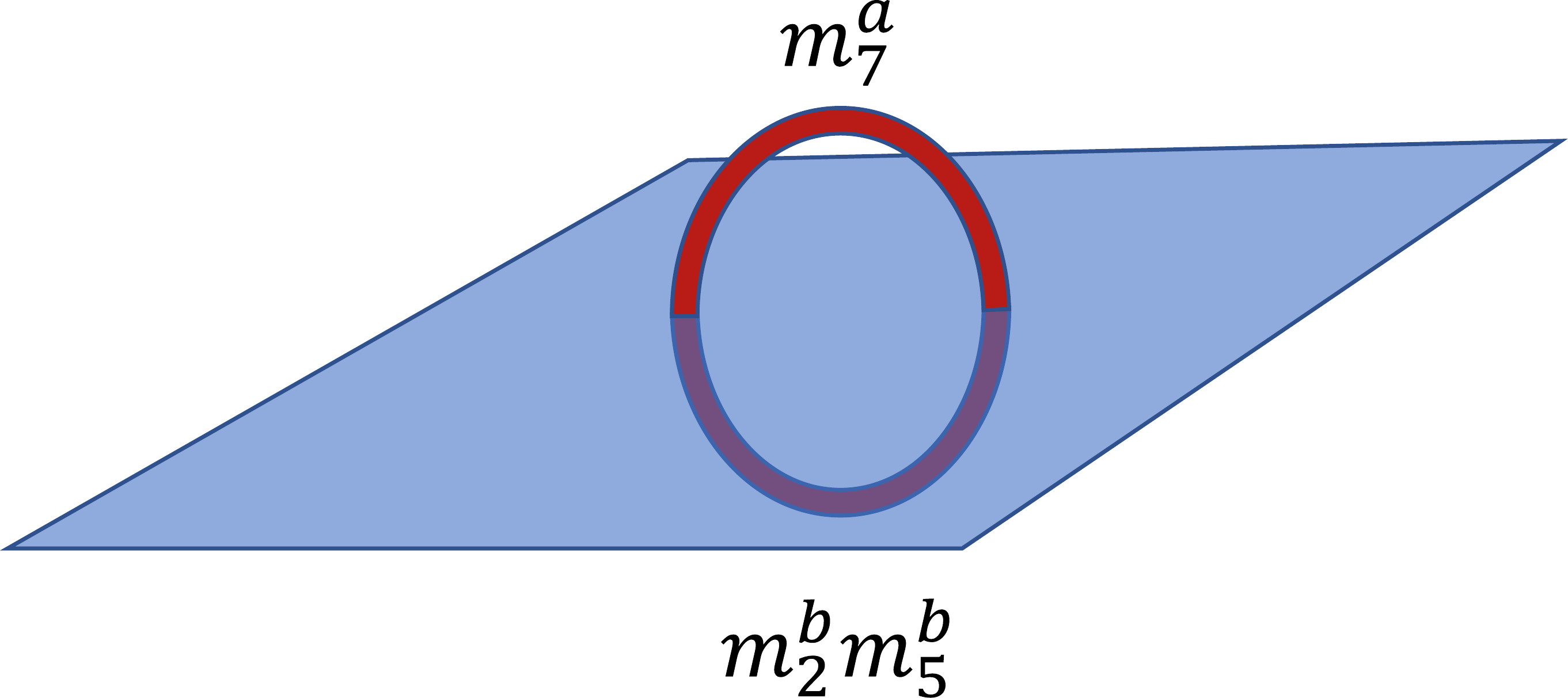}
    \caption{An example of a magnetic condensation on a 2-stratum. We label the adjacent 3-strata $\{a,b\}$. On each 3-stratum, there are 8 layers of 3D toric code labeled $\{1,2,...,8\}$. We depict a cluster of magnetic flux loops $m_7^a m_2^b m_5^b$ that  condenses on the topological defect on the 2-stratum, $m_7^a$ from above and $m_2^b m_5^b$ from below.}
    \label{Stratas_3}
\end{figure}

To conclude this section we derive electric and magnetic condensations induced on the various strata of the cubic code TDN we have constructed. 

We first consider the condensation of electric topological charges. 
As defined in Section \ref{sec35}, electric condensations can be created by local operators on the defects. The TDN of Haah's cubic code 1 not only has defects between strata, it also has couplings between different layers in the same stratum. 
So the electric condensations of the cubic code TDN can be created by local operators on defects and the non-trivial gauge qubits between layers. By checking Eq.~(\ref{Haahga3}), Eq.~(\ref{Haahga2}) and Eq.~(\ref{Haahga1}), we find the following electric condensations on 3-, 2- and 1-strata
\begin{align*}
    \text{3-strata} : \big\langle &e_1 e_2 e_4 e_5, e_1 e_3 e_6 e_8\big\rangle \\
    \text{2-strata($yz$)} : \big\langle &e_2^a e_3^a e_6^a e_1^b, e_2^a e_7^a e_4^b e_5^b  \\&+\text{four 3-strata electric condensations}\big\rangle\\
    \text{1-strata($z$)} : \big\langle &e_3^a e_7^a e_4^b e_2^d, e_3^a e_8^b e_1^c e_6^d \\&+\text{eight 3-strata electric condensations} \\
    & + \text{eight 2-strata electric condensations}
    \big\rangle.
    \label{eq:Haahcond}
\end{align*}
Each 2-stratum has two neighboring 3-strata, hence the condensations on the neighboring 3-strata also condense on this 2-stratum. 
Similarly, electric condensations on adjacent 2- and 3-strata also condense on a 1-strata. 
The superscript labels on excitations denote the 3-strata that contains them, this is depicted in Figs.~\ref{Stratas_1} \& \ref{Stratas_3}. 

The magnetic condensations, as described in Section \ref{sec35}, must braid trivially with the electric ones. 
They are also described by collections of magnetic excitations in the neighboring strata. 
Due to the electric condensations within the 3-strata certain combinations of $m$ excitations are confined, leaving six deconfined copies of toric code with deconfined $m$ loops generated by the following set
\begin{align}
    \big\langle &m_1 m_5 m_8, m_2 m_5, m_3 m_8, m_4 m_5, m_6 m_8, m_7\big\rangle.
\end{align}
We provide a complete list of electric and magnetic condensations below.\footnote{A condensible algebra (object) of bosons $A$ is Lagrangian if $(\mathrm{dim}\ A)^2 = \mathrm{dim}\ \mathcal{D}(\mathbb{Z}_2^n)$. Here $n$ is the number of $\mathbb{Z}_2$ layers and $\mathcal{D}(\mathbb{Z}_2)$ stands for the toric code topological order, or quantum double of $\mathbb{Z}_2$. 
In the cubic code TDN, on 3-strata $\mathrm{dim}\ A = 2^8$ and one can verify $(\mathrm{dim}\ A)^2 = 2^{16} = 4^8 = \mathrm{dim}\  \mathcal{D}(\mathbb{Z}_2^8)$. On 2-strata we have $(\mathrm{dim}\ A)^2 = 2^{32} = 4^{16} = \mathrm{dim}\ \mathcal{D}(\mathbb{Z}_2^{16})$. On 1-strata we have $(\mathrm{dim}\ A)^2 = 2^{64} = 4^{32} = \mathrm{dim}\ \mathcal{D}(\mathbb{Z}_2^{32})$. Hence the defects that induce these condensations fully gap out the relevant strata. } 
\begin{align*}
    \text{3-strata} : \big\langle &e_1 e_2 e_4 e_5, e_1 e_3 e_6 e_8\big\rangle \\
    \text{2-strata ($yz$)} : \big\langle &e_2^a e_3^a e_6^a e_1^b, e_2^a e_7^a e_4^b e_5^b, m_1^a m_5^a m_8^a,\\ 
    &m_2^a m_5^a m_1^b m_5^b m_8^b, m_2^a m_3^a m_8^a m_1^b m_8^b,\\ 
    &m_6^a m_8^a m_1^b m_2^b m_8^b, m_4^a m_5^a,m_7^a m_2^b m_5^b,\\ 
    &m_3^b m_8^b, m_4^b m_5^b, m_6^b m_8^b, m_7^b\\ 
    &+\text{four 3-strata electric condensations}\big\rangle\\
    \text{1-strata ($z$)} : \big\langle &e_3^a e_7^a e_4^b e_2^d, e_3^a e_8^b e_1^c e_6^d, m_1^a m_5^a m_8^a m_1^d m_4^d m_8^d,\\ 
    &m_2^a m_5^a m_1^b m_5^b m_8^b m_1^c m_4^c m_8^c m_3^d m_7^d m_8^d, \\&m_3^a m_8^a m_1^b m_2^b m_8^b m_1^c m_4^c m_8^c m_2^d m_3^d m_5^d m_6^d,\\ 
    &m_4^a m_5^a m_1^d m_5^d m_8^d, m_7^a m_2^b m_5^b m_2^d m_3^d m_4^d m_7^d m_8^d, \\
    &m_6^a m_8^a m_1^b m_2^b m_8^b m_1^c m_4^c m_8^cm_1^d m_3^d m_4^d m_7^d,\\ &m_3^b m_8^b m_3^d m_6^d, m_4^b m_5^b m_1^c m_5^c m_8^c m_2^d m_3^d m_5^d m_6^d, \\
    &m_6^b m_8^b m_1^c m_4^c m_8^c m_3^d m_7^d m_8^d, m_7^b m_4^c m_5^c,\\ 
    &m_2^c m_5^c m_7^d, m_3^c m_8^c, m_6^c m_8^c, m_7^c, \\
    &+ \text{eight 3-strata electric condensations}\\ 
    &+\text{eight 2-strata electric condensations}
    \big\rangle
\end{align*}
Due to the order three rotation about $(1,1,1)$ symmetry of the cubic code, one can straightforwardly find the condensations on the remaining 1- and 2-strata by relabeling the indices with the permutation $(2 4 5)(3 8 6)$. The definition of this notation is introduced at the end of Sec.~\ref{UngaugedCubicCodeTDN}. 
Again, the superscript labels on excitations denote the 3-strata that contains them, as depicted in Figs.~\ref{Stratas_1} \& \ref{Stratas_3}. 

Similar to the X-cube TDN, we pick a complete set of eigenvalues to diagonalize the operator algebra supported in the vicinity of the 0-strata~\cite{Aasen2020}, which is again abelian in this example. This corresponds to the inclusion of flux terms on the 0-strata in the lattice Hamiltonian. After lifting this local degeneracy, nothing further condenses at the 0-strata. 

\section{TDN representations of all topological CSS stabilizer models} \label{generalapproach}

In this section, we describe a general method to construct a TDN representation of any topological CSS stabilizer model. 
To perform our construction, we require the original topological CSS stabilizer model to satisfy a certain locality condition described below. This can always be ensured by coarse graining the model until the stabilizer generators are local to a single cube, and then performing one additional step of coarse graining by a factor of two along each axis. 

\subsection{The ungauged defect network}
We consider a topological CSS stabilizer model, whose $\mathbb{Z}_2$ matrix representation is given by
\begin{align}
    \sigma = \left(\begin{array}{cc}
         \sigma_X & 0\\
         0 & \sigma_Z
    \end{array}\right) \label{CSS_1}
\end{align}
By ungauging the model~\cite{Williamson_cubic_code} as described in Section \ref{Gauging1}, we find a generalized Ising model with a set of constraints 
\begin{align}
    \sigma' = \left(\begin{array}{cc}
          0\\
          \sigma_c
    \end{array}\right) \label{CSS_2} \, .
\end{align}
The matrix $\sigma_c = \sigma_X^T$ is the constraint map introduced in Eq.~(\ref{con}). 
The construction introduced in this section applies in any spatial dimension, for clarity we present it specifically for three dimensional space as we expect this to be the case of most immediate interest. 
In three dimensions, the entries of $\sigma_c$ are given by polynomials that depend on $1, x, y, z$ and higher order monomials such as $xyz$, $x^2y, x^3y^4z^5$, etc. 
We are working under the assumption that the original CSS stabilizer code has terms that are local to a single cube (this can always be ensured after sufficient coarse graining). Hence we can choose generators that do not involve any powers of $x,y,z$ that are higher than 1. 

We let $r$ denote the relation map for these constraints, as defined in Eq. (\ref{chaincomplex}). 
Then Eq.~(\ref{zrelation1}) can be written as
\begin{align}
    \sigma_c r = 0. \label{relations}
\end{align}
The relations determine the flux terms of the gauged model, as described in Eq.~(\ref{puregauge}). 
Hence if the flux terms of the gauged model, corresponding to the original topological CSS stabilizer code, are assumed to be local to a cube then a generating set of relations are also local to a cube. 
That is, the entries of the map $r$ do not involve any powers of $x,y,z$ that are higher than 1. 

For our construction to produce a valid TDN we require that there is a generating set of relations that are not only cube local, but satisfy a \textit{stronger locality condition}. 
This condition corresponds to having \textit{all operators} involved in the generating relations being cube local. 
This is a stronger condition than simply requiring the relation generators themselves to be cube local, as such relations can include nearest neighbor constraints which themselves can involve nearest neighbor operators. 
Overall this can lead to next nearest neighbor operators being involved in a single relation. 
This results in non ultra-local terms in the TDN, as we encountered in the failed first attempt at constructing a TDN for cubic code, described in Section~\ref{UngaugedCubicCodeTDN}. 
Coarse graining the original model until the generators are cube local, followed by an addition coarse graining step by a factor of two along each axis is enough to ensure that the desired strict locality condition is satisfied. 
This is simply because any next nearest neighbor operators that may be involved in a relation become nearest neighbor after the additional coarse graining. 
We remark that the additional coarse graining is sufficient, but may not be necessary since models such as X-cube and Haah's cubic code 1B~\cite{haah2014bifurcation} directly satisfy the strict locality condition. 
In fact all the models considered in Ref.~\cite{Aasen2020} satisfy the strict locality condition without further coarse graining. 

To convert the generalized Ising model into a defect network 
we follow Section~\ref{Plaquette} and introduce a much \textit{finer} lattice scale that is shifted by half a lattice spacing along each axis, such that the vertices of the original \textit{coarser} lattice are contained within cubes of the finer lattice. 
The unit cells of the coarser lattice define the 3-strata, the faces define the 2-strata, the edges define the 1-strata and the vertices define the 0-strata. 
If there are multiple qubits on each vertex on the coarse-grained lattice we label them $\{a,b,c,...\}$. Then on the finer lattice we add the same number of qubits to each vertex and label them by $\{a,b,c,...\}$ as well. 
The qubits that have the same label are considered to lie within the same layer. 
For each layer, we add qubits governed by a trivial Hamiltonian to the vertices of the finer lattice in the ground state of 
\begin{align}
    H_{trivial} = -\sum_v Z(v) - J X(v), \label{trivialqubits}
\end{align}
in which $J$ is a tunable coupling factor and $v$ labels vertices of the finer lattice. 

In the polynomial matrix representation, adding trivial qubits is equivalent to adding blocks of identity matrices to $X$ and $Z$-sectors of Eq.~(\ref{CSS_2}). The Pauli-$Z$ constraint sector of the new matrix is given by
\begin{align}
    \sigma_c' \rightarrow \left(\begin{array}{cc}
            \sigma_c & 0 \\
           0 & \mathbb{1} 
    \end{array}\right).
\end{align}
In this equation $\sigma_c$ now describes the constraints on the coarser lattice scale and the identity matrix describes the newly added qubits on the finer lattice. 

Next, we apply CNOT gates from the original qubits to the corresponding trivial qubits within the same layer and 3-stratum. The CNOT gates are defined in Eq.~(\ref{cnotz}). This results in 
\begin{align}
    \sigma_c' \rightarrow \left(\begin{array}{ccccc}
            \sigma_c & I & I & ... & I\\
           0 & I & 0 & ... & 0\\
           0 & 0 & I & ... & 0\\
           \vdots & \vdots & \vdots & \ddots & \vdots\\
           0 & 0 & 0 & ... & I
    \end{array}\right)_{n_v L \times n_v L}, \label{add_trivial_qubits}
\end{align}
where $I$ are $n_v \times n_v$ identity matrices, $n_v$ is the number of qubits per vertex on the finer lattice, and $L$ is the number of vertices of the finer lattice in each 3-stratum. In Eq.~(\ref{add_trivial_qubits}) we have a complete set of $ZZ$ constraint terms between the qubits on each layer of each 3-stratum. We next add redundant $ZZ$ terms to all edges of the finer lattice within the 3-strata. This corresponds to adding linearly dependent columns to $\sigma_c'$. 
This also introduces additional relation generators, which corresponds to adding columns to the relation map $r$. 
Then we change the basis of constraint generators to make all the $ZZ$ terms ultra-local i.e. nearest-neighbor on the finer lattice. This is implemented by applying column operations, as defined in Eq.~(\ref{column}), to the constraint map $\sigma_c'$. 
We similarly choose a basis of relation generators that is ultra-local on the finer lattice. 
Schematically, the resultant matrix for the Pauli-$Z$ constraints is then
\begin{align}
    \sigma_c^{'} = \left(\begin{array}{c|c}
          \sigma_c & \sigma_{Ising}
    \end{array}\right). \label{3strata}
\end{align}
The left part, $\sigma_c$, represents the generalized Ising constraint terms on the coarse lattice, 
while the right part, $\sigma_{Ising}$ represents the standard Ising terms on the finer lattice within each 3-stratum. 
To summarize, we have $n_v$ layers of the standard Ising model within each 3-stratum which are coupled by the constraint terms $\sigma_c$ on the larger lattice scale.

To make the $\sigma_c$ couplings in Eq. (\ref{3strata}) ultra-local, we follow a similar strategy as we did for the $ZZ$ constraint terms within the 3-strata.
We first move each coupling term to become local at an appropriate  3-, 2-, 1-, or 0-strata, depending on its form, by multiplying with $ZZ$ edge terms within the 3-strata. This is implemented by column operations on the $\sigma_c'$ map.  
Next we add redundant copies of each term along the corresponding 3-, 2-, 1-, or 0-strata, to make the couplings homogeneous along these strata. This corresponds to adding linearly dependent columns to $\sigma_c'$. 
This also introduces linearly dependent columns to the relation map $r$, we can further choose a basis of relation generators that are ultra-local, which is implemented by column operations on $r$. 
At this point we have arrived at a Hamiltonian for the ungauged generalized Ising defect network that is ultra-local, and each step was simply implemented by column and row operations on $\sigma$ and $r$, which correspond to local unitary circuits, the addition of auxiliary qubits and the redefinition of generators for a stabilizer group, all of which preserve the quantum phase of matter. 
This completes the construction of the ungauged defect network.

\subsection{Gauging the ungauged defect network}

To obtain a TDN in the same topological phase of matter as the original CSS stabilizer model, we gauge the ungauged defect network constructed in the previous subsection. 
Following the formalism reviewed in Section~\ref{GaugingSpinModels}, we assign a gauge degree of freedom to each constraint term. 
The Gauss's law operators are then given by Eq.~(\ref{gaugecommu2}) and Eq.~(\ref{gaugesymmetry}). 
The flux operators are determined by the ultra-local relations $r_l \in r$ we defined in Eq.~(\ref{relations}).  
In general, three different kinds of flux terms appear: 
\begin{enumerate}
    \item \textbf{The relations of the Ising models within 3-strata:} After gauging, each layer of the standard Ising model becomes a layer of toric code, as described in Section \ref{toriccode}. 
    
    \item \textbf{The relations of the ungauged stabilizer model:} Following the construction outlined above, the constraints of the ungauged model now have ultra-local relations 
    that can be realized near individual 0-strata. 
    The flux terms that are associated with these relations are essentially the same as the flux terms of the original stabilizer model. 
    For example, see the flux term for Haah's cubic code A in Eq.~(\ref{gauged_H0}), Section \ref{HaahCC1}. 
    
    \item \textbf{The relations involving $ZZ$ constraint terms and constraint terms of the ungauged stabilizer model:}
    The constraint terms of the ungauged stabilizer model on 0-, 1-, 2-, and 3- strata can multiply with the neighboring standard Ising $ZZ$ constraint terms in the 3-strata, to form a non-trivial set of relations on lower dimensional strata. 
    Some of the unconventional mobility properties of fractons that occur in TDNs are consequences of the flux terms derived from these relations. 
    Example of such terms are given in Eq.~(\ref{H1XFlux}) for the X-Cube model and Eq.~(\ref{Haahga3}), Eq.~(\ref{Haahga2}) and Eq.~(\ref{Haahga1}) for Haah's Cubic Code A.
\end{enumerate}
The two-step coarse-graining process assumed throughout this section guarantees all of the above relations are ultra-local, as explained in the previous subsection. 

The electric topological excitations that condense on different strata can be found by applying local $Z$ operators on the gauge qubits corresponding to non-trivial constraint terms. These operators commute with the corresponding flux terms in the Hamiltonian, but anti-commute with the Gauss's law terms in the neighboring 3-strata. 
This process allow the clusters of electric charge that condense in the vicinity of each strata to be calculated using the TDN Hamiltonian. 
For examples see Eq.~(\ref{X0Strata4}) for the X-cube model and Eq.~(\ref{Haahga3}), Eq.~(\ref{Haahga2}) and Eq.~(\ref{Haahga1}) for Haah's Cubic Code A. 

As explained in Section \ref{sec35}, the magnetic excitations that condense on each stratum must braid trivially with the electric condensate. 
They are, of course, generated by collections of magnetic excitations in the neighboring strata. 
One can find the magnetic condensate by finding all such collections that braid trivially with the electric condensate, which boils down to simple matrix algebra over $\mathbb{F}_2$, the field with two elements. Another method to construct the magnetic condensate is by finding all ultra-local operators that commute with the Hamiltonian on a certain stratum, while anti-commuting with flux terms on the neighboring 3-strata. These two methods are equivalent. 

Finally, we explain how the phase equivalence operations we implemented on the ungauged TDN, imply phase equivalence between the original CSS stabilizer model and the final gauged TDN. The chain complex of the gauged and ungauged models, in their strong coupling limits, can be summarized as (see Section~\ref{GaugingSpinModels}) 
\begin{align}
    R \xrightarrow{\quad r \quad} &C \xrightarrow{\quad \sigma_c \quad} Q, \quad (\text{Ungauged}) \notag\\
    C_Z \xrightarrow{\quad \hat{\sigma}_Z \quad} &G \xleftarrow{\quad \hat{\sigma}_X \quad} C_X, \quad (\text{Gauged})
\end{align}
in which $\hat{\sigma_Z} = r$ and $\hat{\sigma_X} = \sigma_c^{\dagger}$. $R$ is isomorphic to $C_Z$, $C$ is isomorphic to $G$, and $Q$ is isomorphic to $C_X$. The relation between phase equivalences of the ungauged model and phase equivalences of the gauged model essentially follows from realizing that each step in such equivalences can be rephrased as a sequence of operations on the isomorphic spaces $R\cong C_Z,C\cong G,Q\cong C_X$ that can be interpreted as phase equivalences on both gauged and ungauged models.

When we construct the TDN from an ungauged model, we apply several operations:
\begin{itemize}
    \item \textbf{Adding trivial qubits.} On the ungauged model, we add trivial qubits in the ground state of Hamiltonian Eq.~(\ref{trivialqubits}), thus the constraint map changes to $\sigma_Z \oplus \mathbbm{1}$. On the gauged model, gauge qubits are coupled to the matter qubits. The matrix of the gauged Hamiltonian also changes to $\sigma_Z^{\dagger} \oplus \mathbbm{1}$. At the same time, $C$ and $Q$ are enlarged.
    
    \item \textbf{Applying CNOT gates.} On the ungauged model, applying CNOT gates is implemented by row operation on $\sigma_c$, see Eq.~(\ref{cnotz}). 
    This corresponds to changing the basis on $Q$. 
    On the gauged model, these row operations map to column operations on $\hat{\sigma}_X$, which correspond to changing the basis on $C_X$. 
    
    \item \textbf{Adding redundant constraints.} On the ungauged model, this corresponds to adding redundant columns to $\sigma_c$, which is equivalent to enlarging $C$. On the gauged model, these operations correspond to adding redundant rows (i.e. gauge qubits). Adding redundant constraints generates additional redundant relations, which correspond to adding more columns to $r$ ($\hat{\sigma}_Z$) and $\hat{\sigma}_X$.
    \item \textbf{Applying column operations.} We apply column operations, see Eq.~(\ref{column}), to the ungauged model to make all the constraints ultra-local. 
    This corresponds to changing the basis of $C$. 
    This is equivalent to changing the basis of $G$. 
    Furthermore, changing the basis of $C$ induces a change of basis on $R$. We pick a new basis on $R$ that induces $r$ to be ultra-local. 
    This is equivalent to making $\hat{\sigma}_Z$ ultra-local. 
\end{itemize}
All the steps we perform in the construction of the TDN (gauged and ungauged) are phase preserving as they correspond to local unitary operators, the addition of auxiliary qubits, and the redefinition of the set of generators for a stabilizer group.
Hence the whole construction is also phase preserving, and therefore the resulting gauged TDN lies in the same topological quantum phase of matter as the original topological CSS stabilizer model. 

\section{TDN representations of non-CSS stabilizer codes}
\label{nonCSSTDN}

In this section, we demonstrate that the techniques discussed for CSS codes can be generalized to construct TDN representations for a large class of non-CSS fracton codes characterized by emergent gauge theory. We provide three illustrative examples.
The first example is the fermionic Haah's code, which is one member of a large family of non-CSS fracton codes obtained by gauging subsystem fermion parity symmetries of an atomic insulator state~\cite{Tantivasadakarn2020,Shirley2020}. Our method applies to all members of this class.
The second example is the semionic X-cube model~\cite{PhysRevB.95.245126}, which is a twisted (meaning the lineonic gauge fluxes have nontrivial exchange and braiding statistics) fractonic gauge theory that can be represented as a non-CSS stabilizer code~\cite{Wang2019}.
The final example is Chamon's model, which is a prototypical type I fracton model~\cite{chamon2005quantum}. It has recently been shown to exhibit an emergent fractonic gauge theory with fermionic gauge charge (i.e. it is obtained by gauging fermion parity subsystem symmetries) and, like the semionic X-cube model, nontrivial (twisted) statistics in the gauge flux sector~\cite{Shirley_Chamon}.

\subsection{Fermionic Haah's code}

The construction of the Haah code TDN proceeded by ungauging the model, coarse-graining the lattice, then expanding individual Ising spins on the ungauged level into macroscopic blocks of Ising paramagnets, which become the 3-strata of the resulting TDN upon gauging. An analogous procedure can be carried through for the fermionic Haah's code, except that instead of Ising spins, individual fermionic orbitals are expanded into macroscopic blocks of atomic insulator states. Upon gauging, these blocks become 3-strata occupied by fermionic $\mathbb{Z}_2$ topological order (i.e. with fermionic electric charge). Therefore, the TDN of the fermionic Haah's code is identical to that of the Haah code, except with the toric code 3-strata replaced by fermionic toric code 3-strata. The electric condensations of Sec. \ref{sec:Haahcond} are consistent with fermionic electric charge since an even number of charges is contained in each condensed excitation.

\subsection{Semionic X-cube model}

The semionic X-cube model is a variant of the X-cube model obtained by coupling three stacks of 2D doubled semion topological order~\cite{PhysRevB.95.245126}. The original Hamiltonian can be transformed into a non-CSS stabilizer code model via a generalized local unitary transformation~\cite{Wang2019}. The excitation content of the model is identical to that of the ordinary X-cube model in terms of fusion and mobility, however it differs in the self-exchange statistics of the elementary lineons --- a given lineon has statistics identical to those of a bound state of 2D semions living in orthogonal planes. Like the X-cube model, the semionic X-cube model exhibits an emergent fractonic gauge theory, which is obtained by gauging $\mathbb{Z}_2$ planar subsystem symmetries of a nontrivial SSPT phase. The nontrivial statistics  of the gauge flux lineons are a manifestation of the nontriviality of this SSPT.

\subsubsection{Weak SSPT}

The ungauged semionic X-cube model is in fact a \textit{weak} SSPT in the sense of Refs.~\cite{subsystemphaserel,Devakul2020}, meaning roughly that it can be obtained from a trivial symmetric product state by stacking 2D SPT states. For our purposes, it is useful to consider the following construction of a Hamiltonian $H_\text{SSPT}$ realizing this phase. The model is defined in reference to a cubic lattice $\Lambda$ and its dual lattice $\Lambda'$. The Hilbert space has the form $\mathcal{H}=\mathcal{H}_\text{bulk}\bigotimes_{P\in\Lambda'}\mathcal{H}_P$ where $P$ runs over all $x$, $y$, and $z$-oriented planes of the dual lattice. $\mathcal{H}_\text{bulk}$ contains one qubit on each site of the direct lattice $\Lambda$, whereas $\mathcal{H}_P$ is a tensor product Hilbert space whose degrees of freedom lie in the plane $P$. The specific microscopic form of $\mathcal{H}_P$ is unimportant. The Hamiltonian is
\begin{equation}
    H_\text{SSPT}=-\sum_{i\in\Lambda}X_i+\sum_{P\in\Lambda'}H_P
\end{equation}
where $H_P$ is the Hamiltonian for a Levin-Gu $\mathbb{Z}_2$ SPT state~\cite{levin2012braiding} living in $\mathcal{H}_P$. The Hamiltonian $H_P$ is symmetric under a $\mathbb{Z}_2$ operator $S_P$, however we do not consider this operator to be a physical symmetry of the system. Instead, there is one $\mathbb{Z}_2$ planar subsystem symmetry for each plane $Q$ of the direct lattice $\Lambda$, which has the form
\begin{equation}
    S_Q=S_PS_P'\prod_{i\in Q}X_i
\end{equation}
where $P$ and $P'$ are the dual lattice planes adjacent to $Q$. When all $S_Q$ symmetries are gauged, the resulting model is equivalent to the semionic X-cube under generalized local unitary transformation~\cite{Devakul2020}.

\subsubsection{TDN}

We can now construct a TDN for the semionic X-cube model by first fine-graining the ungauged model $H_\text{SSPT}$ to produce an ungauged TDN, and then gauging the planar subsystem symmetries. Hence, each qubit in $\mathcal{H}_\text{Ising}$ is expanded into a paramagnetic 3-stratum, and the Levin-Gu state in each $\mathcal{H}_P$ becomes highly fine-grained with respect to the spacing of $\Lambda$. Upon gauging the fine-grained $S_Q$ symmetries, the 3-strata are occupied by blocks of 3D toric code. Because each $S_Q$ acts on the adjacent Hilbert spaces $\mathcal{H}_P$ and $\mathcal{H}_{P'}$, each of which hosts a Levin-Gu SPT under $S_Q$, the 3-strata toric codes have boundary condition on 2-strata corresponding to a condensate of $m$ loops in which each $m$ string endpoint is attached to an otherwise confined semion. On the other hand, in the ungauged TDN a quadruple of symmetry charges in each of the four paramagnetic 3-strata adjacent to a given 1-strata is uncharged under all of the fine-grained $S_Q$ symmetries, as in the case of the ungauged X-cube TDN. Therefore, in the gauged TDN, the 1-strata are characterized by condensation of $e_1e_2e_3e_4$ composites. In summary, the 
semionic X-cube TDN is identical to the X-cube TDN of Sec. \ref{sec:XcubeExample}, except that the $m$ loop-condensing boundary conditions on 2-strata are replaced by their twisted counterparts~\cite{TGTboundary}. As in the X-cube TDN, the fracton excitation of the semionic X-cube model in this TDN construction can be identified as a single electric charge excitation of a particular 3-stratum. The lineon excitation is likewise identified as a short gauge flux string segment in the vicinity of a particular 1-stratum. However, in this case a flux loop can only terminate on a 2-stratum if it is bound to a semion. Therefore the lineon inherits nontrivial statistics arising from the bound state of two semions in orthogonal layers.

An alternative way to understand this TDN is via the following condensation procedure. We begin with the ordinary X-cube TDN, and stack a 2D doubled semion layer onto each plane of the stratification. Then, we condense all excitations of the form $e_-be_+$ on the 2-strata. Where $e_\pm$ are the electric charges living on either side of a given 2-stratum, and $b$ is the boson of the doubled semion layer on that 2-stratum. The state prior to condensation can be understood as a TDN composed of blocks of 1) 3D toric code on 3-strata and 2) 2D doubled semion layers on 2-strata, glued together via condensation of the following excitations:
\begin{align*}
    \text{2-strata} : \big\langle &m_-,m_+ \big\rangle \\
    \text{1-strata} : \big\langle &e_1e_2e_3e_4, m_1m_2, m_2m_3, m_3m_4,\\
    & s_{12}s_{34}, \bar{s}_{12}\bar{s}_{34}, s_{23}s_{41}, \bar{s}_{23}\bar{s}_{41} \big\rangle
    %\{&e_1e_2e_3e_4, m_1m_2, m_2m_3, m_3m_4, m_4m_1,\\
    %&b_{12}b_{34}, b_{23}b_{41}, s_{12}s_{34}, \bar{s}_{12}\bar{s}_{34}, s_{23}s_{41}, \bar{s}_{23}\bar{s}_{41}\}
\end{align*}
where $m_\pm$ are the gauge flux loops living above or below the given 2-stratum, $e_i$ and $m_i$ the excitations of the $i$th 3-stratum adjacent to the given 1-stratum ($i=1,2,3,4)$, and $s_{ij}$ ($\bar{s}_{ij}$) the semion (anti-semion) on the 2-stratum between the $i$th and $j$th 3-strata. The 2-strata condensations and first line of 1-strata condensation simply encode the X-cube TDN, whereas the second line of 1-strata condensation simply encodes the fact that the blocks of 2D doubled semion order within a given plane are all connected to form a single doubled semion layer.\footnote{Note that $s_{12}s_{34}$ becomes a boson after folding the layers meeting at a 1-strata to form a gapped boundary to vacuum~\cite{Aasen2020}.}

The condensation of all $e_-be_+$ excitations on 2-strata has the following consequences. First, all individual semions $s$ and antisemions $\bar{s}$ become confined due to the nontrivial braiding with $e_-be_+$. However, bound states of semions (or antisemions) attached to the endpoints of an $m_-$ or $m_+$ loop terminating on a 2-stratum survive as deconfined excitations. In fact, $m$ loops must be bound to semions (or antisemions) at their endpoints due to the nontrivial braiding of a bare loop with $e_-be_+$, hence modifying the statistics of the lineon excitations. Finally, the 1-strata condensate is reduced to the subset of topological excitations that braid trivially with $e_-be_+$:
\begin{align*}
    \big\langle e_1e_2e_3e_4,m_1m_2s_{41}s_{23},m_2m_3s_{12}s_{34},&m_3m_4s_{23}s_{41} \big\rangle \, .
\end{align*}
Due to the condensation of $e_-be_+$ excitations on 2-strata, the bosons $b$ from the doubled semion layers may still pass through the 1-strata, i.e. $b_{12}b_{34}, b_{23}b_{41},$ are condensed there. 
This further implies that composite excitations formed by pairs of magnetic flux lines with antisemions replacing the semions in the condensate above, such as $m_1m_2\bar{s}_{41}\bar{s}_{23}$, condense on the 1-strata.

\subsection{Chamon's model}

\begin{figure}
    \centering
    \includegraphics[width=5cm]{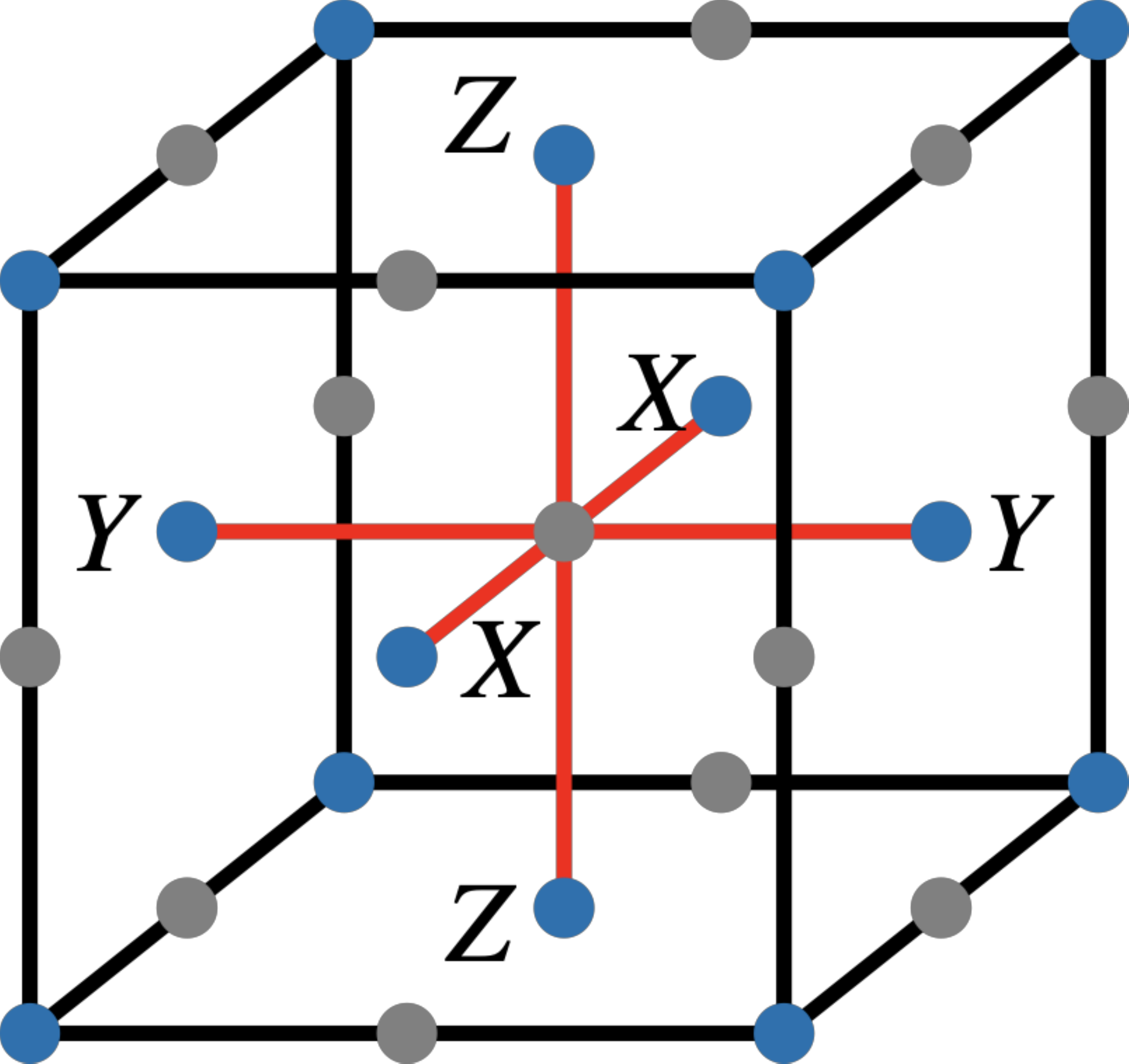}
    \caption{Chamon's fracton model on the FCC lattice. There is one qubit on every blue dot and the 6-body stabilizer term, as shown, is defined on every vertex indicated by a gray dot.}
    \label{fig:Chamon}
\end{figure}

Chamon's model~\cite{chamon2005quantum} was the first topological fracton model to appear in the literature. 
It is a qubit model defined on an FCC lattice with one qubit per site. The Hamiltonian takes the form $H=-\sum_i O_i$, where $O_i$ is the multi-qubit Pauli operator depicted in Fig. \ref{fig:Chamon}. These terms mutually commute and have an unfrustrated ground space, hence this Hamiltonian constitutes a non-CSS stabilizer code. The basic properties of this model, such as ground state degeneracy and types of fractonic excitations, are discussed in detail in Ref.~\cite{bravyi2011topological}.

\begin{figure}
    \centering
    \includegraphics[width=5cm]{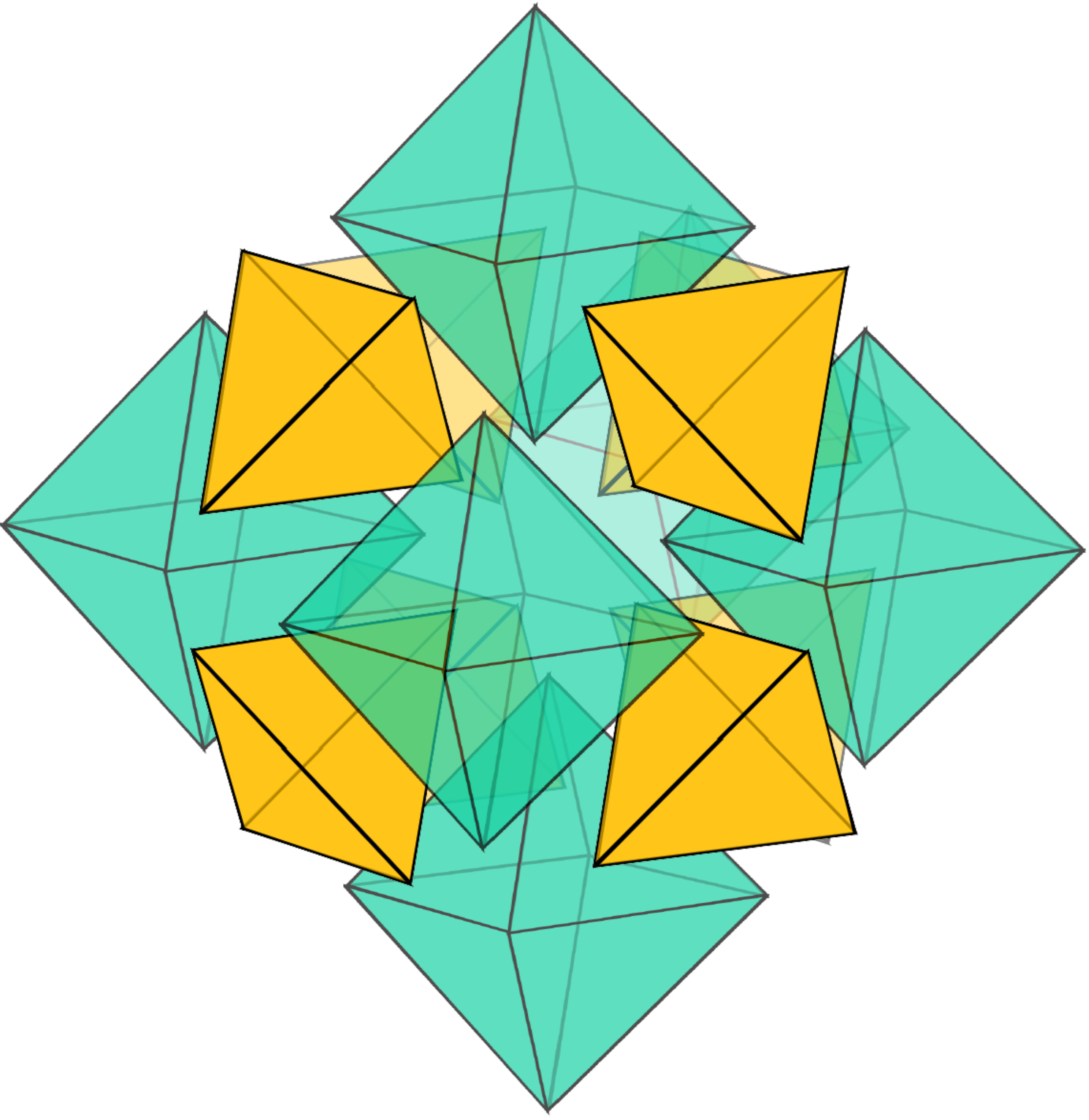}
    \caption{Tetrahedral-octahedral honeycomb stratification used in the construction of a TDN for Chamon's model.}
    \label{fig:TetraOcta}
\end{figure}

Recently, a generalized local unitary equivalence was discovered~\cite{Shirley_Chamon} between Chamon's model and a 4-foliated X-cube model~\cite{Slagle18,shirley2018fractional} variant $H_\text{4-fol}$ characterized by an emergent fractonic gauge theory with fermionic gauge charges and nontrivial (twisted) lineonic gauge flux exchange statistics. 
Similar to the case of the semionic X-cube model, this emergent gauge theory is dual to a weak SSPT state, meaning it can be obtained by stacking 2D invertible states onto an trivial symmetric product state. This structure enables the construction of a TDN representation as we describe below.

\subsubsection{Weak SSPT}

We will consider the following construction of a Hamiltonian $H_\text{SSPT}$ dual to $H_\text{4-fol}$~\cite{Shirley_Chamon}. The model is defined in reference to a tetrahedral-octahedral honeycomb cellulation $\Lambda$ (see Fig. \ref{fig:TetraOcta}). The Hilbert space has the form $\mathcal{H}=\mathcal{H}_\text{bulk}\bigotimes_{P\in\Lambda}\mathcal{H}_P$ where $P$ runs over all planes of the honeycomb. $\mathcal{H}_\text{bulk}$ contains a qubit attached to each tetrahedral 3-cell of $\Lambda$ and a fermionic orbital attached to each octahedral 3-cell of $\Lambda$, whereas $\mathcal{H}_P$ is a fermionic tensor product Hilbert space whose degrees of freedom lie in plane $P$ of the honeycomb. The specific microscopic form of $\mathcal{H}_P$ is unimportant. The Hamiltonian is
\begin{equation}
    H_\text{SSPT}=-\sum_{c\in T}X_c-\sum_{c\in O}i\gamma_c\gamma'_c+\sum_{P\in\Lambda}H_P
\end{equation}
where $T$ ($O$) runs over all tetrahedral (octahedral) 3-cells, $\gamma_c$ and $\gamma'_c$ refer to the two Majorana operators on orbital $c$, and $H_P$ is the Hamiltonian of an invertible fermionic topological order living in $\mathcal{H}_P$. The invertible topological order described by $H_P$ is dual to the $\nu=4$ ($\nu=-4$) Kitaev sixteenfold way state~\cite{kitaev2006anyons} for even (odd) plane $P$. There is one $\mathbb{Z}_2$ planar subsystem symmetry for each slice $Q$ of the honeycomb $\Lambda$, which has the form
\begin{equation}
    S_Q=\mathbb{Z}_2^P\mathbb{Z}_2^{P'}\prod_{c\in Q\cap T}X_c\prod_{c\in Q\cap O}i\gamma_c\gamma_c'
\end{equation}
where $P$ and $P'$ are the honeycomb planes enclosing the slice $Q$, and $\mathbb{Z}_2^P$ refers to the fermion parity of $\mathcal{H}_P$. When all $S_Q$ symmetries are gauged, the resulting model is equivalent to $H_\text{4-fol}$ and hence Chamon's model under generalized local unitary transformation.\footnote{The set of $S_Q$ symmetries generate the global fermion parity, hence upon gauging the model no longer has physical fermions.}

\subsubsection{TDN}

We can now construct a TDN for Chamon's model by first fine-graining the ungauged model $H_\text{SSPT}$ to produce an ungauged TDN, and then gauging the planar subsystem symmetries. The stratification of space is thus given by the tetrahedral-octahedral honeycomb on which $H_\text{SSPT}$ is defined. This TDN combines nontrivial aspects of both of the previous examples: fermionic 3D toric code in (some of) the 3-strata, and twisted boundary conditions along 2-strata. Note that each unit cell of the honeycomb actually corresponds to a $2\times2\times2$ unit cell of Chamon's model~\cite{Shirley_Chamon}. Upon gauging the fine-grained $S_Q$ symmetries, the tetrahedral (octahedral) 3-strata are occupied by blocks of bosonic (fermionic) 3D toric code. Because each $S_Q$ acts on the adjacent Hilbert spaces $\mathcal{H}_P$ and $\mathcal{H}_{P'}$, each of which hosts an invertible fermionic topological order dual to the $\nu=\pm4$ Kitaev state, the 3-strata toric codes have boundary condition on 2-strata corresponding to a condensate of $m$ loops in which each $m$ string endpoint is attached to an otherwise confined semion or anti-semion (depending on whether $P$ is an even or odd layer). On the other hand, in the ungauged TDN a quadruple of symmetry charges in each of the four paramagnetic 3-strata adjacent to a given 1-strata is uncharged under all of the fine-grained $S_Q$ symmetries. 
After gauging, this leads to a quadruple of gauge charges that has trivial total gauge charge, i.e. a cluster that is topologically trivial and can be created locally. 
Hence in the gauged TDN the 1-strata are characterized by condensation of $e_1e_2e_3e_4$ composites, where $e_i$ is either fermionic or bosonic depending on the 3-stratum.

The intuition behind this TDN is similar to that of the semionic X-cube model. Individual electric charge excitations of a particular 3-stratum are fractons, which are either bosonic or fermionic depending on the 3-stratum. Short gauge flux string segments in the vicinity of a particular 1-stratum are lineons. In this case a flux loop can only terminate on a 2-stratum if it is bound to a semion or anti-semion of the 2D topological order on that stratum. Therefore the lineon inherits nontrivial statistics arising from the bound state of two anyons in orthogonal layers. It can be shown that the topological excitations of this TDN, in terms of fusion rules and statistics, are identical to those of Chamon's model (see Ref.~\cite{Shirley_Chamon} for details).

We now briefly sketch a condensation procedure that reproduces this TDN. First, we construct a TDN corresponding to the untwisted version of $H_\text{4-fol}$, i.e. with trivial lineon statistics. This TDN is identical to the X-cube TDN except that 1) it is defined on the tetrahedral-octahedral honeycomb rather than a simple cubic stratification, and 2) the octahedral 3-strata are occupied by fermionic 3D toric code rather than ordinary 3D toric code. We then stack onto this TDN a $\nu=4$ Kitaev state on each even honeycomb plane, and a $\nu=-4$ Kitaev state on each odd honeycomb plane. Then, we condense all excitations of the form $e_-fe_+$ where $e_\pm$ are the electric charges living on either side of a given 2-stratum, and $f$ is the fermion of Kitaev state on that 2-stratum (recall that the $\nu=4$ state is the semion-fermion topological order, whereas the $\nu=-4$ state is the antisemion-fermion topological order). The $e_-fe_+$ excitations are always bosonic since each 2-stratum is straddled by one fermionic and one bosonic 3D toric code. The result of this condensation is to 1) confine all individual semions or antisemions, such that these Kitaev states no longer exist as separate topological orders but rather as twisted boundary conditions for the adjacent toric codes, and 2) attach a semion or antisemion to the endpoints of each $m$ loop terminating on a given 2-stratum, thus yielding the TDN described above. 
The condensate on 1-strata is of the same form as the semionic X-cube described above, with some $e$ charges being replaced by fermionic $e$ charges, and the semions being replaced by the semions in the $\nu=\pm 4$ Kitaev states. 

\section{Conclusion}
\label{sec:Conclusion}

In this work, we have proposed a general method to construct topological defect networks for a wide range of topological lattice Hamiltonians. 
We provided a general recipe that produces a phase equivalent TDN from any topological CSS stabilizer Hamiltonian. 
Our general recipe was applied to produce a TDN for Haah's cubic code, the canonical type-II example for which no TDN was previously known. 
We additionally proposed a TDN for Chamon's non-CSS fracton code, the first topological fracton model to be discovered. 
Our focus was restricted to models with prime-dimensional qudits but we forsee no obstacle to generalizing our TDN constructions to models with qudits of non-prime dimensions. 

Our results provides a new point of view on topological CSS stabilizer codes that brings the TDN framework to bear on important questions about their structure and code properties. 
We believe this will lead to valuable insights, just as ideas from TQFT have for 2D codes~\cite{qdouble,haah}. 

Interestingly, our construction generalizes directly to any phase of matter that can be obtained by gauging (potentially fermionic) subsystem symmetries of stacked (possibly subdimensional) topological quantum field theory layers\footnote{More generally, the TQFT layers can be replaced by any states that are entanglement renormalization group fixed points.}. 
To the best of our knowledge this covers almost all known gapped fracton models whose excitations have finite order under fusion, including non-Abelian models~\cite{vijay2017generalization,prem2018cage,song2018twisted,Bulmash2019,Prem2019,Williamson2020Designer,Stephen2020Subsystem,Sullivan2021Planar,Tantivasadakarn2021Non}. 
This general construction supports the following refinement of the conjecture posed in Ref.~\cite{Aasen2020}: \textit{TDNs can realize all zero temperature gapped phases of matter whose excitations have finite order under fusion}. 
It is notable that even for a relatively simple non-CSS Pauli stabilizer Hamiltonian such as Chamon's model, our general recipe may produce a TDN that involves far more complex ingredients that are beyond Pauli stabilizer models. 
This differs from the CSS case we studied, in which the general construction produces a CSS Pauli stabilizer Hamiltonian for the TDN which has essentially the same level of complexity as the original Hamiltonian, only more degrees of freedom. 

There are, however, models that are expected to lie beyond the class for which our recipe works and hence provide an important testing ground for the above conjecture. The gauged strong subsystem symmetry protected phases introduced in Ref.~\cite{Devakul2020} provide explicit examples, which have so far not been realized via a TDN. 

For the description of models with topological excitations that have infinite order under fusion, and/or local degrees of freedoms that have an infinite dimensional local Hilbert space, we expect a generalization of the TDN construction to be necessary. 
We envision such a construction to be built from well understood components, such as gauge theories based on continuous Lie groups, the simplest example being U(1), on the 3-strata, and appropriate defects therein on the lower dimensional strata. 
Working out the details of such a framework, and whether this is capable of describing all gapped phases in three (and higher) spatial dimensions, specifically those with infinite order excitations, remains an open question. 

\section*{Acknowledgments}
A.D.~acknowledges discussions with Meng Cheng, Qing-Rui Wang and Kevin Slagle. 
D.W.~acknowledges useful discussions with Daniel Bulmash and support from the Simons Foundation. 
A.D. and W.S. acknowledge support from the Institute for Quantum Information and Matter, an NSF Physics Frontiers Center (PHY-1733907) and from the Simons Collaboration on Ultra-Quantum Matter which is a grant from the Simons Foundation (651438, AD; 651444, WS).

\appendix
\section{Further TDN examples}

Our TDN construction can be applied to an arbitrary topological CSS stabilizer Hamiltonian. 
Yoshida's fractal spin liquid models~\cite{yoshida2013exotic} provide an infinite family of fracton CSS stabilizer Hamiltonians to which our construction applies. 
In this appendix we describe TDNs for two specific fractal spin liquid models, a simple example, previously considered in Ref.~\cite{Aasen2020}, and a more nontrivial type-II example, that is equivalent to Haah's cubic code~\cite{yoshida2013exotic}. 

\subsection{Yoshida's first order fractal spin liquid}

The Hamiltonian of the simplest of Yoshida's first order fractal spin liquids is given by,
\begin{align}\label{fsl}
    H_{FSL} = &- \sum_{c} \adjincludegraphics[width=2cm,valign=c]{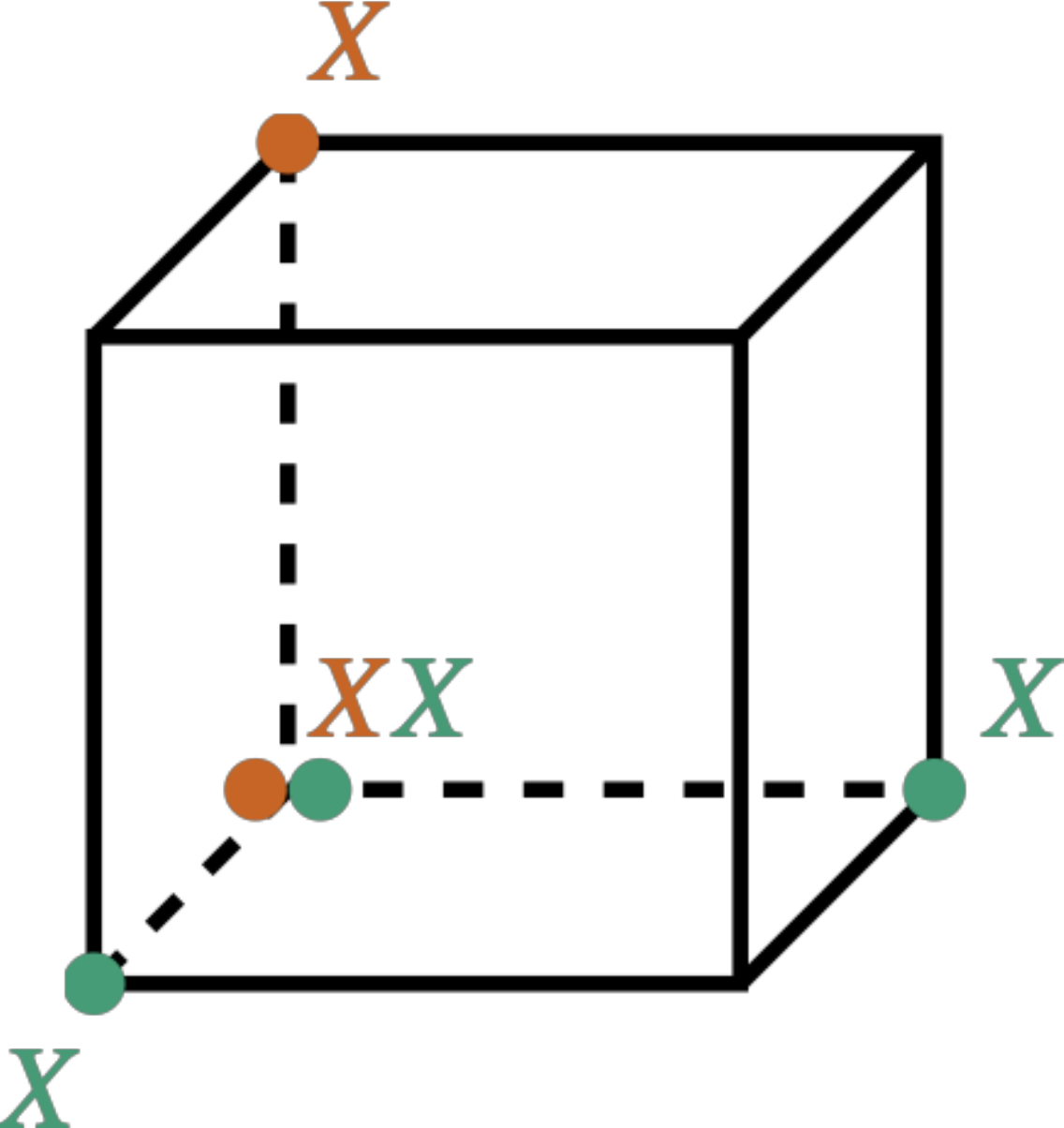} - \sum_{c} \adjincludegraphics[width=1.8cm,valign=c]{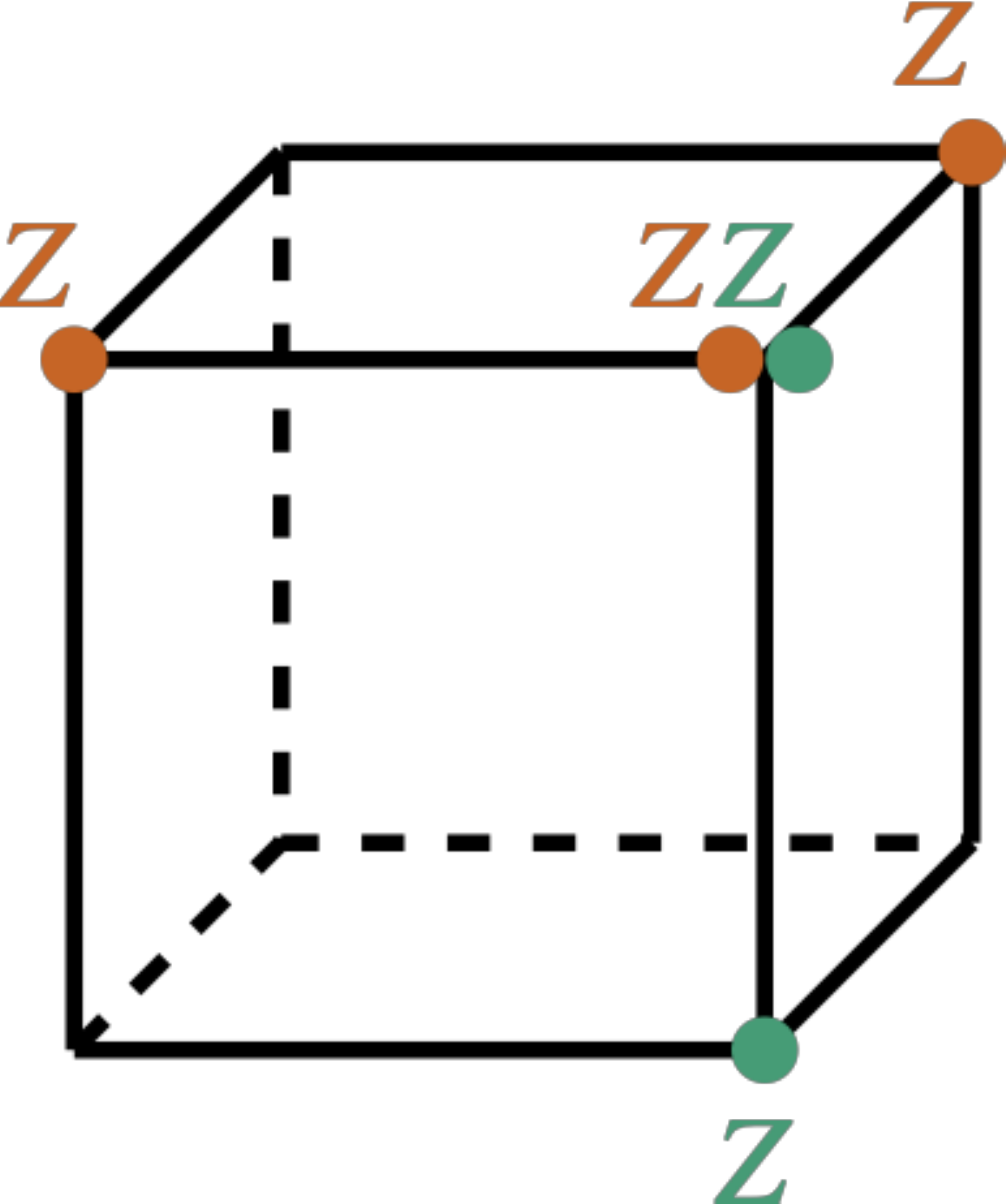}
\end{align}
The sums are over all cubes. The constraints of the ungauged FSL are thus given by
\begin{align}
\adjincludegraphics[width=5cm,valign=c]{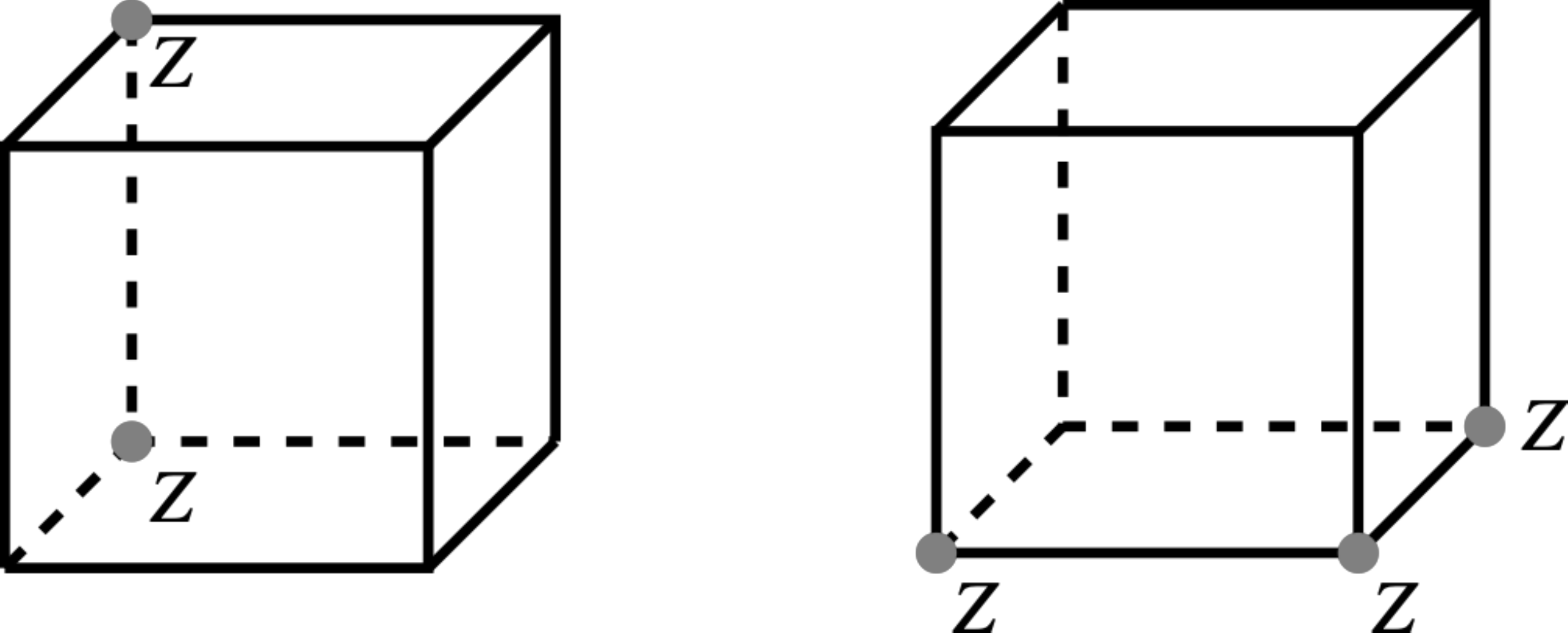}
\end{align}
By applying the gauging map we introduce in Section \ref{Gauging1} and the general recipe of constructing TDNs in Section \ref{generalapproach}, we get the non-trivial Gauss's law terms near the $z$-oriented 1-Strata, which are given by
\begin{align}
\adjincludegraphics[width=6cm,valign=c]{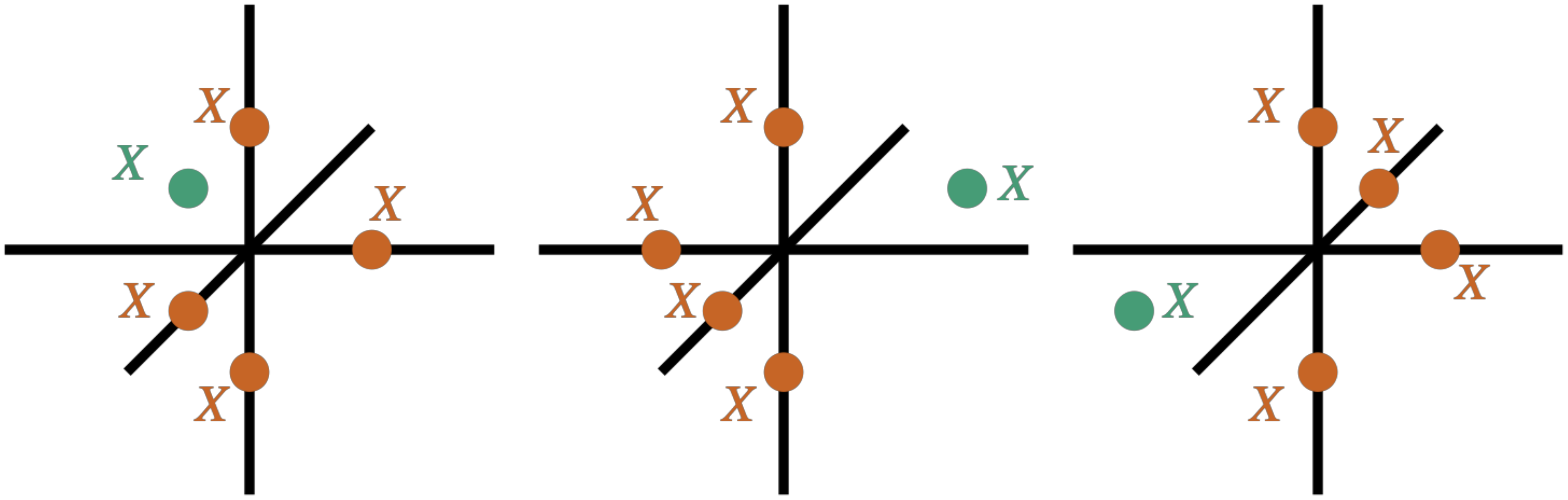}
\end{align}
The green gauge qubits are on the center of plaquettes. The flux terms are given by
\begin{align}
\adjincludegraphics[width=2cm,valign=c]{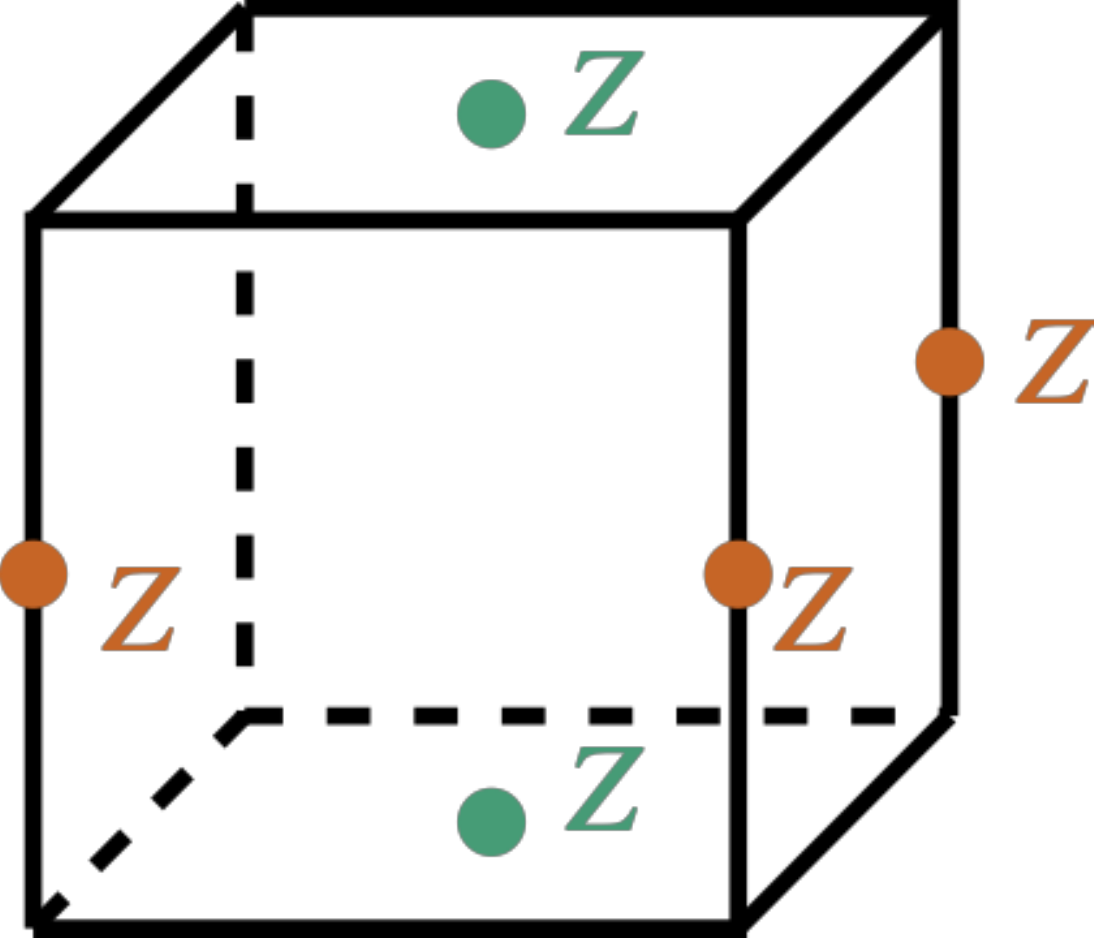},
\end{align}
which are exactly the same as the flux terms of Eq.~(\ref{fsl}).

By applying the method we introduce in Section \ref{sec35}, the condensations on the $z$-oriented 1-Strata are given by
\begin{align}
    \big\langle e_2 e_3 e_4, m_2 m_3, m_3 m_4, m_1\big\rangle. \notag
\end{align}
On the $xy$-oriented 2-Strata, all Hamiltonian terms are the same as the bulk. Therefore, the condensation are given by
\begin{align}
    \big\langle e_+ e_-, m_+ m_-\big\rangle. \notag
\end{align}
$+$ and $-$ are the labels for two neighbouring 3-Strata. As we don't have constraint terms on other 2-Strata, the condensations there are  simply given by
\begin{align}
    \big\langle m_+, m_-\big\rangle.\notag
\end{align}
There are no further condensations on other 1-Strata.

\subsection{Cubic code in FSL form}

\begin{figure}
    \centering
    \vspace{2mm}
    \includegraphics[width=6cm]{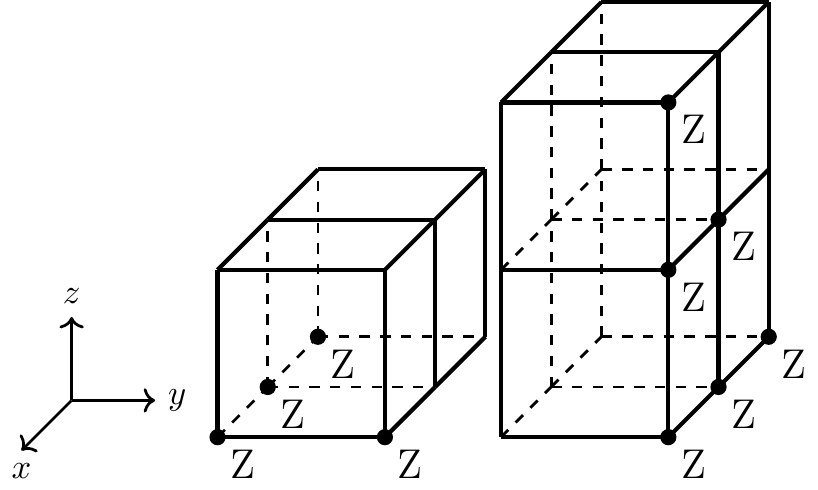}
    \caption{The constraint terms of the ungauged CC1 in FSL form}
    \label{Z_FSL}
\end{figure}

One can also write Haah's Cubic Code 1 A in FSL form. The constraint terms of the ungauged model are given by the matrix,
\begin{align}
    \sigma_c = \left(\begin{array}{cc}
           1 + x + x^2 + x^2 y & 1 + x + x^2 + xz + x^2 z + x^2 z^2
    \end{array}\right),
\end{align}
which are shown in FIG~\ref{Z_FSL}. We notice these terms are not nearest-neighbour. So to get the TDN, according to Section \ref{generalapproach}, we first coarse-grain it along $x$ and $z$ directions as shwon in FIG~\ref{Z_FSL}. 
The coarse-grained constraint terms are given by
\begin{widetext}
\begin{align}
    \sigma_c \to \left(\begin{array}{cccccccc}
           1+x & 0 & x+xy & 0 & 1+x+xz & 0 & x+xz & xz \\
           0 & 1+x & 0 & x+xy & 0 & 1+x+xz & x & x+xz \\
           1+y & 0 & 1+x & 0 & 1+z & z & 1+x+xz & 0 \\
           0 & 1+y & 0 & 1+x & 1 & 1+z & 0 & 1+x+xz
    \end{array}\right).
\end{align}
\end{widetext}
We notice that the first 4 columns only depend on $x$ and $y$, so these terms become ultra-local on $z$-oriented 1-strata. By applying the method we introduce in Section \ref{sec35}, electric condensations on $z$-oriented 1-strata are given by
\begin{align*}
    \big\langle &e_1^c e_1^d e_3^c e_3^d, e_1^b e_1^d e_3^c e_3^d, e_2^c e_2^d e_4^c e_4^d, e_2^b e_2^d e_4^c e_4^d\big\rangle.
\end{align*}
Same as the labels we use in Section \ref{sec:Haahcond}. $\{a,b,c,d,e,f,g,h\}$ are the labels of 3-strata and $\{1,2,3,4\}$ are the labels of the layers inside each 3-stratum. Here the reason why we only have 4 layers rather than 8 is that we only coarse-grain the lattice along two different directions $x$ and $z$.

The last 4 columns only depend on $x$ and $z$, so these terms become ultra-local on $y$-oriented 1-strata. Similarly, the corresponding electric condensations are given by
\begin{align*}
    \big\langle &e_1^a e_1^b e_1^f e_3^a e_3^e e_4^a, e_2^a e_2^b e_2^f e_3^e e_4^a e_4^e, e_1^b e_1^f e_2^b e_3^a e_3^b e_3^f,\notag \\
    &e_1^f e_2^b e_2^f e_4^a e_4^b e_4^f\big\rangle. \notag
\end{align*}
According to Section \ref{sec35} we know that the magnetic condensations braid with the electric condensations trivially. Thus for the $z$-oriented 1-strata, the magnetic condensations are given by
\begin{align*}
    \big\langle &m_1^d m_3^a m_3^d, m_2^d m_4^a m_4^d, m_3^a m_3^c m_3^d, m_4^a m_4^c m_4^d, m_1^c m_3^a, \nonumber \\
    &m_1^b m_3^d, m_2^c m_4^a, m_2^b m_4^b, m_1^a, m_2^a, m_3^b, m_4^b \big\rangle.
\end{align*}
For the $y$-oriented 1-strata, they are given by
\begin{align*}
    \big\langle &m_1^b m_3^f m_4^a m_4^e m_4^f, m_3^a m_3^f m_4^a m_4^e m_4^f, m_1^a m_4^a m_4^e m_4^f,\nonumber\\
    &m_1^f m_3^f m_4^a m_4^e, m_2^b m_3^f m_4^e m_4^f, m_2^f m_4^e m_4^f, m_3^e m_4^a m_4^f,\nonumber\\
    & m_2^a m_4^e, m_3^b m_3^f, m_4^b m_4^f, m_1^e, m_2^e\big\rangle.
\end{align*}
Since there is no constraint terms become ultra-local on 2-strata, similar to the X-cube model in Section \ref{sec35}, the condensations on 2-strata are just trivial magnetic fluxes, which are given by
\begin{align*}
    \big\langle m_1^+,m_2^+,m_3^+,m_4^+,m_1^-,m_2^-,m_3^-,m_4^-\big\rangle,
\end{align*}
where $+$($-$) refer to two neighboring 3-strata and $\{1,2,3,4\}$ are the labels of layers inside each 3-stratum.

In conclusion, the condensations of CC1 in FSL form are given by 
\begin{align*}
    \text{1-strata ($y$)} : \big\langle &e_1^a e_1^b e_1^f e_3^a e_3^e e_4^a, e_2^a e_2^b e_2^f e_3^e e_4^a e_4^e, e_1^b e_1^f e_2^b e_3^a e_3^b e_3^f,\\ &e_1^f e_2^b e_2^f e_4^a e_4^b e_4^f, m_1^b m_3^f m_4^a m_4^e m_4^f, m_1^a m_4^a m_4^e m_4^f,\\ 
     &m_3^a m_3^f m_4^a m_4^e m_4^f, m_1^f m_3^f m_4^a m_4^e, m_2^b m_3^f m_4^e m_4^f, \\&m_2^f m_4^e m_4^f, m_3^e m_4^a m_4^f, m_2^a m_4^e, m_3^b m_3^f, m_4^b m_4^f,\\ &m_1^e, m_2^e\big\rangle \\
    \text{1-strata ($z$)} : \big\langle &e_1^c e_1^d e_3^c e_3^d, e_1^b e_1^d e_3^c e_3^d, e_2^c e_2^d e_4^c e_4^d, e_2^b e_2^d e_4^c e_4^d,m_1^c m_3^a,\\ &m_1^d m_3^a m_3^d, m_2^d m_4^a m_4^d, m_3^a m_3^c m_3^d, m_4^a m_4^c m_4^d, \\
    &m_1^b m_3^d, m_2^c m_4^a, m_2^b m_4^b, m_1^a, m_2^a, m_3^b, m_4^b\big\rangle\\
    \text{2-strata} : \big\langle&m_1^+,m_2^+,m_3^+,m_4^+,m_1^-,m_2^-,m_3^-,m_4^-\big\rangle.
\end{align*}
There is no condensations on $x$ oriented 1-strata, 3-strata and 0-strata.

\bibliography{mainBib}

\end{document}